\documentclass[12pt]{article}

\usepackage{tikz-feynman}
\usepackage{jheppub}
\usepackage{hyperref}
\usepackage{footmisc}
\usepackage{comment}
\usepackage{subcaption}
\usepackage{cancel}

\newcommand{\intnsinf}{\int_{-\infty}^{0}}

\newcommand{\hd}{\hat \Delta}

\usepackage{bm, bbm, bbold}
\usepackage{latexsym}
\usepackage{dcolumn}
\usepackage{amsmath,amsfonts,amssymb, mathrsfs, amsthm}
\usepackage[T1]{fontenc}

\usepackage{graphicx,epsfig}
\usepackage{environ}
\usepackage{mathtools}
\usepackage{braket}
\usepackage{fancyhdr}
\usepackage{hyperref}
\usepackage{graphicx,epstopdf}
\usepackage{tikz}
\usepackage{float}
\usepackage{framed}
\usepackage{soul}
\usetikzlibrary{positioning,decorations.markings, decorations.pathmorphing, calc}
\tikzset{snake it/.style={decorate, decoration=snake}}
\usetikzlibrary{angles,quotes}

\usepackage{stmaryrd}
\usepackage{slashed}
\usepackage{array,multirow}

\usepackage[framemethod=default]{mdframed}
\newmdenv[skipabove=7pt,
skipbelow=7pt,
rightline=false,
leftline=false,
topline=false,
bottomline=false,
backgroundcolor=gray!10,
linecolor=gray,
innerleftmargin=5pt,
innerrightmargin=5pt,
innertopmargin=5pt,
innerbottommargin=5pt,
leftmargin=0cm,
rightmargin=0cm,
linewidth=4pt]{eBox}

\newcommand{\dagg}{\dagger}

\def\a{\alpha}
\def\b{\beta}

\def\e{\epsilon}

\def\T{\Theta}
 
\def\p{\partial}

\newcommand{\be}{\begin{equation}}
\newcommand{\ee}{\end{equation}}
\newcommand{\bes}{\begin{equation*}}
\newcommand{\ees}{\end{equation*}}

\NewEnviron{derivation}{
\begin{framed}
\begin{center}
{\bf-: Derivation :-}\\
\end{center}
  \BODY

\end{framed}
}

\NewEnviron{eqn}{
\begin{align}
\begin{split}
  \BODY
\end{split}
\end{align}
}

\NewEnviron{eqn*}{
\begin{align*}
\begin{split}
  \BODY
\end{split}
\end{align*}
}

\newcommand{\nno}{\nonumber}
\newcommand{\intinf}{\int_{-\infty}^{\infty}} 
\newcommand{\intsinf}{\int_{0}^{\infty}} 

\newcommand{\Res}[1]{\underset{#1}{\mbox{Res}}}

\usepackage{cancel}
\usepackage{tikz, pgf}
\usetikzlibrary{shapes.misc}
\usetikzlibrary{shapes}
\usetikzlibrary{fit}
\usetikzlibrary{arrows}

\usetikzlibrary{decorations.pathmorphing}	
\usetikzlibrary{decorations.markings}
\tikzset{
    sugra/.style={decorate, decoration={snake}, draw=black},
    scalarphi/.style={dashed,draw=black, postaction={decorate},
        },
    hwbou/.style={draw=blue, postaction={decorate}, ultra thick
        },
    vector/.style={draw=blue,decorate, decoration={snake}, draw},
	provector/.style={decorate, decoration={snake,amplitude=2.5pt}, draw},
	antivector/.style={decorate, decoration={snake,amplitude=-2.5pt}, draw},
   	 fermion/.style={draw=cyan, postaction={decorate},
        decoration={markings,mark=at position .55 with {\arrow[draw=black]{>}}}},
    fermionbar/.style={draw=cyan, postaction={decorate},
        decoration={markings,mark=at position .55 with {\arrow[draw=black]{<}}}},
    fermionnoarrow/.style={draw=black},
    gluon/.style={decorate, draw=red,
        decoration={coil,amplitude=4pt, segment length=5pt}},
    scalar/.style={dashed,draw=black, postaction={decorate},
        decoration={markings,mark=at position .55 with {\arrow[draw=black]{>}}}},
    scalarbar/.style={dashed,draw=black, postaction={decorate},
        decoration={markings,mark=at position .55 with {\arrow[draw=black]{<}}}},
    electron/.style={draw=black, postaction={decorate},
        decoration={markings,mark=at position .55 with {\arrow[draw=black]{>}}}},
    scalarnoarrow/.style={dashed, draw=black},
    electron/.style={draw=black, postaction={decorate},
        decoration={markings, mark=at position .55 with {\arrow[draw=black]{>}}}},
	bigvector/.style={decorate, decoration={snake, amplitude=4pt}, draw},
    photon/.style={draw=red, decorate, decoration={snake}, draw},
    higgs/.style={dashed, draw=black, postaction={decorate},
        },	
        goldstone/.style={draw=brown, postaction={decorate},
        },    
          ghost/.style={dashed, draw=magenta, postaction={decorate},
        decoration={markings, mark=at position .55 with {\arrow[draw=black]{>}}}
        },  
          antighost/.style={dashed, draw=magenta, postaction={decorate},
        decoration={markings, mark=at position .55 with {\arrow[draw=black]{<}}}
        },  
          mphoton/.style={decorate, decoration={snake}, draw=violet},
            realscalar/.style={draw=black}, 
           mgluon/.style={decorate, draw=blue,
        decoration={coil,amplitude=4pt, segment length=5pt}},
         weylfermion/.style={draw=orange, postaction={decorate},
        decoration={markings,mark=at position .55 with {\arrow[draw=black]{>}}}},
         weylfermionbar/.style={draw=orange, postaction={decorate},
        decoration={markings,mark=at position .55 with {\arrow[draw=black]{<}}}}, 
   	wboson/.style={draw=blue,decorate, decoration={snake,amplitude=4pt}, draw},  
    zboson/.style={draw=violet, decorate, decoration={snake}, draw},   
    lepton/.style={draw=black, postaction={decorate},
        decoration={markings,mark=at position .55 with {\arrow[draw=black]{>}}}},
    leptonbar/.style={draw=black, postaction={decorate},
        decoration={markings,mark=at position .55 with {\arrow[draw=black]{<}}}}, 
        graviton/.style={draw=black,decorate, decoration={snake, amplitude=.7mm, segment length=1.5mm, pre length=.5mm, post length=.5mm}, double},  
        gravitino/.style={draw=red, postaction={decorate},
        decoration={snake, markings, mark=at position .55 with {\arrow[draw=black]{>}}}},
    gravitinobar/.style={draw=red, postaction={decorate},
        decoration={snake, markings,mark=at position .55 with {\arrow[draw=black]{<}}} },    
}

\newcommand{\vertex}[1]{\begin{tikzpicture}[baseline]
\node at (0,0) {\textbullet};
\node at (0,-0.25) {$#1$};
\end{tikzpicture}}
\allowdisplaybreaks 

\begin{document} 

\title{On in-in correlators for spinning theories and their shadow formulation}  
\author[a]{Chandramouli Chowdhury,}
\author[b]{Arthur Lipstein,}
\author[b]{Joe Marshall, }
\author[b]{Alex Jiayi Zhang}
\affiliation[a]{Mathematical Sciences and STAG Research Centre, University of Southampton, Highfield,
Southampton SO17 1BJ, United Kingdom}
\affiliation[b]{Department of Mathematical Sciences, Durham University, Stockton Road, DH1 3LE, Durham, United Kingdom}

\abstract{In-in correlators are the basic observables in cosmology and are traditionally computed using the Schwinger-Keldysh formalism. In this paper we revisit this formalism for photons, gluons, and gravitons coupled to scalars in four dimensional de Sitter space and provide a novel treatment of boundary gauge-fixing of the underlying path integral. We also derive effective actions in Euclidean Anti-de Sitter space whose Feynman rules compute the in-in correlators of these theories in de Sitter space, generalising the shadow formalism recently obtained for scalar theories. We illustrate this formalism in a number of examples at tree-level and 1-loop, demonstrating its relative simplicity compared to other approaches. Moreover, we initiate the study of color/kinematics duality and the double copy for in-in correlators using dressing rules which uplift flat space Feynman diagrams to de Sitter space.}

\maketitle

\noindent
\flushbottom
\allowdisplaybreaks

\section{Introduction}
Recent years have revealed deep connections between mathematical structures of scattering amplitudes and analogous objects in cosmology known as wavefunction coefficients. These include bootstrap methods based on factorisation \cite{Raju:2012zr,Raju:2012zs,Arkani-Hamed:2015bza,Arkani-Hamed:2018kmz,Goodhew:2020hob,Melville:2021lst}, symmetries \cite{Maldacena:2002vr,Maldacena:2011nz,Bzowski:2013sza,Bzowski:2019kwd,Bzowski:2020kfw,Sleight:2019mgd}, differential equations \cite{Arkani-Hamed:2023kig,Arkani-Hamed:2023bsv,Caloro:2023cep}, geometric formulations \cite{Arkani-Hamed:2017fdk,Arkani-Hamed:2024jbp}, scattering equations \cite{Roehrig:2020kck,Eberhardt:2020ewh,Gomez:2021qfd}, and the double copy \cite{Farrow:2018yni,Armstrong:2020woi,Albayrak:2020fyp,Zhou:2021gnu,Herderschee:2022ntr,Armstrong:2023phb}. More recently, it has become evident that in-in correlators of four dimensional de Sitter space (dS$_4$) are often mathematically simpler than wavefunction coefficients and are more closely related to scattering amplitudes in flat space \cite{Chowdhury:2023arc,Donath:2024utn,Chowdhury:2025ohm}. Building on earlier work which mapped the calculation of in-in correlators to Witten diagrams in Euclidean Anti de Sitter space (EAdS) \cite{Sleight:2020obc,Sleight:2021plv,DiPietro:2021sjt}, it was found that the in-in correlators of various scalar theories can be directly lifted from flat space Feynman diagrams by dressing them with certain auxiliary propagators \cite{Chowdhury:2025ohm}. This in turn provides a new point of view for uplifting mathematical structures and techniques from scattering amplitudes to cosmology. The goal of this paper will be to generalise these developments to in-in correlators involving spinning fields in dS$_4$, notably photons, gluons, and gravitons coupled to scalar fields. We note there has also been recent progress in mapping spinning in-in correlators to Witten diagrams in EAdS \cite{Sleight:2020obc,Sleight:2021plv,Schaub:2023scu,MdAbhishek:2025dhx,Sleight:2025dmt}.

In-in correlators are traditionally computed either by squaring the cosmological wavefunction and performing a path integral over the boundary conditions of the bulk fields \cite{Maldacena:2002vr,Ghosh:2014kba}, or using the Schwinger-Keldysh (SK) formalism \cite{Calzetta:1986ey,Weinberg:2005vy}. Most calculations in the literature for spinning correlators have been performed by integrating out non-transverse-traceless modes (which we will refer to collectively as longitudinal modes), leading to a non-local action that retains only the physical degrees of freedom \cite{Seery:2006vu,Seery:2008ax, Bonifacio:2022vwa, Ansari:2024pgq}. While this approach is very useful, it can become cumbersome because it requires solving for the unphysical modes via constraint equations of the gauge fields order by order in perturbation theory.
In this paper, we develop an alternative method for computing in-in correlators of spinning fields that avoids integrating out the unphysical modes. Instead, we keep them off-shell and perform the in-in path integral. This has the advantage that the action remains manifestly local. 
Careful gauge-fixing of the path integral is required at the turning point of the SK contour \cite{Witten:2022xxp, Chakraborty:2023los, Chakraborty:2025izq}, which ultimately implies that the longitudinal modes obey Dirichlet boundary conditions. We then Wick rotate to EAdS$_4$ and perform certain field redefinitions to unmix the SK 2-point functions. This gives an action with double the number of fields in EAdS$_4$, generalising the action for scalar fields obtained in \cite{DiPietro:2021sjt,Heckelbacher:2022hbq}, which we refer to as the shadow action.

As we will see later on, our formulation has certain advantages over more traditional approaches. For example, the number of Feynman diagrams contributing to an in-in correlator is generically smaller than in the SK or wavefunction-based formalisms. In fact, at tree-level there is a one-to-one correspondence with Feynman diagrams contributing to the corresponding scattering amplitudes. Moreover, using the shadow formalism we find a simple set of dressing rules for lifting flat space Feynman diagrams with internal photons, gluons, and gravitons to in-in correlators. These dressing rules distinguish between transverse-traceless and longitudinal modes and take a more canonical form for the former. We then use the dressing rules to explore how colour/kinematics (CK) duality and the double copy of in-in correlators is inherited from flat space amplitudes (see \cite{Bern:2019prr} for a review of these concepts in flat space). For colour-ordered tree-level 4-point gluon correlators, we find that the kinematic numerators obey a Jacobi relation analogous to those in flat space after integrating over the energies flowing through the propagators. Moreover, we find a simple prescription for mapping the transverse part of the gluon exchange correlators into the transverse-traceless part of graviton exchange correlators, which is essentially inherited from flat space.        

This paper is organised as follows. In section \ref{review} we review some basic concepts such as in-in correlators for scalar theories in dS$_4$, and various approaches for computing them. In section \ref{sec:gauge} we reformulate the SK formalism for gauge fields coupled to conformally coupled scalar fields without integrating out the longitudinal degrees of freedom and derive an associated shadow action in EAdS$_4$. We illustrate this formalism in a number of examples at tree- and loop-level, and use it to derive dressing rules and investigate CK duality for in-in correlators. In section \ref{sec:ininGrav} we then carry out similar steps for Einstein gravity coupled to massless scalars and explore the double copy for in-in correlators. Whereas sections \ref{sec:gauge}-\ref{sec:ininGrav} are primarily focused on correlators with external scalars and spinning fields propagating internally, in section \ref{sec:spinning} we turn our attention to spinning correlators. We present our conclusions and future directions in section \ref{conclusion_sec}. There are also a number of Appendices with further technical details about the SK formalism, the relation between in-in correlators and wavefunction coefficients, graviton 2-point functions, the double copy in flat space, and the relation of our approach to previous formulations.

\section{Preliminaries} \label{review}
In this section we will review some important facts about cosmological correlators that will be useful later on. First we review the definitions of wavefunction coefficients and in-in correlators. Then we will review the Schwinger-Keldysh, shadow formalism, and dressing rules for computing in-in correlators for scalar bulk theories. The goal of the remaining part of the paper will be to generalise these approaches to gauge theory and gravity in the bulk.

\subsection{Basic definitions}
We work in the Poincar\'e patch of de Sitter space, with the metric 
\begin{eqn}
    ds^2=  \frac{-d\eta^2+d\vec{x}^2}{\eta^2}
\end{eqn}
\noindent where we set the de Sitter radius to unity, $\eta \in (-\infty, 0)$ is the conformal time, and $x^i$ with $i=1,2,3$ are the boundary coordinates. Let us consider a scalar field in the bulk. We then define the cosmological wavefunction as follows:
\begin{eqn}
    \Psi[\varphi(\vec x)]\equiv \langle \varphi(\vec x)| \Omega \rangle = \int_{\Phi(-\infty)=0}^{\Phi(0)=\varphi(\vec x)} \mathcal{D}\Phi  e^{i S[\Phi]}
\label{wavepath}
\end{eqn}
where $\varphi(\vec x)$ is the field configuration at the boundary $\eta=0$, we choose $|\Omega \rangle$ to be the Bunch-Davies vacuum, and $S$ denotes the action of the field which is given by
\begin{eqn}\label{action}
S[\Phi] = -\int \frac{d^{3}xd\eta }{\eta^4}
\left[\frac{1}{2}(\partial\Phi)^{2}+m^{2}\Phi^{2}+V(\Phi)\right]
\end{eqn} 
where derivatives are contracted with the Minkowski metric. The wavefunction can be perturbatively expanded as follows:
\begin{eqn}
    \Psi[\varphi]=\exp\bigg(-\sum_{n=2}^{\infty}\frac{1}{n!}\int\prod_{a=1}^{n}\frac{d^{3}k_{a}}{(2\pi)^{3}}\varphi(\vec{k}_{a})\psi_{n}(\vec{k}_{1},...,\vec{k}_{n})\delta^{3}(\vec{k}_{T})  \bigg)
\label{eq:WFPerturbativeExpansion}
\end{eqn}
where $\psi_n$ are known as the wavefunction coefficients and we have Fourier transformed the boundary directions to momentum space. Note that momentum is conserved along the boundary, which is why we include a delta function in the sum over all boundary momenta $\vec k_T$. The wavefunction coefficients can be treated like CFT correlators in the boundary and computed by analytic continuation of AdS Witten diagrams \cite{McFadden:2009fg,Maldacena:2002vr,Maldacena:2011nz,Raju:2012zr,Bzowski:2013sza}. For scalar fields of mass $m$, the scaling dimensions of the dual operators are given by $\Delta_+$ where
\begin{equation}
\Delta_{\pm}=\frac{d}{2}\pm\sqrt{\left(\frac{d}{2}\right)^{2}-m^{2}}
\end{equation}
and $d=3$ is the boundary dimension. We will use the following notation for various quantities throughout the paper:
\begin{align}
    k_a = |\vec{k}_{a}| \quad \quad \quad &\text{energy of $a$\textsuperscript{th} external leg.} \nonumber \\
    k_{ab\cdots} \quad \quad \quad & k_a+ k_b + \cdots \nonumber \\ 
    \vec{y} \quad \quad \quad &\vec{k}_1 + \vec{k}_2 \nonumber \quad  \text{in exchange diagrams}\\
    y \quad \quad \quad &|\vec{y}| 
    \label{eq:Definitions}
\end{align}
In general, the total energy of an n-point wavefunction coefficent $k_{1\cdots n}$ is not conserved and the flat space limit is reached by taking the total energy to zero. In this limit, a flat space scattering amplitude is obtained by taking the residue of the total energy pole \cite{Raju:2012zr}.

Finally, let us recall that in-in correlators are obtained by taking the square modulus of the wavefunction and integrating over all boundary field configurations \cite{Maldacena:2002vr}: 
\begin{eqn}
    \braket{ \varphi(\vec{k}_{1})...\varphi(\vec{k}_{n})}  = \frac{\int \mathcal{D} \varphi \; \varphi(\vec{k}_{1})...\varphi(\vec{k}_{n}) |\Psi(\varphi)|^2}{{\int \mathcal{D} \varphi | \Psi(\varphi)|^2 }}.
    \label{eq:WFtoCorrelator}
\end{eqn}
Using the expansion in \eqref{eq:WFPerturbativeExpansion}, it is straightforward to show that  
\begin{equation}
\braket{ \varphi(\vec{k}_{1})\varphi(\vec{k}_{2})} =\frac{1}{2{\rm Re}\psi_{2}(\vec{k}_{1},\vec{k}_{2})},\,\,\, \braket{\varphi(\vec{k}_{1})\varphi(\vec{k}_{2})\varphi(\vec{k}_{3})} =\frac{{\rm Re}\psi_{3}(\vec{k}_{1},\vec{k_{2}},\vec{k}_{3})}{\Pi_{a=1}^{3}{\rm Re}\psi_{2}(\vec{k}_{a},-\vec{k}_{a})}
\end{equation}
at leading order. In practice, we will neglect the prefactor of $1/k_a^{2 \Delta_+-d}=k_a^{2 \Delta_--d}$ for each external leg and focus on the remaining part of the in-in correlator. Alternatively, in-in correlators can be computed using Schwinger-Keldysh formalism \cite{Weinberg:2005vy}, which we now turn to.

\subsection{Scalar correlators} \label{scalarcorr}
\begin{figure}[H]
    \centering
    \begin{subfigure}[t]{0.45\textwidth}
    \centering
        \begin{tikzpicture}[baseline=(b.base)]
            \begin{feynman}
                \vertex (a);
                \vertex [right = 75pt of a](b);
                \vertex [right = 75pt of b](c);
                \vertex [above = 75pt of b](y1);
                \vertex [below = 75pt of b](y2);

                \vertex [above = 10pt of a](i1);
                \vertex [below = 10pt of a](i2);
                \vertex [above right = 60pt and 50pt of b](l1);
                \vertex [above=10pt of l1](l2);
                \vertex [right =10pt of l1](l3);
                \vertex [above right =-1pt and -1pt of l1](l4){\(\eta\)};

                \diagram*{
                    (a)--(c),
                    (a) -- [boson,red](b),
                    (y1)--(y2),
                    (i1) -- [fermion] (b) -- [fermion](i2),
                    (l2)--(l1)--(l3),

                };
            \end{feynman}
            \draw[->,dashed] (0.5,0.25) to[bend left] (1.25,1.5);
            \draw[->,dashed] (0.5,-0.25) to[bend right] (1.25,-1.5);
        \end{tikzpicture}
        \caption{The standard Schwinger-Keldysh integration contour.\label{sfig:SKContoura}}
    \end{subfigure}
    \begin{subfigure}[t]{0.45\textwidth}
    \centering
        \begin{tikzpicture}[baseline=(b.base)]
            \begin{feynman}
                \vertex (a);
                \vertex [right = 75pt of a](b);
                \vertex [right = 75pt of b](c);
                \vertex [above = 75pt of b](y1);
                \vertex [below = 75pt of b](y2);

                \vertex [above = 10pt of a](i1);
                \vertex [below = 10pt of a](i2);
                \vertex [above right = 60pt and 50pt of b](l1);
                \vertex [above=10pt of l1](l2);
                \vertex [right =10pt of l1](l3);
                \vertex [above right =-1pt and -1pt of l1](l4){\(\eta\)};

                \diagram*{
                    (a)--(c),
                    (a)--[boson,red](b),
                    (y1)--[fermion](b)--[fermion](y2),
                    (l2)--(l1)--(l3),
                    
                };
            \end{feynman}
        \end{tikzpicture}
        \caption{The Wick rotated contour.\label{sfig:SKContourb}}
    \end{subfigure}
    \caption{The diagrammatic prescription for going from Schwinger-Keldysh to the shadow formalism. In the process of Wick rotating, the two branches must be rotated in opposite directions to avoid crossing the branch cut, shown by the dashed arrows.}
\end{figure}
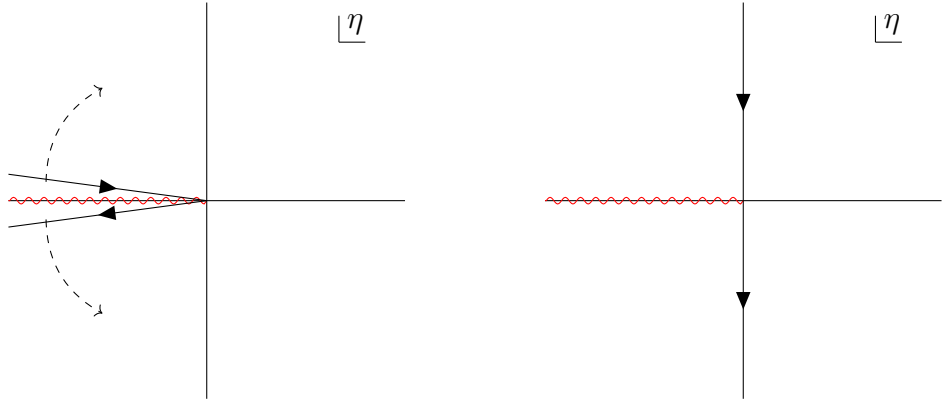

In this subsection, we will review how to compute scalar in-in correlators using the  Schwinger-Keldysh formalism, shadow formalism \cite{Sleight:2020obc, Sleight:2021plv, DiPietro:2021sjt}, and dressing rules \cite{Chowdhury:2025ohm,Chowdhury:2023arc}. Inserting the path integral expression for the wavefunction in \eqref{wavepath} into \eqref{eq:WFtoCorrelator} gives
\begin{eqn}\label{scalarmatching}
    \langle Q(\varphi) \rangle = \int  \mathcal{D} \Phi^> \mathcal{D} \Phi^< \; Q(\varphi) \delta\left(\left.\left(\Phi^{<}-\Phi^{>}\right)\right|_{\eta=0}\right)e^{i S[\Phi^>]-i S[\Phi^<]}
\end{eqn}
where we drop the normalisation, $Q(\varphi)$ denotes the insertion of operators at the boundary $\eta=0$ and $>, <$ denote fields in the forward and backward contour as shown in figure \ref{sfig:SKContoura}. The Schwinger-Keldysh action is given by 
\begin{eqn}\label{SKaction}
    i S_{SK}[\Phi^>, \Phi^<] = i S[\Phi^>]-i S[\Phi^<].
\end{eqn}
Note that the boundary values of the two fields are identified at the turning point of the contour which results in non-zero two-point functions between $\Phi^>$ and $\Phi^<$ (see Appendix \ref{app:SKProps} for a  derivation). The Schwinger-Keldysh two-point functions for conformally coupled and massless scalars then take the form
\begin{eqn}
\braket{\phi^>(\eta_1) \phi^>(\eta_2)} &=
i\frac{(\eta_1 \eta_2)^\frac32}{2} \intinf dp\frac{ p H^{(1)}_\nu(p \eta_1) H_{\nu}^{(2)}(p \eta_2)}{p^2 - k^2 + i \e}
, \\
\braket{\phi^{<}(\eta_1) \phi^{<}(\eta_2)} &= 
-i\frac{(\eta_1 \eta_2)^\frac32}{2} \intinf dp\frac{ p H^{(2)}_\nu(p \eta_1) H_{\nu}^{(1)}(p \eta_2)}{p^2 - k^2 - i \e}
, \\
\braket{\phi^{>}(\eta_1) \phi^{<}(\eta_2)} &= 
-i\frac{(\eta_1 \eta_2)^\frac32}{2}\int_{C_1} dp\frac{ p H^{(1)}_\nu(p \eta_1) H_{\nu}^{(2)}(p \eta_2)}{p^2 - k^2 }
,\\
\braket{\phi^{<}(\eta_1) \phi^{>}(\eta_2)} &= i\frac{(\eta_1 \eta_2)^\frac32}{2} \int_{C_2} dp\frac{ p H^{(1)}_\nu(p \eta_1) H_{\nu}^{(2)}(p \eta_2)}{p^2 - k^2 }
\label{eq:SKScalarProps}
\end{eqn}
where $\nu = \Delta_+ - \frac{d}{2}$ and $\nu=\frac{1}{2},\frac{3}{2}$ for conformally coupled and massless scalars respectively, and $C_1$ and $C_2$ denote the anti-clockwise closed contour as in Fig.\ \ref{fig:WightmanContours}.
\begin{figure}[H]
\centering
\begin{tikzpicture}
\draw (-2,0) -- (2,0);
\draw (0,-2) -- (0,2);
\node at (1,0) {$\times$};
\node at (-1,0) {$\times$};
\draw[scalar,solid,black, thick] (-1,0) circle (0.25);
\draw[scalar,solid,black, thick] (1,0) circle (0.25);
\node at (-1,-0.45) {$-k$};
\node at (1,-0.45) {$k$};
\node at (-1,0.55) {\scriptsize$C_2$};
\node at (1,0.55) {\scriptsize$C_1$};
\node at (1.75,1.75) {$ p$};
\draw (1.5,2) --(1.5,1.5) --(2,1.5);
\end{tikzpicture}
\caption{Contours of the $p$ integral in propagators. \label{fig:WightmanContours}}
\end{figure}
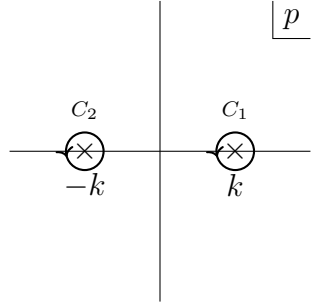

\noindent Next, we Wick-rotate to EAdS$_4$ with the prescription
\begin{eqn}\label{eq:WickRotationConvention}
    \eta^> \to e^{-i\frac{\pi}{2}} z \qquad \eta^< \to e^{i\frac{\pi}{2}} z 
\end{eqn}
which maps the SK contour to the one shown in figure \ref{sfig:SKContourb}. Moreover, we perform the following field redefinition:
\begin{eqn}
    \phi^> &= (-i)^{\Delta_+}\phi^+ + (-i)^{\Delta_-}\phi^- \\
    \phi^< & = (i)^{\Delta_+}\phi^+ + (i)^{\Delta_-}\phi^- ~.
    \label{eq:shadowBasisChange}
\end{eqn}
which unmixes the 2-point functions in \eqref{eq:SKScalarProps} to give
\begin{eqn}
    \braket{\phi^+(z_1) \phi^+(z_2)}&=G_\nu(z_1,z_2,\vec k)\\ 
    \braket{\phi^-(z_1) \phi^-(z_2)}&=G_{-\nu}(z_1,z_2,\vec k)\\
    \braket{\phi^+(z_1) \phi^-(z_2)}&=0\\
    \braket{\phi^-(z_1) \phi^+(z_2)}&=0
\end{eqn}
where the bulk-to-bulk propagator is given by
\begin{equation}
G_\nu(z, z'; \vec k) =\sin\left(\pi \nu  \right)\int dp \, \frac{p (z\, z')^{d/2}J_\nu (p z) J_\nu(p z')}{p^2+k^2}.
\label{eq:EAdSBBProps}
\end{equation}
for $\nu = \frac12, \frac32$. From this we can deduce the following bulk-to-boundary propagators by taking $z' \rightarrow 0$ and performing an appropriate rescaling by $z'$:
\begin{equation}
\mathcal B_\nu(z, \vec k) = \sin\left(\pi \nu  \right)z^{d/2} \frac{2^{ -\nu} k^{\nu} }{\Gamma(1 + \nu)}  K_{\nu  }(k z)
\label{eq:EAdSBbProps}
\end{equation}
Note that the $\phi_{-}$ and $\phi_{+}$ fields obey Neumann and Dirichlet boundary conditions, respectively. To compute wavefunction coefficients, we can simply use the propagators for $\phi_{+}$ fields \cite{Benincasa:2022gtd}. Applying \eqref{eq:WickRotationConvention} and \eqref{eq:shadowBasisChange} to \eqref{SKaction} finally gives the following action: 
\begin{eqn}
     S_{\rm{shadow}} &= \int\limits_0^{\infty}\frac{{\rm d} z {\rm d}^dx}{z^{d+1}}\left[\sin\left(\pi(\frac{d}{2}-\Delta_+)\right)\left((\partial\phi^+)^2\!-\!m^2{\phi_+}^2\right)\right.\\
     &+\sin\left(\pi(\frac{d}{2}-\Delta_- )\right)\left((\partial\phi_-)^2\!-\!m^2{\phi_-}^2\right)\\
	&\left.+e^{i\pi\frac{d-1}{2}}V\left(e^{-i\frac{\pi}{2}\Delta_+}\phi_+ +e^{-i\frac{\pi}{2}\Delta^-}\phi_-\right)
	+e^{-i\pi\frac{d-1}{2}}V\left(e^{i\frac{\pi}{2}\Delta_+}\phi_+ +e^{i\frac{\pi}{2}\Delta_-}\phi_-\right)\right],
\end{eqn}
which we refer to as the shadow action. The Feynman rules for $\phi_{\pm}$ can easily be read off from this action and used to compute in-in correlators. We will restrict our attention to Witten diagrams with external $\phi^-$ fields since these provide the dominant contribution to the correlator at late times. However, note that both types of fields propagate through the internal lines of the Feynman diagrams. 
In summary, the calculation of in-in correlators can be mapped to computing Witten diagrams in EAdS using an action with double the number of bulk fields that appeared in de Sitter space.  

In \cite{Chowdhury:2025ohm,Chowdhury:2023arc}, it was found that after summing over all diagrams contributing to an in-in correlator using the shadow formalism, remarkable simplifications occur which reduce the in-in correlator to a sum over flat space Feynman diagram dressed with theory-dependent auxiliary propagators attached to the interaction vertices. To illustrate how this works, consider the simple example of conformally coupled $\phi^4$ theory in dS$_4$. In this case, the auxiliary propagators are simply given by 
\begin{eqn}
\begin{tikzpicture}[baseline]
\node at (-1, 0) {\textbullet};
\draw[dashed] (-1,0) -- (0,0);
\node at (-0.5, 0.25) {$\longrightarrow$};
\node at (-0.5, 0.75) {$p_{tot}$};
\end{tikzpicture}
=  \frac{k_{ext}}{p_{tot}^2 + k_{ext}^2}
\end{eqn}
where the variable $k_{ext}$ denotes the sum of the external non-auxiliary energies at the vertex and $p_{tot}$ denotes the sum of the internal non-auxiliary energies. For a tree-level 4-point correlator there are no internal propagators and the auxiliary propagator simply gives a total energy pole 
\begin{eqn}
\begin{tikzpicture}[baseline]
\draw (-1, 1) -- (1,-1);
\draw (1, 1) -- (-1,-1);
\node at (-1, 1.25) {$\vec k_1$};
\node at (1, 1.25) {$\vec k_2$};
\node at (-1, -0.5) {$\vec k_4$};
\node at (1, -0.5) {$\vec k_3$};
\draw[dashed] (0,0) -- (0,-1.75);
\draw[->] (0.15, -1.5) -- (0.15, -2);
\node at (0.75, -1.85) {\small $p = 0 $};
\end{tikzpicture}
= \frac{1}{k_{1234}}.
\end{eqn}
Note that, as claimed above, the residue of the total energy pole is none other than the flat space amplitude. 

More generally, when there are multiple auxiliary propagators we must attach them at a common point, conserve the total energy at that point, and integrate over all the independent auxiliary energies as illustrated in the 1-loop example below:
\begin{eqn}
 & \begin{tikzpicture}[baseline=(b.base)]
        \begin{feynman}
            \vertex (a);
            \vertex [right = 75pt of a       ] (b) ;
            \vertex [right = 75pt of b        ] (c) ;
            \vertex [above left = 26.39pt and 63.71pt of b](l1) {$\vec{k_1}$};
            \vertex [below left = 26.39pt and 63.71pt of b](l2) {$\vec{k_2}$};
            \vertex [above right = 26.39pt and 63.71pt of b](r1) {$\vec{k_3}$};
            \vertex [below right = 26.39pt and 63.71pt of b](r2) {$\vec{k_4}$};
            \vertex [right = 40pt of a       ] (i1);
            \vertex [left = 40pt of c       ](i2);
            \vertex [above = 50pt of b](e) ;
    
            \diagram* {
                (l1) -- (i1),
                (l2) -- (i1),
                (r1) -- (i2),
                (r2) -- (i2),
                (i1) --[ momentum' = $L$, half right,looseness=0.8  ] (i2),
                (i2) --[ momentum= $L+K$, half right, looseness=0.8  ] (i1),
                (e) --[scalar,momentum = $p$](i2),
                (i1) --[scalar, momentum = $p$](e),
            };
        \end{feynman}
    \end{tikzpicture}
=\intinf \frac{dp k_{12} k_{34}}{(p^2 + k_{12}^2) (p^2 + k_{34}^2)} \int \frac{d^4 L }{L^2 (L + K)^2}, 
\end{eqn}
where the four vector $K^\mu = (p, \vec k)^\mu$. For other theories, such as conformally coupled $\phi^3$ or massless theories the dressing rules are more intricate but nonetheless can be systematically derived, as was done in \cite{Chowdhury:2025ohm}.

The goal of this paper will be to generalise these ideas to bulk theories with photons, gluons, and gravitons. Roughly speaking, we will find that conformally coupled $\phi^4$ theory provides a toy model for photons and gluons while massless scalar theories provide a toy model for graviton, but care must be taken when gauge-fixing the Schwinger-Keyldysh path integral. While the transverse traceless degrees of freedom will be described by two set of fields with different boundary conditions as we found above, the remaining degrees of freedom will be described by Dirichlet boundary conditions only. We will then derive a shadow formalism for in-in correlators of spinning fields coupled to scalars and use it to deduce dressing rules and explore properties known as the colour/kinematics duality and the double copy, which have played an important role in the study of scattering amplitudes.

\section{Gauge theory}\label{sec:gauge}
In this section, we will present a novel formulation of the SK formalism for scalar QED in dS$_4$. Previous studies of gauge theory in-in correlators typically integrate out the longitudinal modes which results in a non-local action \cite{Liu:1998ty,Maldacena:2002vr, Seery:2006vu, Seery:2006vu, Maldacena:2011nz, Bzowski:2011ab, Ghosh:2014kba}, however this approach requires one to integrate out the fields order by order in the coupling constant which is cumbersome for higher point functions. We review this approach in Appendix \ref{app:nonLocal}. Instead, we keep the longitudinal degrees of freedom and derive a local action in EAdS$_4$, whose Witten diagrams compute in-in correlators of scalar QED in dS$_4$, which we refer to as the shadow action. In doing so, we must carefully treat the boundary gauge fixing of the SK path integral, finding that the longitudinal modes satisfy Dirichlet boundary conditions. We then use this approach to compute various examples (notably a 1-loop example and tree-level examples up to six points) of in-in correlators in scalar QED and match the results to the SK and wavefunction approach, illustrating the relative simplicity of the shadow approach. Finally, we show that tree-level 4-point correlator with conformally coupled external scalars exchanging a photon can be directly uplifted from flat space Feynman diagram after applying certain dressing rules and show how colour/kinematics duality is realised for this correlator. Note that this correlator can be obtained from dimensional reduction of a spinning correlator \cite{Cachazo:2014xea}. In section \ref{sec:spinning} we will explain how to uplift this to a spinning correlator in Yang-Mills theory.

\subsection{SK formalism} \label{sksection}

We use the following action for conformally coupled scalar QED in de Sitter
\begin{eqn}
S = -\int \frac{d^3 x d\eta}{\eta^4} \Big[ \frac{1}{4} F_{ \mu\nu} F^{\mu\nu} +(D_\mu \phi)^\dagger (D^\mu \phi) + m^2 \phi^\dagger \phi  \Big]
\end{eqn}
where $m^2= 2$, the gauge covariant derivative is defined as $D_\mu \phi = \p_\mu + i g A_\mu$, and field strength $F_{\mu \nu} = \p_\mu A_\nu - \p_\nu A_\mu$.  
For this choice of the mass, correlators with external scalars and internal photons can be obtained from spinning correlators (which we consider in section \ref{sec:spinning}) by replacing inner products of polarisations with one and inner products of polarisations with external momenta with zero in a process called generalised dimensional reduction \cite{Cachazo:2014xea}. Additionally, we obtain a massless scalar after performing a Weyl transformation to half of flat space as follows:
\begin{eqn}
g_{\mu\nu} \to \eta^2 g_{\mu\nu} = \eta_{\mu\nu}; \qquad \phi \to \frac{\phi}{\eta}.
\end{eqn}
For simplicity, we therefore work in half of flat space in the remainder of this section, and continue to refer to the time coordinate as $\eta$.

For concreteness, we will focus on boundary correlators of scalars fields (which we collectively denote as $\Phi$) although it is straightforward to adapt the analysis to gauge fields. Following the discussion for scalar fields in section \ref{scalarcorr}, the path integral for the in-in correlator is 
\begin{eqn}\label{pathint1}
\braket{Q(\Phi)} \sim \int D \vec{\mathcal A} D\Phi |\Psi(\vec{\mathcal A}, \Phi)|^2  Q(\Phi)
\end{eqn}
where $Q(\Phi)$ denotes insertions of scalar fields at $\eta =0$ and $\vec{\mathcal A}$ denotes the boundary value of the gauge field. Note that we may use the bulk gauge symmetry to impose temporal gauge $A_{\eta}=0$. While $Q(\Phi)$ generically breaks gauge invariance at the boundary because the scalar fields transform non-trivially under gauge transformations, the wavefunction enjoys a residual gauge symmetry
\begin{eqn}\label{resgaugeinv}
\vec{\mathcal A}(\vec x) \to \vec{\mathcal A}(\vec x) + \vec{\partial} \Lambda(\vec x).
\end{eqn}
This symmetry exists after imposing temporal gauge because we can still perform $\eta$-independent gauge transformations. The invariance of the wavefunction under such residual gauge transformations is a consequence of the Gauss law 
\cite{DeWitt:1967yk, Kuchar:1970mu, Chowdhury:2021nxw, Chakraborty:2023los, Chakraborty:2023yed} 
and can be realized in practice by noting that the wavefunction has a perturbative expansion whose coefficients can be interpreted at CFT correlators which can be reconstructed from gauge-theory Ward identities \cite{Maldacena:2011nz, Bzowski:2013sza, Armstrong:2020woi}. 

After fixing the residual gauge symmetry of the wavefunction, we then integrate over all boundary configurations consistent with the gauge-fixing to compute in-in correlators. To fix this residual gauge invariance, let us decompose the boundary gauge field into a transverse and longitudinal part as follows:
\begin{equation}
\vec{\mathcal{A}}=\vec{\mathcal{A}}_{T}+\vec{\mathcal{A}}_{L}
\end{equation}
where the transverse part satisfies $\vec \p \cdot \vec{\mathcal A}_T =0$ and the longitudinal part can be written as the gradient of a scalar. This is simply the statement of the Helmholtz decomposition. In momentum space, this can be written:
\begin{equation}
\mathcal{A}_{i}=\left(\eta_{ij}-\frac{k_{i}k_{j}}{k^{2}}\right)\mathcal{A}_{j}+\frac{k_{i}k_{j}}{k^{2}}\mathcal{A}_{j}=\pi_{ij}\mathcal{A}_{j}+\frac{k_{i}k_{j}}{k^{2}}\mathcal{A}_{j}
\end{equation}
where all indices are downstairs for ease of notation, since the boundary directions are Euclidean. The boundary gauge transformation therefore leaves the transverse part invariant and shifts the longitudinal part, since translating \eqref{resgaugeinv} into momentum space shows that the gauge transformation is a shift proportional to $k_j$ and $\pi_{ij} k_j=0$ by construction. Hence, boundary gauge freedom can be used to set the longitudinal part to zero, which amounts to inserting $\delta(\vec{\mathcal A}_L)$ into the path integral \cite{Hatfield:1992rz}\footnote{More formally, this can be derived using the Faddeev-Popov gauge-fixing procedure, but we will not need to consider Fadeev-Popov ghosts since they decouple in QED and will not play a role in the examples that we consider later on.}:
\begin{eqn}\label{resgaugefixing}
\braket{Q(\Phi)} = \int D \vec{\mathcal A}_T D \vec{\mathcal A}_L D\Phi \delta(\vec{\mathcal A}_L) |\Psi(\vec{\mathcal A}_T, \Phi)|^2  Q(\Phi),
\end{eqn}
where the dependence of the wavefunction on the longitudinal part of the boundary field drops out. Ultimately, we will see that in-in correlators computed using this definition agree with those computed using previous formulations.

We can now write the Schwinger-Keldysh path integral \cite{Schwinger:1960qe, Keldysh:1964ud, Weinberg:2005vy, Haehl:2016pec, Haehl:2025zfn} more explicitly by expressing the wavefunction as
\begin{eqn}
\Psi(\vec{\mathcal A}_T, \Phi) = \int_{A^>_T(\eta = - \infty) = 0}^{A^>_T(\eta = 0, \vec x)= \vec{\mathcal A}_T(\vec x)} D A^>_T \int_{A^>_L(\eta = - \infty) = 0}^{A^>_L(\eta = 0, \vec x )= 0} D A^>_L \int_{\phi^>(\eta = - \infty) = 0}^{\phi^>(0, \vec x) = \Phi(\vec x)} D \Phi^> e^{ i S[\vec A^>_T, \vec A^>_L, \Phi^>]}
\label{eq:GaugeWF}
\end{eqn}
where we have set the upper limit of the integral over $A_L$ to zero using the residual gauge fixing condition. $>,<$ once again denote time-ordered and anti-time ordered fields living on the forwards and backwards parts of the SK contour, respectively. There is a similar expression for $\Psi^*(\vec{\mathcal A}_T, \Phi)$ in terms of a contour going in the opposite direction:
\begin{eqn}
\Psi^*(\vec{\mathcal A}_T, \Phi) = \int_{A^<_T(\eta = - \infty) = 0}^{A^<_T(\eta = 0, \vec x)= \vec{\mathcal A}_T(\vec x)} D A^<_T \int_{A^<_L(\eta = - \infty) = 0}^{A^<_L(\eta = 0, \vec x )= 0} D A^<_L \int_{\phi^<(\eta = - \infty) = 0}^{\phi^<(0, \vec x) = \Phi(\vec x)} D \Phi^< e^{- i S[\vec A^<_T, \vec A^<_L, \Phi^<]}
\label{eq:GaugeWFConj}
\end{eqn}
Plugging these expressions into \eqref{resgaugefixing} then gives
\begin{eqn}
\braket{Q(\Phi)} = \int_C D\vec A_T^> D \vec A_T^< D\vec A_L^> D \vec A_L^<  D\Phi^> D\Phi^<   
Q\big(\Phi \big) e^{i S[\vec A_T^>, \vec A_L^>, \Phi^>] - i S[\vec A_T^<, \vec A_L^<, \Phi^<]}
\label{corrSKform1}\end{eqn}
where $C$ denotes the SK contour and includes integration over boundary field configurations at $\eta=0$ with boundary conditions encoded by the limits of integration in 
\eqref{eq:GaugeWF} and \eqref{eq:GaugeWFConj}. From this we see that the scalar fields and the transverse part of the gauge field have 2-point functions of the form given in \eqref{eq:SKScalarProps}, while the longitudinal part of the gauge field obeys Dirichlet boundary conditions. Moreover there is no mixing between the time-ordered and anti-time-ordered longitudinal fields since their path integrals decouple. We refer the reader to Appendix \ref{app:SKProps} for more details and a derivation of the 2-point functions.

In summary, we have the following SK 2-point functions in scalar QED with momentum $\vec y$: 
\begin{enumerate}
\item Real Scalar Field:
\begin{eqn}\label{SKscalarcorr}
\braket{\phi^>(t_1) \phi^>(t_2)} &= 
\frac{i}{\pi}\intinf \frac{dp e^{i p t_{12}}}{p^2 - y^2 - i \e}
\equiv 
\begin{tikzpicture}[baseline]
\draw[solid] (-0.5, 0) -- (0.5, 0);
\node at (-0.5, -0.25) {$t_1$};
\node at (0.5, -0.25) {$t_2$};
\node at (-0.5, 0.25) {$>$}; 
\node at (0.5, 0.25) {$>$};
\end{tikzpicture}
, \\
\braket{\phi^{<}(t_1) \phi^{<}(t_2)} &= 
\frac{-i}{\pi}\intinf \frac{dp e^{-i p t_{12}}}{p^2 - y^2 + i \e}
\equiv 
\begin{tikzpicture}[baseline]
\draw[  solid] (-0.5, 0) -- (0.5, 0);
\node at (-0.5, -0.25) {$t_1$};
\node at (0.5, -0.25) {$t_2$};
\node at (-0.5, 0.25) {$<$}; 
\node at (0.5, 0.25) {$<$};
\end{tikzpicture}
, \\
\braket{\phi^{>}(t_1) \phi^{<}(t_2)} &= 
\frac{-i}{\pi}\int_{C_1} \frac{dp e^{i p t_{12}}}{p^2 - y^2 }
\equiv 
\begin{tikzpicture}[baseline]
\draw (-0.5, 0) -- (0, 0);
\draw (0.5, 0) -- (0, 0);
\node at (-0.5, -0.25) {$t_1$};
\node at (0.5, -0.25) {$t_2$};
\node at (-0.5, 0.25) {$>$}; 
\node at (0.5, 0.25) {$<$};
\end{tikzpicture}, \\
\braket{\phi^{<}(t_1) \phi^{>}(t_2)} &= 
\frac{i}{\pi}\int_{C_2} \frac{dp e^{i p t_{12}}}{p^2 - y^2 }
\equiv 
\begin{tikzpicture}[baseline]
\draw (-0.5, 0) -- (0, 0);
\draw (0.5, 0) -- (0, 0);
\node at (-0.5, -0.25) {$t_1$};
\node at (0.5, -0.25) {$t_2$};
\node at (-0.5, 0.25) {$<$}; 
\node at (0.5, 0.25) {$>$};
\end{tikzpicture}
\end{eqn}
which is just the $\nu=\frac12$ case of \eqref{eq:SKScalarProps}.
While these are the propagators for a real scalar as in \eqref{eq:SKScalarProps}, those for complex scalars are equal. We use the notation $t_{12}=t_1-t_2$.

\item Transverse Gauge Field: 
\begin{eqn}\label{gaugetransverseprop}
\braket{A_T^{>i}(t_1) A_T^{>j}(t_2)} &= i\frac{\pi^{ij}}{\pi} \intinf
\frac{dp e^{i p t_{12}}}{p^2 - y^2 - i \e}
\\
\braket{A_T^{<i}(t_1) A_T^{<j}(t_2)} &=-i\frac{\pi^{ij}}{\pi} \intinf 
\frac{dp e^{-i p t_{12}}}{p^2 - y^2 + i \e}
\\
\braket{A_T^{>i}(t_1) A_T^{<j}(t_2)} &= -i\frac{\pi^{ij}}{\pi}
\int_{C_1} \frac{dp e^{i p t_{12}}}{p^2 - y^2}, 
\\
\braket{A_T^{<i}(t_1) A_T^{>j}(t_2)} &= i\frac{\pi^{ij}}{\pi} 
\int_{C_2} \frac{dp e^{i p t_{12}}}{p^2 - y^2}.
\end{eqn}
\noindent Up to the tensor structure $\pi^{ij}$, these are exactly the same as the scalar field \cite{Liu:1998ty}. 

\item Longitudinal Gauge Field: 
\begin{eqn}
\braket{A_L^{>i}(t_1) A_L^{>j}(t_2)} &= i \frac{y^i y^j}{y^2 \pi} 
\intinf \frac{dp \sin(p t_1) \sin(p t_2)}{p^2 - y^2 - i \e}
, \\
\braket{A_L^{<i}(t_1) A_L^{<j}(t_2)} &= -i \frac{ y^i y^j}{y^2 \pi} 
\intinf \frac{dp \sin(p t_1) \sin(p t_2)}{p^2 - y^2 + i \e}
= - \braket{A_L^{>i}(t_1) A_L^{>j}(t_2)} , \\
\braket{A_L^{>i}(t_1) A_L^{<j}(t_2)} &= 0, \\
\braket{A_L^{<i}(t_1) A_L^{>j}(t_2)} &= 0.
\label{eq:SKLongProps}
\end{eqn}
\end{enumerate}

\noindent Noting that the transverse and longitudinal part of the gauge field have vanishing 2-point functions and combining them into a single field $A_i^{>,<}$ finally gives 
\begin{eqn}
\braket{A_i^>(t_1) A_j^>(t_2)}
 &\equiv \begin{tikzpicture}[baseline][baseline]
\begin{feynman}
\draw[boson, black] (-0.5, 0) -- (0.5, 0);
\node at (-0.5, -0.25) {$i$};
\node at (0.5, -0.25) {$j$};
\node at (-0.5, 0.25) {$>$}; 
\node at (0.5, 0.25) {$>$}; 
\end{feynman}
\end{tikzpicture}, \\
\braket{A_i^<(t_1) A_j^<(t_2)} 
 &\equiv \begin{tikzpicture}[baseline][baseline]
\begin{feynman}
\draw[boson, black] (-0.5, 0) -- (0.5, 0);
\node at (-0.5, -0.25) {$i$};
\node at (0.5, -0.25) {$j$};
\node at (-0.5, 0.25) {$<$}; 
\node at (0.5, 0.25) {$<$}; 
\end{feynman}
\end{tikzpicture}, \\
\braket{A_i^>(t_1) A_j^<(t_2)}
 &\equiv \begin{tikzpicture}[baseline][baseline]
\begin{feynman}
\draw[boson, black] (-0.5, 0) -- (0.5, 0);
\node at (-0.5, -0.25) {$i$};
\node at (0.5, -0.25) {$j$};
\node at (-0.5, 0.25) {$>$}; 
\node at (0.5, 0.25) {$<$};
\end{feynman}
\end{tikzpicture}, \\
\braket{A_i^<(t_1) A_j^>(t_2)}
 &\equiv \begin{tikzpicture}[baseline][baseline]
\begin{feynman}
\draw[boson, black] (-0.5, 0) -- (0.5, 0);
\node at (-0.5, -0.25) {$i$};
\node at (0.5, -0.25) {$j$};
\node at (-0.5, 0.25) {$<$}; 
\node at (0.5, 0.25) {$>$};
\end{feynman}
\end{tikzpicture}.
\end{eqn}
The bulk-boundary-propagators can be then obtained from the expressions above by carefully taking $\eta\rightarrow 0$ at one of the endpoints:
\begin{eqn}
\begin{tikzpicture}[baseline
]
\draw[very thick] (-0.5, 0.5) -- (0.5, 0.5);
\draw[scalar, solid, black] (-0.25, 0.5) -- (0,-0.5);
\node at (0, -0.75) {$>$};
\end{tikzpicture} 
= 
\begin{tikzpicture}[baseline
]
\draw[very thick] (-0.5, 0.5) -- (0.5, 0.5);
\draw[scalarbar, solid, black] (-0.25, 0.5) -- (0,-0.5);
\node at (0, -0.75) {$<$};
\end{tikzpicture} = e^{i k t}, 
\qquad 
\begin{tikzpicture}[baseline
]
\draw[very thick] (-0.5, 0.5) -- (0.5, 0.5);
\draw[scalar, solid, black] (-0.25, 0.5) -- (0,-0.5);
\node at (0, -0.75) {$>$};
\end{tikzpicture} 
= 
\begin{tikzpicture}[baseline
]
\draw[very thick] (-0.5, 0.5) -- (0.5, 0.5);
\draw[scalarbar, solid, black] (-0.25, 0.5) -- (0,-0.5);
\node at (0, -0.75) {$<$};
\end{tikzpicture} = e^{-i k t}, \qquad
\begin{tikzpicture}[baseline
]
\begin{feynman}
\draw[very thick] (-0.5, 0.5) -- (0.5, 0.5);
\draw[boson, solid, black] (-0.25, 0.5) -- (0,-0.5);
\node at (0, -0.75) {$>$};
\end{feynman}
\end{tikzpicture} 
 = \e_i e^{i k t}, 
 \begin{tikzpicture}[baseline
]
\begin{feynman}
\draw[very thick] (-0.5, 0.5) -- (0.5, 0.5);
\draw[boson, solid, black] (-0.25, 0.5) -- (0,-0.5);
\node at (0, -0.75) {$<$};
\end{feynman}
\end{tikzpicture} 
 = \e_i e^{-i k t}
\end{eqn}

\noindent where $\vec k$ is the external momentum.

\noindent Finally, from the SK path integral \eqref{corrSKform1} we have the following SK action for scalar QED:
\begin{eqn}\label{SKQEDaction}
 S_{SK} = \int d^3 x d\eta \Big[ -\frac{1}{4} F_{ \mu\nu}^{>} F^{>  \,\mu\nu}  + \frac{1}{4} F_{ \mu\nu}^{<} F^{<  \,\mu\nu} &-  (D_\mu \phi^>)^\dagger (D^\mu \phi^>) - m^2 \phi^{>}{}^\dagger \phi^{>} \\
&+   (D_\mu \phi^<) (D^\mu \phi^<)+m^2 \phi^{<}{}^\dagger \phi^{<}  \Big]
\end{eqn}
where we recall that $A_\eta=0$. From this, we deduce the following 3-point vertices:
\begin{eqn}
\begin{tikzpicture}[baseline]
\begin{feynman}
\draw[scalar, solid, black] (-0.5, 0.5) -- (0,0);
\draw[scalarbar, solid, black] (-0.5, -0.5) -- (0,0);
\draw[boson, black] (0,0) -- (1,0);
\node at (0, -0.25) {$>$};
\end{feynman}
\end{tikzpicture} = - i g (\vec k_1 - \vec k_2)_i ,
\qquad 
\begin{tikzpicture}[baseline]
\begin{feynman}
\draw[scalar, solid, black] (-0.5, 0.5) -- (0,0);
\draw[scalarbar, solid, black] (-0.5, -0.5) -- (0,0);
\draw[boson, black] (0,0) -- (1,0);
\node at (0, -0.25) {$<$};
\end{feynman}
\end{tikzpicture} =  i g (\vec k_1 - \vec k_2)_i.
\end{eqn}

\subsection{Shadow formalism \label{ssec:GaugeShadow}}

Starting from the SK action in \eqref{SKQEDaction}, we can obtain a shadow action in EAdS$_4$ by performing the Wick rotation in \eqref{eq:WickRotationConvention} and the following field redefinitions:
\begin{align}
    \phi^> & \to \phi^- - i \phi^+,  \quad 
    A_i^>\to A_i^- - i A_i^+, \nonumber \\
    \phi^< & \to \phi^- + i \phi^+, \quad 
    A_i^<\to A_i^- + i A_i^+.
\label{spin1redef}
\end{align}
The shadow action then takes the following form:
\begin{eqn}
     S_{\text{shadow}}&=2\int dz d^3 x \bigg\{-\frac{1}{4}F_{ \mu\nu}^{+} F^{+  \,\mu\nu}  + \frac{1}{4} F_{ \mu\nu}^{- } F^{-  \,\mu\nu}
     +(\p_\mu \phi^-)^\dagger \p^\mu \phi^- - (\p_\mu \phi^+)^\dagger \p^\mu \phi^+\\
     &+ig \delta^{ij} \bigg[A_i^+\big({\phi^-}^\dagger\p_j \phi^+-\phi^+ \p_j{\phi^-}^\dagger   + {\phi^+}^\dagger \p_j \phi^- - \phi^- \p_\nu {\phi^+}^\dagger\big)\\
     &+A_i^-\big({\phi^+}^\dagger\p_j \phi^+-\phi^+ \p_j{\phi^+}^\dagger   - {\phi^-}^\dagger \p_j \phi^- + \phi^- \p_j {\phi^-}^\dagger \big) \bigg]\\
     &+g^2 \delta^{ij} \bigg[(A_i^-{A_j}^- -A_i^+{A_j}^+)({\phi^-}^\dagger \phi^- - {\phi^+}^\dagger\phi^+)-2 A_i^-{A_j}^+(\phi^+ {\phi^-}^\dagger+\phi^-{\phi^+}^\dagger)\bigg]\bigg\}.
     \label{eq:SQEDShadowFull}
\end{eqn}

After performing the field redefinitions shown in \eqref{spin1redef}, the 2-point functions of the scalar field and transverse gauge field become unmixed, while the 2-point functions of the longitudinal gauge fields remain unmixed. We obtain the following 2-point functions for the gauge fields: 
\begin{eqn}
G^-_{A, ij}(z_1,z_2,\vec{y})&= \frac{1}{2\pi} \int_{-\infty}^\infty dp \, \big(\cos(p z_1) \cos(p z_2) \frac{\pi_{ij}}{p^2+y^2}+\sin(p z_1) \sin(p z_2) \frac{y_i y_j}{y^2 p^2} \big)\\
 G^+_{A, ij}(z_1,z_2,\vec{y})&= -\frac{1}{2\pi} \int_{-\infty}^\infty dp \, \sin(p z_1) \sin(p z_2) \Big( \frac{\pi_{ij}}{p^2 + y^2} + \frac{y_i y_j}{y^2 p^2} \Big)
 \label{eq:bulk-bulkph}
\end{eqn}
where 
\begin{equation}
\pi_{ij}=\eta_{ij}-\frac{y_{i}y_{j}}{y^{2}}
\end{equation}
is the transverse projection tensor. These propagators are diagrammatically denoted as 
\begin{eqn}
G_{A,ij}^-(z_1, z_2, \vec y) = \begin{tikzpicture}[baseline]
\draw[thick] (0,0) circle (2);
\draw[photon, black, solid] (-1,0) -- (1,0);
\node at (-1, +0.25) {$z_1, i$};
\node at (1, +0.25) {$z_2, j$};
\node at (0,0.35) {$\vec y$};
\node at (0,-0.35) {$-$};
\end{tikzpicture}, \qquad 
G_{A,ij}^+(z_1, z_2, \vec y) = \begin{tikzpicture}[baseline]
\draw[thick] (0,0) circle (2);
\draw[photon, black, solid] (-1,0) -- (1,0);
\node at (-1, +0.25) {$z_1, i$};
\node at (1, +0.25) {$z_2, j$};
\node at (0,0.35) {$\vec y$};
\node at (0,-0.35) {$+$};
\end{tikzpicture},
\end{eqn}
In summary, we find that longitudinal gauge fields have Dirichlet boundary conditions always. Meanwhile, the transverse gauge fields and scalar fields have either Neumann or Dirichlet boundary conditions, if they are $-$ or $+$ fields respectively.

Moreover, the bulk-to-bulk propagators for the scalars are given by
\begin{eqn}\label{scalarbulkbulkshadow}
    G^+_\phi(z_1,z_2,\vec{y})&= -\frac{1}{2\pi}\int_{-\infty}^\infty dp \, \frac{\sin(p z_1) \sin(p z_2)}{p^2+y^2}= \begin{tikzpicture}[baseline]
\draw[thick] (0,0) circle (2);
\draw[scalar, solid] (-1,0) -- (1,0);
\node at (-1, +0.25) {$z_1$};
\node at (1, +0.25) {$z_1$};
\node at (0,0.35) {$\vec y$};
\node at (0,-0.35) {$+$};
\end{tikzpicture}\\
    G^-_\phi(z_1,z_2,\vec{y})&= \frac{1}{2\pi}\int_{-\infty}^\infty dp \, \frac{\cos(p z_1)\cos(p z_2)}{p^2+y^2}=\begin{tikzpicture}[baseline]
\draw[thick] (0,0) circle (2);
\draw[scalar, solid] (-1,0) -- (1,0);
\node at (-1, +0.25) {$z_1$};
\node at (1, +0.25) {$z_1$};
\node at (0,0.35) {$\vec y$};
\node at (0,-0.35) {$-$};
\end{tikzpicture}
\end{eqn}
where $\vec{y}$ denotes the boundary momentum flowing through the propagator. 

From these formulae, we can easily deduce the following bulk-to-boundary propagators:
 \begin{eqn}
    K_\phi^{\pm}(\vec{k},z) &= \mp e^{- k z}
    = 
    \begin{tikzpicture}[baseline]
    \draw[thick] (0,0) circle (1.25);
    \draw ({1.25*cos(60)}, {1.25*sin(60)})-- (0,0);
    \node at (60:1.55) {$\vec k$};
    \node at (0,-0.25) {$z$};
    \end{tikzpicture}\\
    {K_A^{\pm}}_i(\vec{k},z) &= \mp \epsilon_i e^{- k z}
    = 
    \begin{tikzpicture}[baseline]
    \draw[thick] (0,0) circle (1.25);
    \draw [photon,black,solid] ({1.25*cos(60)}, {1.25*sin(60)})-- (0,0);
    \node at (60:1.55) {$\vec k{,} \, \vec \epsilon$};
    \node at (0,-0.25) {$z$};
    \end{tikzpicture}\,; \qquad k^i \e_i=0.
\end{eqn}
Note that the correlators corresponding to in-in correlators in dS are those with external $-$ fields. 

From the action in \eqref{eq:SQEDShadowFull}, we find the following 3-point vertices:
\begin{eqn}
        \begin{tikzpicture}[baseline=(b.base)]
         \begin{feynman}
             \vertex (b);
             \vertex [above left = 20pt and 20pt of b,label ={[xshift=-4pt,yshift=4pt]above left:\(\vec{k}_1, \; - \)}] (l1) ;
             \vertex [below left = 20pt and 20pt of b,label ={[xshift=-4pt,yshift=-4pt]below left:\(\vec{k}_2, \; + \)}] (l2);
             \vertex [right = 28pt of b, label ={[xshift=5.5pt]right:\( +  \)}] (r);
             
             \diagram*{
                (l1) --[fermion](b),
                (b) --[fermion]  (l2),
                (b) --[boson, edge label = \(i\)](r),
             };

             \draw[thick] ($(b)$) circle [radius=34pt];
         \end{feynman}
     \end{tikzpicture}&=\begin{tikzpicture}[baseline=(b.base)]
         \begin{feynman}
             \vertex (b);
             \vertex [above left = 20pt and 20pt of b,label ={[xshift=-4pt,yshift=4pt]above left:\(\vec{k}_1, \; + \)}] (l1) ;
             \vertex [below left = 20pt and 20pt of b,label ={[xshift=-4pt,yshift=-4pt]below left:\(\vec{k}_2, \; - \)}] (l2);
             \vertex [right = 28pt of b, label ={[xshift=5.5pt]right:\( +  \)}] (r);
             
             \diagram*{
                (l1) --[fermion](b),
                (b) --[fermion]  (l2),
                (b) --[boson, edge label = \(i\)](r),
             };

             \draw[thick] ($(b)$) circle [radius=34pt];
         \end{feynman}
     \end{tikzpicture}=2  g (k_1 - k_2)^i \\
     \vspace{1cm}
     \begin{tikzpicture}[baseline=(b.base)]
         \begin{feynman}
             \vertex (b);
             \vertex [above left = 20pt and 20pt of b,label ={[xshift=-4pt,yshift=4pt]above left:\(\vec{k}_1, \; + \)}] (l1) ;
             \vertex [below left = 20pt and 20pt of b,label ={[xshift=-4pt,yshift=-4pt]below left:\(\vec{k}_2, \; + \)}] (l2);
             \vertex [right = 28pt of b, label ={[xshift=5.5pt]right:\( -  \)}] (r);
             
             \diagram*{
                (l1) --[fermion](b),
                (b) --[fermion]  (l2),
                (b) --[boson, edge label = \(i\)](r),
             };

             \draw[thick] ($(b)$) circle [radius=34pt];
         \end{feynman}
     \end{tikzpicture}&=- \Bigg(\;\begin{tikzpicture}[baseline=(b.base)]
         \begin{feynman}
             \vertex (b);
             \vertex [above left = 20pt and 20pt of b,label ={[xshift=-4pt,yshift=4pt]above left:\(\vec{k}_1, \; - \)}] (l1) ;
             \vertex [below left = 20pt and 20pt of b,label ={[xshift=-4pt,yshift=-4pt]below left:\(\vec{k}_2, \; - \)}] (l2);
             \vertex [right = 28pt of b, label ={[xshift=5.5pt]right:\( -  \)}] (r);
             
             \diagram*{
                (l1) --[fermion](b),
                (b) --[fermion]  (l2),
                (b) --[boson, edge label = \(i\)](r),
             };

             \draw[thick] ($(b)$) circle [radius=34pt];
         \end{feynman}
     \end{tikzpicture}\Bigg)=2 g (k_1 -k_2)^i
\label{shadowintervertices3pt}
\end{eqn}

\noindent As we will see shortly, these Feynman rules compute in-in correlators in dS$_4$. Since wavefunction coefficients have Dirichlet boundary conditions, they can essentially be obtained from the above Feynman rules by restricting internal lines to fields with a $+$ superscript (although one should still use the above vertex factor, since the $+++$ vertex vanishes in the shadow formalism). From the shadow action in \eqref{eq:SQEDShadowFull}, we can also read off the following 4-point vertices: 
\begin{eqn}
     \begin{tikzpicture}[baseline=(b.base)]
         \begin{feynman}
             \vertex (b);
              \vertex [above left = 20pt and 20pt of b,label ={[xshift=-4pt,yshift=4pt]above left:\( - \)}] (l1);
             \vertex [below left = 20pt and 20pt of b,label ={[xshift=-4pt,yshift=-4pt]below left:\( - \)}] (l2);
             \vertex [above right = 20pt and 20pt of b,label ={[xshift=4pt,yshift=4pt]above right:\( - \)}] (r1);
             \vertex [below right = 20pt and 20pt of b,label ={[xshift=4pt,yshift=-4pt]below right:\( - \)}] (r2);

             \diagram*{
                (l1) --[fermion](b),
                (b) --[fermion]  (l2),
                (r1) --[boson, edge label' = \(i\)](b),
                (b) --[boson, edge label' = \(j\)]  (r2),
             };
             \draw[thick] ($(b)$) circle [radius=34pt];
         \end{feynman}
     \end{tikzpicture}&=-\Bigg(\begin{tikzpicture}[baseline=(b.base)]
         \begin{feynman}
             \vertex (b);
              \vertex [above left = 20pt and 20pt of b,label ={[xshift=-4pt,yshift=4pt]above left:\( - \)}] (l1);
             \vertex [below left = 20pt and 20pt of b,label ={[xshift=-4pt,yshift=-4pt]below left:\( - \)}] (l2);
             \vertex [above right = 20pt and 20pt of b,label ={[xshift=4pt,yshift=4pt]above right:\( + \)}] (r1);
             \vertex [below right = 20pt and 20pt of b,label ={[xshift=4pt,yshift=-4pt]below right:\( + \)}] (r2);

             \diagram*{
                (l1) --[fermion](b),
                (b) --[fermion]  (l2),
                (r1) --[boson, edge label' = \(i\)](b),
                (b) --[boson, edge label' = \(j\)]  (r2),
             };
             \draw[thick] ($(b)$) circle [radius=34pt];
         \end{feynman}
     \end{tikzpicture} \Bigg)=4g^2 \delta^{ij} 
\label{eq:shadowintervertices4pt}
\end{eqn}
where we only list those with $-$ scalar fields for conciseness.

\subsection{Tree-level examples}\label{ssec:tree}

In this section we will illustrate the formalisms described above in various tree-level examples. We will find that the shadow formalism generally requires the fewest number of diagrams, generalising the results for scalar theories in \cite{Chowdhury:2023arc}. First we will consider the 4-point correlator $\braket{\phi^\dagg(\vec k_1) \phi(\vec k_2) \phi^\dagg(\vec k_3) \phi(\vec k_4)}$ describing scalars exchanging a photon.

\subsubsection*{Wavefunction formalism}
Let us focus on the s-channel contribution to the 4-point correlator. The other two channels can be obtained by permuting the external legs. In the wavefunction formalism this contribution is given by (see appendix \ref{app:wavefunc} for further details):  
\begin{eqn}\label{corrinWF1}
\braket{\phi_1 \phi_2^\dagg \phi_3  \phi_4^\dagg}_{(s)} = \psi_{4,s}^{\phi_1 \phi_2^\dagger \phi_3 \phi_4^\dagger} + \sum_h \frac{\psi_3^{\phi_1 \phi_2^\dagger A}(h)\psi_3^{\phi_3 \phi_4^\dagger A}(-h)}{\psi_2},
\end{eqn}
where $h$ denotes the helicity of the exchanged photon. The wavefunction coefficient $\psi_2$ is fixed from conformal invariance and is given as $\psi_2(\vec k) = k$. The other wavefunction coefficients $\psi_3^{\phi^\dagg \phi A}, \psi_4^{\phi\phi^\dagg\phi\phi^\dagg}$ are given by the following Witten diagrams:
\begin{eqn}
\psi_{4,s}^{\phi \phi^\dagg \phi \phi^\dagg} = 
\begin{tikzpicture}[baseline]
\draw (-2, 1) -- (2, 1);
\draw[scalar, solid, black] (-1.75, 1) -- (-1.5, -1);
\draw[scalarbar, solid, black] (-1.25, 1) -- (-1.5, -1);
\draw[scalar, solid, black]  (1.75, 1) -- (1.5, -1);
\draw[scalarbar, solid, black]  (1.25, 1) -- (1.5, -1);
\draw[photon, black] (-1.5, -1) -- (1.5, -1);
\node at (-1.75, 1.25) {$+$};
\node at (-1.25, 1.25) {$+$};
\node at (1.75, 1.25) {$+$};
\node at (1.25, 1.25) {$+$};
\end{tikzpicture}, 
\qquad 
\psi_{3}^{\phi \phi^\dagg A}(h) = 
\begin{tikzpicture}[baseline]
\draw (-2, 1) -- (2, 1);
\draw[scalar, solid, black] (-1.5, 1) -- (0, -1);
\draw[scalarbar, solid, black] (1.5, 1) -- (0, -1);

\draw[photon, black] (0, 1) -- (0, -1);

\node at (-1.5, 1.25) {$+$};
\node at (1.5, 1.25) {$+$};
\node at (0, 1.25) {$h, +$};

\end{tikzpicture}
\end{eqn}
Note that the propagators for the wavefunction formalism are only those of the $+$ fields introduced in section \ref{ssec:GaugeShadow} as these satisfy the Dirichlet boundary conditions. Hence we label every diagram with those fields.

From the propagator \eqref{eq:bulk-bulkph} we see that the wavefunction coefficient $\psi_4$ has two pieces, a longitudinal piece and a transverse piece \cite{Ghosh:2014kba, Albayrak:2019asr, Baumann:2020dch}. These are given as 
\begin{eqn}\label{psi4qed}
\psi_{4,s} &= \a^i \b^j \intsinf dz_1 dz_2 e^{- k_{12} z_1} e^{- k_{34} z_2} (z_1 z_2)^{1/2} \intinf dp p J_{1/2}(p z_1) J_{1/2}(p z_2) \Big[ \frac{\pi_{ij}}{p^2 + y^2} + \frac{y_i y_j}{y^2} \frac{1}{p^2} \Big]  \\
&\propto \a^i \pi_{ij} \b^j \frac{1}{(k_{12} + y)(k_{34} + y)(k_{12}+ k_{34})} 
+ \a^i \frac{y_i y_j}{y^2} \b^j \frac{1}{k_{12} k_{34} (k_{12} + k_{34})},
\end{eqn}
 where we have ignored some overall unimportant constants and have defined $\a_i = (\vec k_1 - \vec k_2)_i$ and $\b_j = (\vec k_3 - \vec k_4)_j$. The three-point wavefunction $\psi_3^h$ can be written as 
\begin{eqn}\label{psi3qed}
\psi_3^h(\vec k_1, \vec k_2, \vec y) &= g \e^h_i \a^i \intsinf dz e^{- (k_{12} + y)z} = g\frac{\e^h_i \a^i}{k_{12} + y}.
\end{eqn}
The sum over photon helicity in \eqref{corrinWF1} can be performed by using the completeness relation for the polarization tensor
\begin{eqn}
\sum_h \e_{n}^h(\vec k) \e_{m}^{-h}(\vec k)=\pi_{nm}(\vec k)=\eta_{mn}-k_m k_n/k^2
\end{eqn}
From this relation it is clear that the product of 3-point functions in \eqref{corrinWF1} only contributes to the transverse part of the answer and hence the longitudinal piece entirely comes from $\psi_4$. We then express the s-channel contributions to the correlator as a sum of the transverse $T$ and longitudinal piece $L$: 
\begin{eqn}
\braket{\phi_1 \cdots \phi^\dagg_4}_{(s)} = \braket{\phi_1 \cdots \phi^\dagg_4}_{(s)}^T + \braket{\phi_1 \cdots \phi^\dagg_4}^L_{(s)}~,
\end{eqn}
where the transverse piece receives a contribution from both $\psi_4$ and $(\psi_3)^2$ and the longitudinal piece only receives a contribution from $\psi_4$: 
\begin{eqn}\label{QED4ptresult1}
\braket{\phi_1 \cdots \phi^\dagg_4}_{(s)}^T &=   g^2 \frac{\a^i \pi_{ij} \b^j}{(k_{12} + y)(k_{34} + y)} \left( \frac{1}{k_{12} + k_{34}} + \frac{1}{y} \right), \\
\braket{\phi_1 \cdots \phi^\dagg_4}_{(s)}^L &= g^2 \alpha^i \frac{y_i y_j}{y^2}\beta^j \frac{1}{k_{12} k_{34}k_{1234}},
\end{eqn} 
where $\a^i = (\vec k_1 - \vec k_2)^i$ and $\b^i = (\vec k_3 - \vec k_4)^i$.

It is easy to see that this gives the correct flat space limit by considering the residue at the total energy pole \cite{Raju:2012zr}: 
\begin{eqn}
\Res{k_{12} + k_{34} = 0} \braket{\phi_1\phi_2^\dagg \phi_3 \phi_4^\dagg}_{(s)} = g^2 \frac{-(k_1 - k_2) (k_3 - k_4) + \vec \a \cdot \vec \b}{k_{12}^2 - y^2}.
\end{eqn}

\subsection*{SK formalism}

Now let us consider the photon exchange correlator from the point of view of the Schwinger-Keldysh formalism. Using the Feynman rules in section \ref{sksection}, the following diagrams contribute to the s-channel:
\begin{eqn}\label{SKdiag}
\braket{\phi_1\phi_2^\dagg \phi_3 \phi_4^\dagg}_{(s)}
= 
\begin{tikzpicture}[baseline]
\begin{feynman}
\draw[very thick] (-1, 0.5) -- (1, 0.5);
\draw[scalar, solid, black] (-0.75, 0.5) -- (-0.5, -0.5);
\draw[scalarbar, solid, black] (-0.25, 0.5) -- (-0.5, -0.5);
\draw[scalarbar, solid, black] (0.75, 0.5) -- (0.5, -0.5);
\draw[scalar, solid, black] (0.25, 0.5) -- (0.5, -0.5);
\draw[boson, black] (-0.5, -0.5) -- (0.5, -0.5);
\node at (-0.5, -0.75) {$>$};
\node at (0.5, -0.75) {$>$};
\end{feynman}
\end{tikzpicture}
+ 
\begin{tikzpicture}[baseline]
\begin{feynman}
\draw[very thick] (-1, 0.5) -- (1, 0.5);
\draw[scalar, solid, black] (-0.75, 0.5) -- (-0.5, -0.5);
\draw[scalarbar, solid, black] (-0.25, 0.5) -- (-0.5, -0.5);
\draw[scalarbar, solid, black] (0.75, 0.5) -- (0.5, -0.5);
\draw[scalar, solid, black] (0.25, 0.5) -- (0.5, -0.5);
\draw[boson, black] (-0.5, -0.5) -- (0.5, -0.5);
\node at (-0.5, -0.75) {$<$};
\node at (0.5, -0.75) {$>$};
\end{feynman}
\end{tikzpicture}
+ 
\begin{tikzpicture}[baseline]
\begin{feynman}
\draw[very thick] (-1, 0.5) -- (1, 0.5);
\draw[scalar, solid, black] (-0.75, 0.5) -- (-0.5, -0.5);
\draw[scalarbar, solid, black] (-0.25, 0.5) -- (-0.5, -0.5);
\draw[scalarbar, solid, black] (0.75, 0.5) -- (0.5, -0.5);
\draw[scalar, solid, black] (0.25, 0.5) -- (0.5, -0.5);
\draw[boson, black] (-0.5, -0.5) -- (0.5, -0.5);
\node at (-0.5, -0.75) {$>$};
\node at (0.5, -0.75) {$<$};
\end{feynman}
\end{tikzpicture}
+
\begin{tikzpicture}[baseline]
\begin{feynman}
\draw[very thick] (-1, 0.5) -- (1, 0.5);
\draw[scalar, solid, black] (-0.75, 0.5) -- (-0.5, -0.5);
\draw[scalarbar, solid, black] (-0.25, 0.5) -- (-0.5, -0.5);
\draw[scalarbar, solid, black] (0.75, 0.5) -- (0.5, -0.5);
\draw[scalar, solid, black] (0.25, 0.5) -- (0.5, -0.5);
\draw[boson, black] (-0.5, -0.5) -- (0.5, -0.5);
\node at (-0.5, -0.75) {$<$};
\node at (0.5, -0.75) {$<$};
\end{feynman}
\end{tikzpicture}
\end{eqn}
The time integrals can be easily performed to give
\begin{eqn}\label{SK4ptQED}
\braket{\phi_1\phi_2^\dagg \phi_3 \phi_4^\dagg}_{(s)}
= g^2 \a^i \b^j \Bigg[ \frac{\pi_{ij}}{(k_{12}+ y)(k_{34} + y)} \Big( \frac{1}{k_{12} + k_{34}} + \frac{1}{y} \Big) + \frac{y_i y_j}{y^2} \frac{1}{(k_{12} + k_{34}) k_{12} k_{34}} \Bigg],
\end{eqn}
which matches the result obtained above from wavefunctions.

\subsection*{Shadow formalism}

Finally, let us see how things work in the shadow formalism. Since the dS correlator is given by the correlators of the external $-$ fields, only a single Feynman diagram contributes to the s-channel, just like in flat space:
\begin{eqn}\label{C4shadow}
\braket{\phi_1\phi_2^\dagg \phi_3 \phi_4^\dagg}_{(s)}&= 
\begin{tikzpicture}[baseline]
\begin{feynman}
    \draw[thick] (0,0) circle (2.25);
    \draw[scalar, solid] (-2,1) -- (-1, 0);
    \draw[scalarbar, solid] (-2,-1) -- (-1, 0);
    \draw[scalar, solid] (2,1) -- (1, 0);
    \draw[scalarbar, solid] (2,-1) -- (1, 0);
    \draw[boson, black] (-1, 0) -- (1, 0);
    \node at (-2.1, 1.25) {$-$};
    \node at (-2.1, -1.25) {$-$};
    \node at (2.1, 1.25) {$-$};
    \node at (2.1, -1.25) {$-$};
    \node at (0, 0.4) {$-$};
\end{feynman}
\end{tikzpicture}
\end{eqn}
By comparison, we found that four diagrams contribute in the Schwinger-Keldysh approach and the wavefunction approach had two types of contributions (one coming from a 4-point exchange diagram and the other involving the square of a 3-point diagram). In more detail, the diagram in \eqref{C4shadow} is given by
\begin{eqn}\label{FINALANS}
&\braket{\phi_1\phi_2^\dagg \phi_3 \phi_4^\dagg}_{(s)}=g^2 \a^i \b^j \int_0^\infty dz \, dz' \int dp\; e^{-k_{12} z-k_{34} z'} \bigg(\cos(p z) \cos(p z')\frac{\pi_{ij}}{p^2+y^2}+\sin(p z) \sin(p z')\frac{y_i y_j}{y^2 p^2} \bigg) \\
    &=  g^2 \alpha^i \beta^j \left( \frac{\pi_{ij}}{(k_{12}+y)(k_{34}+y)}\Big(\frac{1}{k_{1234}}+\frac{1}{y}\Big)+\frac{y_i y_j}{k_{12} k_{34} k_{1234} y^2}\right)
\end{eqn}
which indeed matches the in-in correlator obtained using SK \eqref{SK4ptQED} and the wavefunction \eqref{QED4ptresult1}. This simple example also demonstrates the efficiency of computing correlators using the shadow formalism. We will explain how this extends to higher points below.

\subsection*{Higher-point functions}\label{sssec:higherpt}

In this section, we will sketch how the above calculations extend to higher points. For concreteness, consider the following 6-point diagram in the shadow formalism:
\begin{eqn}
\braket{\phi_1^\dagg \phi_2 \phi_3^\dagg \phi_4 \phi_5^\dagg \phi_6} =
 \begin{tikzpicture}[baseline]
\begin{feynman}
    \draw[thick] (0,0) circle (2.25);
    \draw[scalar, solid] (-2,1) -- (-1, 0);
    \draw[scalarbar, solid] (-2,-1) -- (-1, 0);
    \draw[scalar, solid] (2,1) -- (1, 0);
    \draw[scalarbar, solid] (2,-1) -- (1, 0);
    \draw[ scalar, solid] (0, 2.25)--(0,0);
    \draw[ scalarbar, solid](0,-2.25)--(0,0);
    \draw[boson, black] (-1, 0) -- (0, 0);
    \draw[boson, black] (0, 0) -- (1, 0);
    \node at (-2.1, 1.25) {$-$};
    \node at (-2.1, -1.25) {$-$};
    \node at (2.1, 1.25) {$-$};
    \node at (2.1, -1.25) {$-$};
    \node at (0, 2.5) {$-$};
    \node at (0, -2.5) {$-$};
    \node at (-1, 0.4) {$-$};
    \node at (1, 0.4) {$-$};
\end{feynman}
\end{tikzpicture}
\end{eqn}
Using the wavefunction formalism, this correlator will involve a 6-point diagram as well as a sum over products of lower-point diagrams (see Appendix \ref{app:wavefunc} for more details):  
\begin{eqn}\label{wf6ptterms}
&\braket{\phi_1^\dagg \phi_2 \phi_3^\dagg \phi_4 \phi_5^\dagg \phi_6}= \psi_6^{\phi^\dagg \phi \phi^\dagg \phi \phi^\dagg \phi}(\vec k_1, \cdots, \vec k_6)  \\
& + \psi_{5i}^{\phi^\dagg\phi\phi^\dagg\phi A_T}(\vec k_1, \cdots, \vec k_4) \psi_{3j}^{\phi^\dagg \phi A_T}(\vec k_5, \vec k_6, \vec y_2) \frac{\pi^{ij}(\vec y_2)}{y_2} + \psi_{5i}^{\phi^\dagg\phi\phi^\dagg\phi A_T}(\vec k_3, \vec k_4, \vec k_5, \vec k_6) \psi_{3j}^{\phi^\dagg \phi A_T}(\vec k_1, \vec k_2, \vec y_1) \frac{\pi^{ij}(\vec y_1)}{y_1} \\
& + \psi_{3i}^{\phi^\dagg \phi A_T}(\vec k_1, \vec k_2, \vec y_1) \psi_{4jm}^{\phi^\dagg \phi A_T A_T}(\vec y_1, \vec k_3, \vec k_4, \vec y_2) 
\psi_{3n}^{\phi^\dagg \phi A_T}(\vec y_2, \vec k_5, \vec k_6) \frac{\pi^{ij}(\vec y_1)}{y_1} \frac{\pi^{mn}(\vec y_2)}{y_2}.
\end{eqn}
where $\vec{y}_1 = \vec{k_1}+\vec{k_2}$ and $\vec{y}_2=-(\vec{k}_5+\vec{k_6})$. Moreover, using the Schwinger-Keldysh formalism, we find that $2^3=8$ diagrams contribute; one for each combination of time- and anti-time-ordered vertices. For an explicit evaluation and check against the wavefunction formalism, see \cite{github}.

More generally, the shadow diagrams needed to capture a tree-level in-in correlator in scalar QED are in one-to-one correspondence with Feynman diagrams in flat space. This relies on two key points. First, the correlator is given by diagrams with external $-$ legs at the boundary. Second, the allowed set of Feynman vertices is restricted in the shadow formalism, as given in \eqref{shadowintervertices3pt} and \eqref{eq:shadowintervertices4pt}. In particular, only vertices with an even number of $+$ legs are allowed. This means that, if one has an internal $+$ leg at any point, the diagram must have an external $+$ leg somewhere, or a closed loop. By contrast, in the wavefunction approach, we will have a Feynman diagram with the same topology as the one in flat space and additional contributions corresponding to all possible cuts of the diagram. Moreover, for a flat space Feynman diagram with $n$ vertices, there will be $2^n$ diagrams arising in the Schwinger-Keldysh approach. Hence, the shadow formalism is much more efficient.

\subsection{1-loop example}

\begin{figure}[H]
    \centering
    \begin{tikzpicture}[baseline=(b.base)]
        \begin{feynman}
            \vertex (b);
            \vertex [below left = 30pt and 30pt of b](a) {$k,-$};
            \vertex [below right = 30pt and 30pt of b](c){$k,-$};
            \vertex [above=30pt of b](v);
            \vertex [below right = 5pt and 12pt of v](v1) {$\vec{l}, +$};
            
            \diagram*{
            (a) -- [fermion] (b) --[fermion] (c),
            b -- [boson, out=135, in=45, loop, min distance=1.7cm] b
            };
        \end{feynman}
        \draw[thick] ($(b)$) circle [radius=42.426pt];
    \end{tikzpicture}
    \hspace{1cm}
    $+$
    \hspace{1cm}
    \begin{tikzpicture}[baseline=(b.base)]
        \begin{feynman}
            \vertex (b);
            \vertex [below left = 30pt and 30pt of b](a) {$k,-$};
            \vertex [below right = 30pt and 30pt of b](c){$k,-$};
            \vertex [above=30pt of b](v);
            \vertex [below right = 5pt and 12pt of v](v1) {$\vec{l}, -$};
            
            \diagram*{
            (a) -- [fermion] (b) --[fermion] (c),
            b -- [boson, out=135, in=45, loop, min distance=1.7cm] b
            };
            \draw[thick] ($(b)$) circle [radius=42.426pt];
        \end{feynman}
    \end{tikzpicture}
    \caption{Tadpole contributions to the one-loop scalar self-energy in the shadow formalism.}
    \label{fig:ShadowTadpole}
\end{figure}
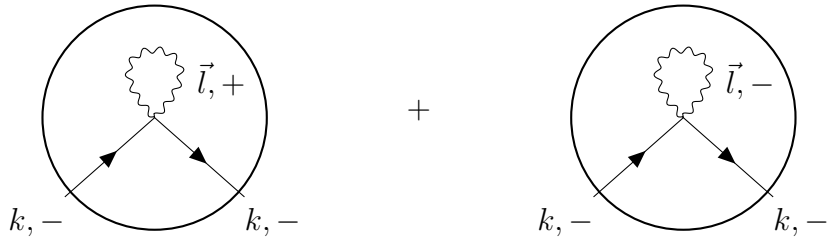

In this section we will calculate one-loop corrections to the scalar self-energy coming from tadpole diagrams in the shadow formalism. Note that other diagrams contribute in which a photon is emitted and reabsorbed via three-point interactions, but we will neglect them for simplicity since they do not add anything conceptually. The two diagrams which contribute to the 1-loop 2-point function are shown in Figure \ref{fig:ShadowTadpole}.
\noindent Note that the $+$ field is ghost-like and has a negative kinetic term, but there is also a relative sign between the $----$ and $++--$ vertices, hence overall the two diagrams add together. The result is
\begin{eqn}\label{tadpoleQED}
    \braket{ \phi^{\dagger}(\vec{k})\phi(-\vec{k})} _{\rm tadpole} &=\delta_{ij} \frac{2}{\pi} \int d^3 l \int dp \int_0^\infty dz \; e^{-2k z}\bigg[\underbrace{\sin^2(p z) \frac{\pi_{ij}}{p^2+l^2}+\sin^2(p z)\frac{l_i l_j}{p^2 l^2}}_{+ \text{ diagram}}\\
    &\hspace{5.5cm}+ \underbrace{\cos^2(p z) \frac{\pi_{ij}}{p^2+l^2}+\sin^2(p z)\frac{l_i l_j}{p^2 l^2}}_{- \text{ diagram}} \bigg]\\
    &= \delta_{ij} \frac{2}{\pi} \int d^3 l \int dp \int_0^\infty dz \; e^{-2k z}\bigg[\underbrace{ \frac{\pi_{ij}}{p^2+l^2}}_{\rm Transverse}+ \underbrace{2 \sin^2(p z)\frac{l_i l_j}{p^2 l^2}}_{\rm Longitudinal }  \bigg]\\
    &=\frac{2\delta_{ij}}{\pi} \int d^4L \frac{1}{2k} \bigg( \frac{\pi_{ij}}{L^2}+ \frac{1}{(L_0)^2+k^2} \frac{L_i L_j}{ L_k L^k}\bigg)\\
    &= \int d^3l \frac{2k+l}{2k^2 l}
\end{eqn}
where $L^\mu =(p, \vec{l})^\mu$ and to obtain the final line we performed the $p$-integral, which can be conveniently done via residues. The answer precisely matches the result obtained via the wavefunction as we shall demonstrate below. Notice that the transverse part of the answer can be elegantly expressed in terms of a flat space diagram with a dressing factor $\frac{1}{2k}$. The longitudinal part however does not have such a simple structure. We leave the regularisation and renormalisation of the loop-level correlator for future work (see \cite{Chowdhury:2023arc,Senatore:2009cf,Bzowski:2023nef,Jain:2025maa} for various approaches).

\begin{figure}[H]
    \centering
\begin{tikzpicture}[baseline]
\draw[thick] (-2, 1) -- (2, 1);
\draw[scalar, solid, black] (-1.5, 1) -- (0, -1);
\draw[scalarbar, solid, black] (1.5, 1) -- (0, -1);

\draw[photon, black] (0, -1.5) circle (0.5);

\node at (-1.5, 1.25) {$k, +$};
\node at (1.5, 1.25) {$k, +$};
\end{tikzpicture}
    \hspace{0.5cm}
    $+$
    \hspace{0.5cm}
    $\frac{1}{2l}$
\begin{tikzpicture}[baseline]
\draw[thick] (-2, 1) -- (2, 1);
\draw[scalar, solid, black] (-1.5, 1) -- (0, -1);
\draw[scalarbar, solid, black] (-0.5, 1) -- (0, -1);

\draw[photon, black] (0.5, 1) -- (0, -1);
\draw[photon, black] (1.5, 1) -- (0, -1);

\node at (-1.5, 1.25) {$k, +$};
\node at (-0.5, 1.25) {$k, +$};
\node at (0.5, 1.25) {$l, +$};
\node at (1.5, 1.25) {$l, +$};
\end{tikzpicture}

    \caption{Wavefunction coefficients encoding tadpole contribution to one-loop two-point function. The relative factor of $\frac{1}{2}$ between the contributions is a symmetry factor and the $\frac{1}{l}$ arises from the inverse of the two point function $\frac{1}{\psi_2(l)}.$}
    \label{fig:WFTadpole}
\end{figure}
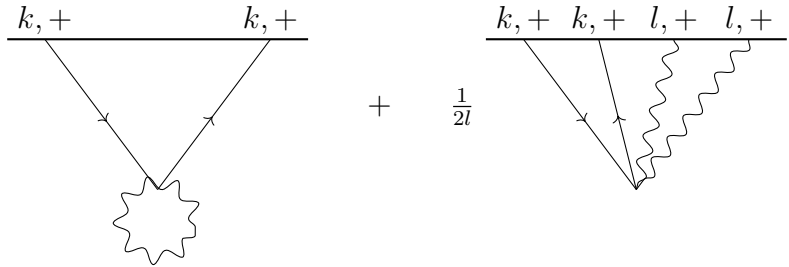

Now consider the same correlator obtained from the wavefunction. The correlator in terms of wavefunction coefficients is shown in Fig.\ \ref{fig:WFTadpole}.
As an equation, these diagrams give (see appendix \ref{app:wavefunc} for details)  
\begin{eqn}\label{tadpolewf}
 \braket{ \phi^{\dagger}(\vec{k})\phi(-\vec{k})} _{\rm tadpole} =\int d^{3}l\;\left(\psi_{2}^{(1)}(\vec k,\vec k,\vec{l})+\frac{\psi_{4}^{(0)}(\vec k, \vec k, \vec l, \vec l)}{2\psi_{2}^{(0)}(l)}\right).
\end{eqn}
where the relative factor of $\frac12$ is a symmetry factor. Using the Feynman rules in section \ref{ssec:GaugeShadow} we find
\begin{eqn}\label{tadpolewfcoeff}
     \braket{ \phi^{\dagger}(\vec{k})\phi(-\vec{k})} _{\rm tadpole} &= \delta_{ij}  \int d^3l \;  \int_0^\infty dz \bigg[ \bigg\{\frac{1}{\pi}\int dp \, e^{-2k z} \sin^2(p z) \big(\frac{\pi_{ij}}{p^2+l^2}+\frac{l_i l_j}{p^2 l^2}\big) \bigg\} +\frac{1}{2} \frac{\pi_{ij}}{l} e^{-2(k+l) z} \bigg] \\
    &= \delta_{ij} \int d^3 l \; \frac{1}{2 k l } \bigg( \pi_{ij}+\frac{l_i l_j}{k l}\bigg) \\
    &= \int d^3l \; \frac{2k+l}{2k^2 l},
\end{eqn}
where the $\pi_{ij}$ in the contribution from the contact diagram term stems from the polarisation sum.

\subsection{Dressing rules and CK duality}\label{ssec:dressingrule}

In this section, we will show how to uplift Feynman diagrams for scalar QED to de Sitter space using certain dressing rules, generalising previous results for scalar theories in \cite{Chowdhury:2025ohm}. Noting that Feynman diagrams of scalar QED can be treated as scalar QCD diagrams after stripping off colour factors (leading to colour-ordered Feynman diagrams). We then use the dressing rules to demonstrate that in-in correlators of scalar QCD inherit a property known as colour/kinematics duality from their flat space counterparts. In flat space, this property underlies a relation between gluon and graviton amplitudes known as the double copy.   

\subsubsection*{Dressing rules} 

Let us revisit the photon exchange diagram in the shadow formalism in \ref{C4shadow}. Restricting to the transverse part of the photon exchange gives
\begin{eqn}
\braket{\phi_{1}\phi_{2}^{\dagger}\phi_{3}\phi_{4}^{\dagger}}_{(s)}^T &= \a^i \pi_{ij} \b^j \intsinf dz_1 dz_2 e^{- k_{12} z_1} e^{- k_{34} z_2} (z_1 z_2)^{1/2} \\
&\times \intinf \frac{dp p }{p^2 + y^2}\Big(J_{-1/2}(p z_1) J_{-1/2}(p z_2)+J_{1/2}(p z_1) J_{1/2}(p z_2) \Big) 
\end{eqn}
Performing the integrals above results in the first line of \eqref{QED4ptresult1}. However if we instead only perform the $z$ integrals we obtain the expression 
\begin{eqn}\label{QEDconfdress1}
\braket{\phi_{1}\phi_{2}^{\dagger}\phi_{3}\phi_{4}^{\dagger}}_{(s)}^T &= \a^i \pi_{ij} \b^j \intinf dp \frac{k_{12}}{p^2 + k_{12}^2} \frac{k_{34}}{p^2 + k_{34}^2} \frac{1}{p^2 + y^2} \\
&= \a^i \pi_{ij} \b^j \intinf dp \hat \Delta_{\gamma}^T(k_{12}, p) \Delta_{\gamma}^T(k_{34}, p) \frac{1}{p^2 + y^2}
\end{eqn}
where
\begin{eqn}
\hat \Delta_{\gamma}^T(k_{ext}, p)
= \frac{k_{ext}}{k_{ext}^2 + p^2}
\label{sqeddressing}
\end{eqn}
with $k_{ext}$ denoting the sum of external energies entering a vertex. This representation allows one to formulate dressing rules for obtaining the transverse part of the in-in correlator from a flat space Feynman diagram by attaching auxiliary propagators, 
\begin{eqn}\label{QEDdress}
\braket{\phi_{1}\phi_{2}^{\dagger}\phi_{3}\phi_{4}^{\dagger}}_{(s)}^T  &=
\begin{tikzpicture}[baseline]
\begin{feynman}
\draw (-2, 1) -- (-1, 0);
\draw (-2, -1) -- (-1, 0);
\draw (2, 1) -- (1, 0);
\draw (2, -1) -- (1, 0);
\draw[boson,->] (-1, 0) -- (1, 0);
\node at (0,0.3) {$p$};
\draw[dashed] (-1, 0) -- (0, -1);
\draw[dashed] (1, 0) -- (0, -1);
\draw[->] (-0.5, -0.75) -- (-0.75, -0.5);
\node at (-0.75, -0.75) {$p$};
\draw[<-] (0.5, -0.75) -- (0.75, -0.5);
\node at (0.75, -0.75) {$p$};
\end{feynman}
\end{tikzpicture}
\end{eqn}
where the dashed propagators are one dimensional auxiliary propagators only carrying energy and are attached to the the vertices of each Feynman diagram. Note that the auxiliary propagator in \eqref{sqeddressing} takes exactly the same form that was previously found for correlators of the conformally coupled $\phi^4$ theory in \cite{Chowdhury:2025ohm}, which we reviewed in section \ref{review}. In section \ref{sec:spinning}, we will find a similar structure for auxiliary propagators of pure Yang-Mills correlators. This reflects the classical conformal invariance of Yang-Mills theory in four dimensions.

To obtain the full correlator we also have to consider the longitudinal part of the photon propagator which is given as,
\begin{eqn}
\braket{\phi_{1}\phi_{2}^{\dagger}\phi_{3}\phi_{4}^{\dagger}}_{(s)}^L &= \a^i \b^j\frac{y_i y_j}{y^2} \intinf dp \frac{p}{p^2 + k_{12}^2} \frac{p}{p^2 + k_{34}^2} \frac{1}{p^2} \\
&= \a^i \b^j\frac{y_i y_j}{y^2} \intinf dp \hat \Delta_{\gamma}^{L}(k_{12}, p) \hat \Delta_{L}^{\gamma}(k_{34}, p) \frac{1}{p^2} 
\label{qeddresslongitudinal}
\end{eqn}
where 
\begin{eqn}\label{sqeddressingL}
\hat \Delta_{\gamma}^{L}(k_{ext}, p) = \frac{p}{k_{ext}^2 + p^2}.
\end{eqn}
Note that the longitudinal dressing rule does not straightforwardly generalise to higher points because when there are multiple bulk-to-bulk propagators at a vertex we get cross terms between the transverse and longitudinal components, which are dressed in different ways due to having different boundary conditions. On the other hand, the dressing rule in \eqref{sqeddressing} can be applied to general flat space diagrams to uplift them to the transverse part of in-in correlators. This in turn provides useful data from which the full correlator can be reconstructed, as we discuss in the conclusion.

\subsubsection*{Colour/kinematics duality}
Let us now explore the implications of the dressing rules for color/kinematics (CK) duality in dS. We briefly review how it works for the example of a 4-point scattering amplitude in flat space. For a more in-depth review, see \cite{Bern:2019prr}. Recall that a tree-level 4-point gluon amplitude takes the form
\begin{eqn}
    \mathcal{A}_{4}= \frac{c_s n_s }{s}+\frac{c_t n_t}{t}+\frac{c_u n_u}{u}
\end{eqn}
where $c_s, c_t, c_u$ are group theoretic factors which add to zero by the standard Jacobi identity. The CK duality implies that the numerators $n_s, n_t, n_u$ (which are constructed from momenta and polarisations) obey an analogous relation known as the kinematic Jacobi identity: 
\begin{eqn}
    n_s+n_t+n_u=0.
    \label{eq:Jacobi}
\end{eqn}
Note that we can reduce the amplitude and associated kinematic Jacobi relation to the case of adjoint scalars exchanging a gluon in scalar QCD by imposing $\epsilon_i\cdot \epsilon_j=1$ and $\epsilon_i \cdot p_j=0$, where $\epsilon_i$ and $p_j$ are external polarisations and momenta (this is referred to as generalised dimensional reduction \cite{Cachazo:2014xea}). The goal of this section will be to uplift this story to de Sitter space using the dressing rules derived above. These were derived for scalar QED, but they can be trivially uplifted to color-ordered scalar QCD correlators, which are defined by stripping off colour factors and considering a fixed ordering of the external legs (see for example \cite{Dixon:1996wi} for review of color-ordering). Given the kinematic numerator in the s-channel (which can be read off from s-channel color-ordered Feynman diagrams), the numerators in the other channels can be obtained by permuting external legs:
\begin{equation}
n_{t}=-\left.n_{s}\right|_{2\leftrightarrow4},\,\,\,n_{u}=-\left.n_{s}\right|_{2\leftrightarrow3}.
\label{numchannels}
\end{equation}
The minus signs in the above equation are tied to the antisymmetry of the associated colour factors.

Our starting point will be the sum of \eqref{QEDconfdress1} and \eqref{qeddresslongitudinal}, which express the s-channel contribution to the photon exchange correlator in terms of dressed Feynman diagrams:
\begin{eqn}
\braket{\phi_{1}\phi_{2}^\dagg \phi_{3}\phi_{4}^\dagg}_{(s)}=\frac{1}{\pi}\int dp \frac{1}{p^2+y^2} \alpha^i \big( \pi_{ij} \frac{k_{12}}{k_{12}^2+p^2} \frac{k_{34}}{k_{34}^2+p^2} +\frac{y_i y_j}{y^2} \frac{p}{k_{12}^2+p^2}\frac{p}{k_{34}^2+p^2} \frac{p^2+y^2}{p^2}\big) \beta^j.
\label{A4sns}\end{eqn}
where $\vec{y}=\vec{k}_1 + \vec k_2$ is the boundary momentum flowing through the propagator. Note that we must integrate over the energy flowing through the propagator $p$ due to the lack of energy conservation.

We then define the kinematic numerator $n_s$ to be the numerator of the Feynman propagator $(p^2+y^2)^{-1}$ in the integrand:
\begin{eqn}
    n_s(p)&=\frac{1}{\pi}\frac{1}{k_{12}^2+p^2} \frac{1}{k_{34}^2+p^2}\big(\alpha\cdot\beta k_{12}k_{34}+\frac{(k_1^2-k_2^2)(k_3^2-k_4^2)}{y^2}(k_{12}k_{34}-p^2-y^2)\big)\\
    &=\frac{1}{\pi}\frac{k_{12}}{k_{12}^2+p^2} \frac{k_{34}}{k_{34}^2+p^2}\big( \alpha\cdot \beta - (k_1-k_2)(k_3-k_4)+ \frac{(k_1-k_2)(k_3-k_4)}{y^2}(k_{12} k_{34} -p^2)\big)\\
    &=\frac{1}{\pi}\frac{k_{12}}{k_{12}^2+p^2} \frac{k_{34}}{k_{34}^2+p^2}\big( \alpha^\mu \eta_{\mu \nu} \beta^\nu+\frac{(k_1-k_2)(k_3-k_4)}{y^2}(k_{12} k_{34} -p^2)\big)
    \label{eq:FlatAxialGluonNumerator}
\end{eqn}
where we define $\alpha^\mu=(k_1-k_2,\vec{k}_1-\vec{k}_2)$ and $\beta^\mu=(k_3-k_4,\vec{k}_3-\vec{k}_4)$. Using \eqref{numchannels}, we can then obtain the kinematic numerators in the other two channels. 

While summing the kinematic numerators over all three channels doesn't vanish, a weaker condition turns out to hold:
\begin{eqn}
    \intinf dp \; (n_s(p)+n_t(p)+n_u(p)) =0.
    \label{eq:dSJacobi}
\end{eqn}
In particular, the second term in \eqref{eq:FlatAxialGluonNumerator} vanishes upon performing the $p$ integral and the remaining term yields 
\begin{eqn}\label{flatns}
    \frac{1}{\pi}\intinf dp \; \frac{k_{12}}{k_{12}^2+p^2}\frac{k_{34}}{k_{34}^2+p^2}\alpha^\mu \eta_{\mu \nu}\beta^\nu = \frac{\alpha^\mu \eta_{\mu \nu} \beta^\nu}{k_{12}+k_{34}}
\end{eqn}
which is precisely the kinematic numerator in we obtain in flat-space in Lorentz gauge along with a total energy pole. Since the energy pole is the same for all three channels, the sum over channels gives precisely the same sum which appears in flat-space amplitudes, which vanishes without requiring energy conservation. Hence, using the dressing rules we see that in-in correlators inherit CK duality from flat space amplitudes. We will extend this result to spinning correlators in Yang-Mills theory in section \ref{sec:spinning}.

A similar story holds for wavefunction coefficients. This can be seen by performing the $z_i$ integrals in the first line of \eqref{psi4qed} resulting in
\begin{eqn}
\psi_{4,s} = \frac{1}{\pi}\int dp \frac{1}{p^2+y^2} \underbrace{\alpha^i \big( \pi_{ij} \frac{p}{k_{12}^2+p^2} \frac{p}{k_{34}^2+p^2} +\frac{y_i y_j}{y^2} \frac{p}{k_{12}^2+p^2}\frac{p}{k_{34}^2+p^2} \frac{p^2+y^2}{p^2}\big) \beta^j}_{n^\psi_s(p)}.
\end{eqn}
We then find that 
\begin{eqn}
\intinf dp \; (n^{\psi}_s(p) + n^{\psi}_t(p) + n^{\psi}_u(p)) = 0.
\end{eqn}
The CK duality for wavefunction coefficients was previously studied in AdS momentum space \cite{Armstrong:2020woi,Albayrak:2020fyp}, where generalised gauge transformations were used to enforce the kinematic Jacobi identity at four points. It was subsequently shown in \cite{Alday:2021odx} that this identity is automatically satisfied in Mellin space without using such transformations. In this paper, we have shown that generalised gauge transformations are also not needed in momentum space if we integrate over the energy flowing through the propagator. It would be interesting to see how this generalises to higher points.

\section{Gravity}\label{sec:ininGrav}
In this section, we will consider gravity coupled to a massless scalar dS$_4$. Following the gauge theory case described earlier, we will derive the SK formalism from a path integral point of view, keeping track of boundary conditions induced by boundary gauge-fixing. Following the blueprint of the gauge field example, we will then perform a Wick rotation and field redefinition which unmixes the SK 2-point functions to obtain a shadow action in EAdS$_4$. We use the shadow formalism to compute a tree-level 4-point and 1-loop 2-point example with external scalars and an internal graviton, matching the results with the wavefunction approach. We provide detailed calculations of these examples for the interested reader at \cite{github}. 

We then conclude this section by showing that the transverse-traceless part of the tree-level 4-point correlator can be uplifted from flat space Feynman diagrams after applying certain dressing rules and relate it to the square of the gluon correlator tensor structure, illustrating how the double copy for in-in correlator may be inherited from the one in flat space.

\subsection{SK formalism}

We will begin with the action for Einstein gravity coupled to a massless scalar,
\begin{eqn}
    S = \frac{1}{2 \kappa } \int d^4 x \sqrt{-\tilde g} (\tilde R-2 \Lambda- \kappa^2 \tilde g^{\mu \nu} \p_ \mu \phi \p_\nu \phi),
    \label{eq:EHFullAction}
\end{eqn}
where $\kappa \sim \sqrt{G_N}$ and the metric is written as
\begin{eqn}
    \tilde{g}_{\mu\nu}=g_{\mu\nu}+\kappa h_{\mu\nu}=\frac{\eta_{\mu\nu}+\kappa \tilde{h}_{\mu\nu}}{\eta^{2}},
    \label{eq:GravitonPerturbation}
\end{eqn}
where we define $\tilde{h}_{\mu\nu}=\eta^2 h_{\mu\nu}$. We work in axial gauge $h_{\eta\mu} = 0$
and decompose the gravitational field as 
\begin{eqn}\label{eq:gravdecom}
h_{ij} =  h_{ij}^{TT} + h_{ij}^L,  
\end{eqn}
where $ h_{ij}^{TT}$ is the transverse-traceless part, which satisfies the the following constraints:
\begin{eqn}
    \p_i h^{TT}_{ij} =0; \qquad \delta_{ij}h^{TT}_{ij}=0,
\end{eqn}
and the remaining part (which we refer to as longitudinal for a shorthand) can be written as follows: 
\begin{eqn}
    h^{L}_{ij} = \p_{(i}\xi_{j)} + \delta_{ij} \chi.
\end{eqn}
Following the analysis in \cite{Chakraborty:2023los}, the modulus of the wavefunction computed using the above action enjoys a residual boundary diffeomorphism and Weyl invariance which can be used to set $\tilde{h}^L_{ij}=0$ at $\eta=0$. To compute in-in correlators, we then insert fields in the boundary and integrate over all boundary field configurations consistent with this gauge-fixing. Since $h_{ij}$ must in general be a linear combination of modes which scale like $\eta^{-2}$ and $\eta$ near the boundary \cite{Baumann:2020dch}, $h^L_{ij}$ must scale like $\eta$ because the other scaling behaviour would not lead to vanishing $\tilde{h}^L_{ij}=\eta^2 h^L_{ij}$ as $\eta \rightarrow 0$.    

Expanding the purely gravitational part of the action in \eqref{eq:EHFullAction} to quadratic order in the metric fluctuation gives
\begin{eqn}\label{eq:axial_graviton}
S_{\text{free}}^{\text{gr}}=
\int d^3xd\eta \;\eta^2\Biggr(\frac{1}{4}\partial_\mu h\partial^\mu h-\frac{1}{4}\partial_\mu h_{ij}\partial^\mu h_{ij} -\frac{1}{2}\partial_j h \partial_i h_{ij}+\frac{1}{2}\partial_i h_{jk}\partial_jh_{ki}\Biggr)
 -\frac{1}{2}h^2+ \frac{1}{2}h_{ij}h_{ij} 
\end{eqn}
\noindent where $h=h_{ii}$. The associated SK action is then given by
\begin{eqn}
S_{\text{free gr}}^{SK}[h_>, h_<] &= S_{\text{free gr}}[h_{>}] - S_{\text{free gr}}[h_{<}]
\end{eqn}
where $>,<$ denote time-ordered and anti-time-ordered fields. We may visualise these fields as living on the two branches of the SK contour depicted in Figure \ref{sfig:SKContoura} and impose the following matching conditions at the turning point $\eta=0$:
\begin{eqn}\label{SKbcgrav}
h^{TT, >}_{ij}(0) = h^{TT, <}_{ij}(0)  , \qquad h^{L,>}_{ij}(0)=h^{L,<}_{ij}(0)=0.
\end{eqn} 
Using these boundary conditions, the graviton two-point functions are then given by (see Appendix \ref{app:GR} for details): 
\begin{eqn}
\label{eq:SKGRProps}
\braket{h_{ij}^>(\eta) h_{mn}^>(\eta')} &= i\frac{(\eta\eta')^{-1/2}}{2} \intinf p dp \Bigg( \frac{H_{3/2}^{(2)}(p \eta) H_{3/2}^{(1)}(p \eta')}{p^2 - y^2 - i \e} \Pi_{ijmn} + J_{3/2}(p \eta) J_{3/2}(p \eta')\tilde T_{ijmn} \Bigg), \\
\braket{h_{ij}^<(\eta) h_{mn}^<(\eta')} &= -i\frac{(\eta\eta')^{-1/2}}{2} \intinf p dp \Bigg(\frac{ H_{3/2}^{(1)}(p \eta) H_{3/2}^{(2)}(p \eta')}{p^2 - y^2 - i \e} \Pi_{ijmn} + J_{3/2}(p \eta) J_{3/2}(p \eta')\tilde T_{ijmn} \Bigg), \\
\braket{h_{ij}^>(\eta) h_{mn}^<(\eta')} &= -i \frac{(\eta\eta')^{-1/2}}{2} \int_{C_1} \frac{p dp }{p^2 - k^2 }\Bigg( H_{3/2}^{(2)}(p \eta) H_{3/2}^{(1)}(p \eta') \Pi_{ijmn} \Bigg), \\
\braket{h_{ij}^<(\eta) h_{mn}^>(\eta')} &= i\frac{(\eta\eta')^{-1/2}}{2} \int_{C_2} \frac{p dp }{p^2 - k^2 }\Bigg( H_{3/2}^{(2)}(p \eta) H_{3/2}^{(1)}(p \eta') \Pi_{ijmn} \Bigg), 
\end{eqn}
where $C_1$ and $C_2$ denote the anti-clockwise closed contour over the pole $p = k$ and $p = - k$, respectively. The tensor structures are given by
\begin{eqn}
\Pi_{ijnm}&=\pi_{in}\pi_{jm}+\pi_{im}\pi_{jn}-\pi_{ij}\pi_{nm}  \\
\tilde{T}_{ijnm} &=
\frac{1}{p^2y^2}G^{\text{TL}}_{ijnm} +\frac{p^2+y^2}{p^4}\frac{y_iy_jy_ny_m}{y^4};\\
G^{\text{TL}}_{ijnm}&= \pi_{in} y_j y_m + \pi_{im} y_j y_n + \pi_{jm} y_i y_n +\pi_{jn} y_i y_m  - \pi_{nm} y_i y_j - \pi_{ij} y_n y_m, \\
\pi_{ij} &= \delta_{ij} - \frac{k_i k_j}{k^2}.
\label{eq:tensorStrucDefs}
\end{eqn}
\noindent Near the boundary, the transverse parts of the 2-point functions scale like $\eta^{-2}$ while the longitudinal parts scale like $\eta^{1}$, as expected.

Let us now turn to interactions. For our purposes, we will only need cubic and quartic interactions vertices with two scalars. Expanding \eqref{eq:EHFullAction} to quadratic order, we find the that the relevant vertices are given by
\begin{eqn}
    V_3 = \frac{\kappa}{2} \big(\frac{1}{2}  \eta^{\mu \nu} h \p_\mu \phi \p_\nu \phi  -h_{ij} \p_i \phi \p_j \phi \big) 
    \label{eq:gravitonScalar3}
\end{eqn} 
\begin{eqn}
  V_4 =  \frac{\kappa^2 \eta^2}{8}   \Bigg[\eta^{\mu\nu} ( h^2 - 2 h_{i j} h_{i j}) \p_\mu \phi \p_\nu \phi  + \big(8 h_{i k}h_{k j} -4 h h_{ij}\big)\p_i \phi \p_j \phi \Bigg] 
\label{eq:gravitonScalar4}
\end{eqn}
where $h = h^\rho_\rho = h^i_i$. The corresponding interaction terms in the SK action are simply obtained by doubling the fields:
\begin{eqn}
\label{eq:SKintAction}
 S_{\text{int gr}}^{\text{SK}}&= \int d^3xd \eta \bigg[h_{ij}^> V_{3 \,ij}^{>>} + h_{ij}^> h_{nm}^>  V_{4 \, ijnm}^{>>}-\big(h_{ij}^< V_{3 \,ij}^{<<} + h_{ij}^< h_{nm}^<  V_{4 \, ijnm}^{<<} \big)\bigg].
\end{eqn}

\subsection{Shadow formalism}

To obtain the shadow action, we perform the Wick rotation in \eqref{eq:WickRotationConvention} and the following field redefinitions: 
\begin{eqn}
\label{eq:grav_shadow_WickRotate}
& h^{>}_{ij} = -i h^+_{ij}-h^-_{ij} \\
& h^{<}_{ij} = +i h^+_{ij}-h^-_{ij}  
\end{eqn}
After doing so, the 2-point functions in \eqref{eq:SKGRProps} become unmixed and we obtain
\begin{eqn}
\braket{ h_{ij}^{ \pm}(z)  h_{mn}^{\pm}(z^\prime)}= \mp \int dp 
p \left(zz^{\prime} \right)^{-\frac{1}{2}}\bigg( J_{ \pm 3/2}\left(p z \right) J_{\pm3/2}\left(p z^{\prime} \right)\frac{\Pi_{ijnm}}{p^2+y^2 }+J_{ 3/2}\left(p z \right) J_{3/2}\left(p z^{\prime} \right) \tilde T_{ijnm}(p) \bigg),
\label{eq:gravLongShadowProp}
\end{eqn}
where the tensor structures are as defined in \eqref{eq:tensorStrucDefs} with $p \rightarrow i p$. 
These are represented diagrammatically as bulk-to-bulk propagators: 
\begin{eqn}
G_{h;ij,mn}^-(z_1, z_2, \vec y) = \begin{tikzpicture}[baseline]
\draw[thick] (0,0) circle (2);
\draw[graviton] (-1,0) -- (1,0);
\node at (-1,-0.25) {$-, z_1$};
\node at (1,-0.25) {$-, z_2$};
\node at (-1.15, +0.25) {$i,j$};
\node at (1.15, +0.25) {$m,n$};
\node at (0,0.35) {$\vec y$};
\end{tikzpicture}, \qquad 
G_{h;ij,mn}^+(z_1, z_2, \vec y) = \begin{tikzpicture}[baseline]
\draw[thick] (0,0) circle (2);
\draw[graviton] (-1,0) -- (1,0);
\node at (-1,-0.25) {$+, z_1$};
\node at (1,-0.25) {$+, z_2$};
\node at (-1.15, +0.25) {$i,j$};
\node at (1.15, +0.25) {$m,n$};
\node at (0,0.35) {$\vec y$};
\end{tikzpicture}.
\end{eqn}
Notice that for the $+$ propagator the tensor structures combine into 
\begin{eqn}
    \frac{\Pi_{ijnm}}{p^2+y^2 }+\tilde T_{ijnm} &= \frac{1}{p^2+y^2 }\bigg(T_{in}T_{jm}+T_{im}T_{jn}-T_{ij}T_{nm}\bigg); \\T_{ij}&=\delta_{ij}+\frac{y_i y_j}{p^2}, 
\end{eqn}
Moreover, after the field redefinition in \eqref{eq:grav_shadow_WickRotate}, the graviton kinetic terms become
\begin{eqn}
\label{shadow_axial_graviton}
S^{shadow}_{\text{free gr}} = -\frac{1}{2} &\int d^3 x dz \left(
z\right)^2\\
\Bigg[ &\Big(\frac{1}{2}\partial_\mu h^-_{ij}\partial_\mu h^-_{ij}-\frac{1}{2}\partial_\mu h^-\partial_\mu h^-
  +\partial_ih^-\partial_jh^-_{ji} -  \partial_ih^-_{ik}\partial_jh^-_{jk}-z^{-2}h^-h^-  +z^{-2}h^-_{ij}h^-_{ij} \Big) \\
- &\Big(\frac{1}{2}\partial_\mu h^+_{ij}\partial_\mu h^+_{ij}   -\frac{1}{2}\partial_\mu h^+\partial_\mu h^+ +\partial_ih^+\partial_jh^+_{ji} -  \partial_ih^+_{ik}\partial_jh^+_{jk}-z^{-2}h^+h^+ +z^{-2}h^+_{ij}h^+_{ij} \Big)  \Bigg]
\end{eqn}
For massless scalars, we use the field redefinition in \eqref{eq:shadowBasisChange} specialised to $\Delta_{+}=3$ and $\Delta_{-}=0$:  
\begin{eqn}
    \phi^> &=i\phi^+ +\phi^- \\
    \phi^< &=-i\phi^+ +\phi^-.\\
\label{scalarredef}
\end{eqn}
After doing so, we obtain the following bulk-to-bulk propagators:
\begin{eqn}
    \label{eq:masslessScalarPropagators}
    \braket{ \phi^{ \pm}(z)  \phi^{\pm}(z^\prime)}= \mp \int_0^\infty dp \frac{p J_{\pm\frac32}(p z) J_{\pm \frac32}(p z^\prime)}{p^2+y^2} (z z^\prime)^\frac{3}{2}
\end{eqn}
which correspond to \eqref{eq:EAdSBBProps} specialised to $\nu=\pm\frac32$. Moreover the bulk-to-boundary propagators are given by
\begin{eqn}
    K^\pm_\phi(z) = \mp(1+ k z)e^{-k z}.
\end{eqn}
Plugging \eqref{eq:grav_shadow_WickRotate} and \eqref{scalarredef} into \eqref{eq:SKintAction} then gives the following interaction terms:
\begin{eqn}\label{eq:shadow_axial_graviton_scalar_int}
S_{\text{int}}^{\text{shadow}}&= \int d^3xdz 
\bigg[h_{ij}^- \big(V_{3 \,ij}^{--}- V_{3 \, ij}^{++} \big) +  2h_{ij}^+  V_{3ij}^{+-}
\\&-z^2 \bigg( h_{ij}^- h_{nm}^- \big( V_{4 \, ijnm}^{--}-V^{++}_{4\, ijnm}\big)+ 2h_{ij}^- h_{nm}^+\big( V_{4 \, ijnm}^{+-}+V^{-+}_{4\, ijnm}\big) - h_{ij}^+ h_{nm}^+ \big( V_{4 \, ijnm}^{--}-V^{++}_{4\, ijnm}\big)\bigg)\bigg] 
\end{eqn}

From this, we can easily read off the following three-point vertex factors 
\begin{eqn}
        \begin{tikzpicture}[baseline=(b.base)]
         \begin{feynman}
             \vertex (b);
             \vertex [above left = 20pt and 20pt of b,label ={[xshift=-4pt,yshift=4pt]above left:\(\vec{k}_1, \; - \)}] (l1) ;
             \vertex [below left = 20pt and 20pt of b,label ={[xshift=-4pt,yshift=-4pt]below left:\(\vec{k}_2, \; + \)}] (l2);
             \vertex [right = 28pt of b, label ={[xshift=5.5pt]right:\( +  \)}] (r);
             \vertex [below = 5pt of b](v) {\(z\)};
             
             \diagram*{
                (l1) -- (b),
                (b) --(l2),
                (b) --[graviton, edge label = \(i{,}j\)](r),
             };

             \draw[thick] ($(b)$) circle [radius=34pt];
         \end{feynman}
     \end{tikzpicture}&=\hspace{-0.8cm}\begin{tikzpicture}[baseline=(b.base)]
         \begin{feynman}
             \vertex (b);
             \vertex [above left = 20pt and 20pt of b,label ={[xshift=-4pt,yshift=4pt]above left:\(\vec{k}_1, \; + \)}] (l1) ;
             \vertex [below left = 20pt and 20pt of b,label ={[xshift=-4pt,yshift=-4pt]below left:\(\vec{k}_2, \; - \)}] (l2);
             \vertex [right = 28pt of b, label ={[xshift=5.5pt]right:\( +  \)}] (r);
             \vertex [below = 5pt of b](v) {\(z\)};
             
             \diagram*{
                (l1) --(b),
                (b) --(l2),
                (b) --[graviton, edge label = \(i{,}j\)](r),
             };

             \draw[thick] ($(b)$) circle [radius=34pt];
         \end{feynman}
     \end{tikzpicture}=\kappa \bigg(k_{1, \,i}k_{2, \, j}+k_{1,\,j}k_{2,\,i}+\big(\p_z P_1 \p_z P_2 - \vec{k}_1 \cdot \vec{k}_2\big) \delta_{ij} \bigg)\\
     \vspace{1cm}
     \begin{tikzpicture}[baseline=(b.base)]
         \begin{feynman}
             \vertex (b);
             \vertex [above left = 20pt and 20pt of b,label ={[xshift=-4pt,yshift=4pt]above left:\(\vec{k}_1, \; - \)}] (l1) ;
             \vertex [below left = 20pt and 20pt of b,label ={[xshift=-4pt,yshift=-4pt]below left:\(\vec{k}_2, \; - \)}] (l2);
             \vertex [right = 28pt of b, label ={[xshift=5.5pt]right:\( -  \)}] (r);
             \vertex [below = 5pt of b](v) {\(z\)};
             
             \diagram*{
                (l1) --(b),
                (b) --(l2),
                (b) --[graviton, edge label = \(i{,}j\)](r),
             };

             \draw[thick] ($(b)$) circle [radius=34pt];
         \end{feynman}
     \end{tikzpicture}&=- \Bigg( \hspace{-0.8cm}\;\begin{tikzpicture}[baseline=(b.base)]
         \begin{feynman}
             \vertex (b);
             \vertex [above left = 20pt and 20pt of b,label ={[xshift=-4pt,yshift=4pt]above left:\(\vec{k}_1, \; + \)}] (l1) ;
             \vertex [below left = 20pt and 20pt of b,label ={[xshift=-4pt,yshift=-4pt]below left:\(\vec{k}_2, \; + \)}] (l2);
             \vertex [right = 28pt of b, label ={[xshift=5.5pt]right:\( -  \)}] (r);
             \vertex [below = 5pt of b](v) {\(z\)};
             
             \diagram*{
                (l1) --(b),
                (b) --(l2),
                (b) --[graviton, edge label = \(i{,}j\)](r),
             };

             \draw[thick] ($(b)$) circle [radius=34pt];
         \end{feynman}
     \end{tikzpicture}\Bigg)=\kappa\bigg(k_{1, \,i}k_{2, \, j}+k_{1,\,j}k_{2,\,i}+\big(\p_z P_1 \p_z P_2 - \vec{k}_1 \cdot \vec{k}_2\big) \delta_{ij} \bigg)\\
\label{shadowGravintervertices3pt}
\end{eqn}
where $P_i$ stands for a generic propagator attached to a vertex with boundary momentum $\vec k_i$. Similarly, we find the following Feynman rules for the graviton-scalar four-point vertices: 
\begin{eqn}
      \begin{tikzpicture}[baseline=(b.base)]
         \begin{feynman}
             \vertex (b);
              \vertex [above left = 20pt and 20pt of b,label ={[xshift=-4pt,yshift=4pt]above left:\( \vec k_1, \, - \)}] (l1);
             \vertex [below left = 20pt and 20pt of b,label ={[xshift=-4pt,yshift=-4pt]below left:\( \vec k_2, \, - \)}] (l2);
             \vertex [above right = 20pt and 20pt of b,label ={[xshift=4pt,yshift=4pt]above right:\( - \)}] (r1);
             \vertex [below right = 20pt and 20pt of b,label ={[xshift=4pt,yshift=-4pt]below right:\( - \)}] (r2);

             \diagram*{
                (l1) --(b),
                (b) -- (l2),
                (r1) --[graviton, edge label' = \(i\)](b),
                (b) --[graviton, edge label' = \(j\)]  (r2),
             };
             \draw[thick] ($(b)$) circle [radius=34pt];
         \end{feynman}
         \end{tikzpicture}=- \Bigg( \hspace{-0.8cm}\begin{tikzpicture}[baseline=(b.base)]
         \begin{feynman}
             \vertex (b);
              \vertex [above left = 20pt and 20pt of b,label ={[xshift=-4pt,yshift=4pt]above left:\( \vec k_1, \, - \)}] (l1);
             \vertex [below left = 20pt and 20pt of b,label ={[xshift=-4pt,yshift=-4pt]below left:\( \vec k_2, \, - \)}] (l2);
             \vertex [above right = 20pt and 20pt of b,label ={[xshift=4pt,yshift=4pt]above right:\( + \)}] (r1);
             \vertex [below right = 20pt and 20pt of b,label ={[xshift=4pt,yshift=-4pt]below right:\( + \)}] (r2);

             \diagram*{
                (l1) --(b),
                (b) --  (l2),
                (r1) --[graviton, edge label' = \(i{,}j\)](b),
                (b) --[graviton, edge label' = \(n{,}m\)]  (r2),
             };
             \draw[thick] ($(b)$) circle [radius=34pt];
         \end{feynman}
     \end{tikzpicture} \Bigg)= \frac{\kappa^2 z^2}{2}\bigg( \big(\p_z P_1 \p_z P_2 - \vec k_1 \cdot \vec k_2 \big) \Delta_{ijnm}  + 2 \tilde G^{TL \; 12}_{ijnm} \big)
\label{eq:shadowGravintervertices4pt}
\end{eqn}
Since we will only consider external scalars we have only listed the rules with $-$ scalars and can make some further simplifications. Note that 
\begin{eqn}
    \frac{\partial}{\partial z}\Big\{ (1+k z)e^{-k z} \Big\} = - k^2 z e^{-k z} = -\frac{k^2 z}{1+k z}K_\phi^\pm(z)
\end{eqn}
\noindent so we can replace $z$-derivatives acting on scalar bulk-to-boundary propagators with algebraic terms. Hence we write the effective three- and four-point vertex factors respectively as 
\begin{align}
    \mathcal{V}_{3\,ij}^{ab}&=\kappa \big(\delta_{ij} W_{ab}+k_{a,i} k_{b,j}+k_{a,j} k_{b,i}\big)
    \label{eq:grav3Vertex}, \\
    \mathcal{V}_{4\, ijnm}^{ab} &= -\frac{\kappa^2 z^2}{2}\big(W_{ab} \Delta_{ijnm}  + 2 \tilde G^{ab}_{ijnm} \big)
    \label{eq:grav4Vertex}
\end{align}
where $a$ and $b$ are here particle labels and
\begin{eqn}\label{Wabdefn}
    W_{ab} &= \frac{k_a^2 k_b^2 z^2}{(1+k_a z)(1+k_b z)}- \vec{k}_a \cdot \vec{k}_b\ \\
    \Delta_{ijnm} &= \delta_{in} \delta_{jm} +\delta_{im} \delta_{jn} - \delta_{ij}\delta_{nm};\\
    \tilde{G}^{ ab}_{ijnm}&=\delta_{in}k_{a,j}k_{b,m}+\delta_{im}k_{a,j}k_{b,n}+\delta_{jn}k_{a,i}k_{b,m}+\delta_{jm}k_{a,i}k_{b,n}-\delta_{ij}k_{a,n}k_{b,m}-\delta_{nm}k_{a,i}k_{b,j}.
\end{eqn}

\subsection{Tree-level example}

In this section, we will compute a tree-level graviton exchange correlator with external scalars using the shadow formalism developed in the previous section. We will focus on the s-channel contribution and split it into two parts which encode the transverse-traceless part (TT) and the longitudinal part (L) of the graviton exchange: 
\begin{eqn}
\left\langle \phi\phi\phi\phi\right\rangle _{(s)}&= 
\begin{tikzpicture}[baseline]
    \draw[thick] (0,0) circle (2.25);
    \draw[ solid] (-2,1) -- (-1, 0);
    \draw[ solid] (-2,-1) -- (-1, 0);
    \draw[ solid] (2,1) -- (1, 0);
    \draw[solid] (2,-1) -- (1, 0);
    \draw[graviton] (-1, 0) -- (1, 0);
    \node at (-2.1, 1.25) {$-$};
    \node at (-2.1, -1.25) {$-$};
    \node at (2.1, 1.25) {$-$};
    \node at (2.1, -1.25) {$-$};
    \node at (0, 0.4) {$-$};
\end{tikzpicture}=\mathcal{M}^{\text{TT}}_4 + \mathcal{M}_4^L. 
\end{eqn}
\noindent Note that only a single diagram contributes to the s-channel in the shadow formalism (in contrast to the SK or wavefunction-based formalisms). The other channels are then obtained by permuting the external legs and the full correlator is obtained by summing over all three channels. 

Using the Feynman rules in the previous section, the transverse-tracless part is given by
\begin{eqn}
\mathcal{M}_4^{\text{TT}}&= \frac{1}{4}(\vec k_1^i \vec k_2^j + \vec k_1^j \vec k_2^i )(\vec k_3^n \vec k_4^m + \vec k_3^m \vec k_4^n ) \Pi_{ijnm}\\
\times &\int dp \int_0^\infty  dz' dz \; (z z')^{-1/2}(1+k_1 z)(1+k_2 z)e^{-k_{12} z} \frac{ p J_{-\frac32}(p z) J_{-\frac32}(p z')}{p^2+y^2}(1+k_3 z')(1+k_4 z')e^{-k_{34} z'}
\label{eq:GravitonUnintegratedTrans}
\end{eqn}
To obtain this expression, we only needed the TT part of the graviton propagator in \eqref{eq:gravLongShadowProp} and we neglected the first term in \eqref{eq:grav3Vertex} since $\delta_{ij} \Pi_{ijnm}=0$. Similarly, using the full interaction vertex in \eqref{eq:grav3Vertex} and the longitudinal part of the propagator in \eqref{eq:gravLongShadowProp}, we obtain the following expression for the longitudinal part:
\begin{eqn}
 \mathcal{M}_4^L&= \frac{1}{4} \int dp \int_0^\infty dz dz' (z z')^{-1/2}(1+k_1 z)(1+k_2 z)e^{-k_{12} z} p J_{\frac32}(p z) J_{\frac32}(p z')(1+k_3 z')(1+k_4 z')e^{-k_{34} z'}\\
 \times \Bigg[&\frac{1}{p^2y^2}\bigg( V^{12}_{ij} V^{34}_{nm} \big(G^{TL}_{ijnm}+\frac{y_i y_j y_n y_m}{y^2} \big)-2y^2\big(\vec k_1 \cdot \vec k_2 W_{34} + \vec k_3 \cdot \vec k_4 W_{12} \big)- 3 W_{12} W_{34} \bigg)\\
 + &\frac{1}{p^4}\bigg( V^{12}_{ij}V^{34}_{nm} \frac{y_i y_j y_n y_m}{y^2} + y_i y_j\big(\vec k_1 \cdot \vec k_2 W_{34} + \vec k_3 \cdot \vec k_4 W_{12} \big)+y^2 W_{12} W_{34}\bigg)\Bigg].
\end{eqn}
where 
\begin{eqn}
    V_{ij}^{12}&=(\vec k_1^i \vec k_2^j + \vec k_1^j \vec k_2^i );\\
    V_{nm}^{34}&=(\vec k_3^n \vec k_4^m + \vec k_3^m \vec k_4^n ).
\end{eqn}

After performing the integrals above\footnote{Note that the $z$ integrals in the transverse-traceless part are finite in dimensional regulation. Additionally, the $p$ integral contour is defined to avoid the pole at $p=0$. See appendix A of \cite{Chowdhury:2025ohm} for details.} we find that the transverse-traceless part is given as
\begin{eqn}\label{gravityTT}
\mathcal M_4^{TT}= \tilde f_{22} y^4 \Pi_{22}
\end{eqn}
where
\begin{eqn}
\tilde f_{22} &= f_{22} + \frac{g_3(\vec k_1, \vec k_2, \vec y) g_3(\vec k_3, \vec k_4, \vec y)}{y^3} \\
f_{22} &= \frac{1}{E} - \frac{y^2}{E E_L E_R} + \frac{y k_1 k_2}{E E_L^2 E_R} + \frac{y k_3 k_4}{E E_R^2 E_L} + \frac{2y k_1 k_2 k_3 k_4}{E^2 E_L^2 E_R^2} - \frac{y(k_1 k_2 + k_3 k_4)}{E^2 E_L E_R} \\
&\quad + \frac{k_1 k_2}{E^2 E_L} + \frac{k_3 k_4}{E^2 E_R} + \frac{2 k_1 k_2 k_3 k_4}{E^3 E_L E_R} , \\
g_3(\vec k_1, \vec k_2, \vec k_3) &= \frac{2 k_1 y^2+2 k_2 y^2+2 k_1^2 y+2 k_2^2 y+2 k_1 k_2 y+k_1^3+k_2^3+2 k_1 k_2^2+2 k_1^2 k_2+y^3}{\left(k_1+k_2+y\right){}^2} \\
\Pi_{22} &= \frac{24}{y^2} \Pi^{ijlm} k_i^1 k_j^2 k_l^3 k_m^4; \\
E&=k_{12}+k_{34}, \qquad E_L = k_{12}+y, \qquad E_R = k_{34}+y, \qquad y=|\vec{k}_{1}+\vec{k}_{2}|. \end{eqn}
Moreover the longitudinal part can be expressed as 
\begin{eqn}\label{eq:grav_s_channel_long_part}
 \mathcal M_{4}^{L}=  f_{21} k^2 \Pi_{21} + f_{20} (E_L E_R - k E) \Pi_{20} + f_c
 \end{eqn}
where
\begin{align}
f_{21} &= -\frac{2 k_1 k_2 k_3 k_4}{E^3}-\frac{k_3 k_4 k_{12}+k_1 k_2 k_{34}}{E^2}-\frac{k_{12} k_{34}}{E} = - f_{20}, \nno \\
\Pi_{21} &= \frac{3 \left(k_1-k_2\right) \left(k_3-k_4\right) \left(k^2 \left(t^2-u^2\right)+\left(k_1-k_2\right) \left(k_3-k_4\right) k_{12} k_{34}\right)}{k^6}, \nno \\
t&=\vec{ k}_1 \cdot \vec{k}_4; \quad u= \vec{k}_1 \cdot \vec{k}_3; \nno \\
\Pi_{20} &= \frac{1}{4} \left(1-\frac{3 \left(k_1-k_2\right){}^2}{y^2}\right) \left(1-\frac{3 \left(k_3-k_4\right){}^2}{y^2}\right),  \\
f_c &= -\frac{2 k_1 k_2 k_3 k_4 \left(y^2+k_{12} k_{34}\right)}{E^3}-\frac{9 E^3}{8}-\frac{3 \left(k_1-k_2\right){}^2 \left(k_3-k_4\right){}^2 \left(k_1 k_2+k_3 k_4\right)}{4 E y^2}\nno\\
&-3 \left(k_1 k_2 k_3+k_1 k_4 k_3+k_2 k_4 k_3+k_1 k_2 k_4\right), \nno \\
&-\frac{\left(k_1 k_2 k_3+k_1 k_4 k_3+k_2 k_4 k_3+k_1 k_2 k_4\right) \left(y^2+k_{12} k_{34}\right)}{E^2}\nno \\
&+\frac{1}{8} E \Bigg(-\frac{6 \left(k_1-k_2\right){}^2 \left(k_3-k_4\right){}^2}{y^2}+3 \left(2 y^2+k_{12}^2+k_{34}^2\right)+34 \left(k_1 k_3+k_2 k_3+k_1 k_4+k_2 k_4\right)\nno \\
&\qquad +21 \left(k_1 k_2+k_3 k_4\right)\Bigg) 
-\frac{1}{4E} \Bigg(8 \left(k_1 k_3+k_2 k_3+k_1 k_4+k_2 k_4\right){}^2+24 k_1 k_2 k_3 k_4  \nno\\
&+\frac{1}{2} \left(4 \left(k_1 k_3+k_2 k_3+k_1 k_4+k_2 k_4\right)+3 \left(k_1 k_2+k_3 k_4\right)\right) \left(2 y^2+k_{12}^2+k_{34}^2\right)\nno \\
&\qquad +\left(k_1 k_2+k_3 k_4\right) \left(9 \left(k_1 k_3+k_2 k_3+k_1 k_4+k_2 k_4\right)-12 \left(k_1 k_2+k_3 k_4\right)\right) \Bigg)\nno
\end{align}
The answer matches with known computations in the literature \cite{Seery:2006vu, Seery:2008ax, Ghosh:2014kba, Bonifacio:2022vwa, Armstrong:2023phb}.

\subsection{1-loop example}
Next let us the consider the 1-loop corrections to the scalar self-energy. As in scalar QED, we will only consider tadpole diagrams for simplicity.

\begin{figure}[H]
    \centering
    \begin{tikzpicture}[baseline=(b.base)]
        \begin{feynman}
            \vertex (b);
            \vertex [below left = 30pt and 30pt of b](a) {$k,-$};
            \vertex [below right = 30pt and 30pt of b](c){$k,-$};
            \vertex [above=30pt of b](v);
            \vertex [below right = 5pt and 12pt of v](v1) {$\vec{l}, -$};
            
            \diagram*{
            (a) --  (b) --(c),
            b -- [graviton, out=135, in=45, loop, min distance=1.7cm] b
            };
        \end{feynman}
        \draw[thick] ($(b)$) circle [radius=42.426pt];
    \end{tikzpicture}
    \hspace{1cm}
    $+$
    \hspace{1cm}
    \begin{tikzpicture}[baseline=(b.base)]
        \begin{feynman}
            \vertex (b);
            \vertex [below left = 30pt and 30pt of b](a) {$k,-$};
            \vertex [below right = 30pt and 30pt of b](c){$k,-$};
            \vertex [above=30pt of b](v);
            \vertex [below right = 5pt and 12pt of v](v1) {$\vec{l}, +$};
            
            \diagram*{
            (a) --  (b) -- (c),
            b -- [graviton, out=135, in=45, loop, min distance=1.7cm] b
            };
            \draw[thick] ($(b)$) circle [radius=42.426pt];
        \end{feynman}
    \end{tikzpicture}
    \caption{The two contributions to the tadpole correlator from the shadow formalism in gravity. }
    \label{fig:ShadowTadpoleGR}
\end{figure}
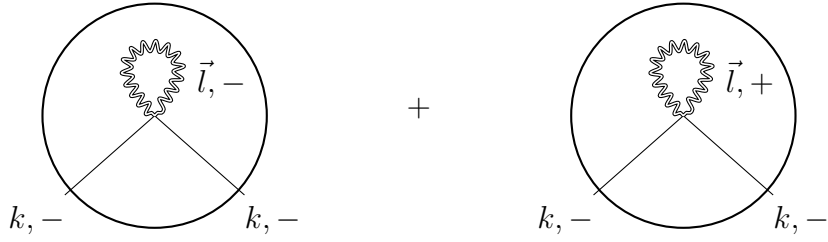

These are depicted in Figure \ref{fig:ShadowTadpoleGR}.
\noindent 

Using the Feynman rules in \eqref{eq:grav3Vertex} and \eqref{eq:grav4Vertex} we find that
\begin{eqn}
   \braket{ \phi(\vec{k})\phi(-\vec{k})} _{\rm tadpole}  &= \int d^3 l \intinf dp \int_0^\infty dz \; 2\mathcal{V}_{4,ijnm}(z) \frac{(1+k z)^2 e^{-2k z}}{z}\\
    &\hspace{3cm}\times\bigg[ \frac{\Pi_{ijnm}}{p^2+l^2} \bigg(J_{-\frac32}(p z)^2+J_{\frac32}(pz)^2 \bigg)
    + 2 J_{\frac32}(pz)^2 \tilde T_{ijnm}  \bigg]  \\
    &= \int d^3l \Bigg( 10+\frac{80 k^3}{l^3}+\frac{56 k}{l}-\frac{3 l^2}{k^2}-\frac{16(6k^5+5k^3 l^2-3k^2l^3-l^5)(\vec{k}\cdot \vec{l})^2}{k^4 l^5}\Bigg)
    \label{shadow_tadpole}
\end{eqn}
Note that integral requires regularisation but we will not consider that in this paper.

Let us now verify that we obtain the same result using the wavefunction based approach. As explained in Appendix \ref{app:wavefunc}, the are two wavefunction coefficients which contribute:
\begin{eqn}
   \braket{ \phi(\vec{k})\phi(-\vec{k})} _{\rm tadpole}=\int d^{3}l\;\left(\psi_{2}^{(1)}(k,k,\vec{l})+\frac{\psi_{4}^{(0)}(k,k,l,l)}{2\psi_{2}^{(0)}(l)}\right)
\end{eqn}
where $\psi_{2}^{(0)}(l)=-\frac{l^3}{2}$ \cite{McFadden:2011CC3, Maldacena:2011ngi} and the relative factor of $\frac{1}{2}$ is a symmetry factor. The corresponding wavefunction coefficients are given by the following Feynman diagrams: 
\begin{eqn}
\psi_2^{(1)} =
\begin{tikzpicture}[baseline]
\draw[thick] (-2, 1) -- (2, 1);
\draw[scalar, solid, black] (-1.5, 1) -- (0, -1);
\draw[scalarbar, solid, black] (1.5, 1) -- (0, -1);

\draw[graviton, black] (0, -1.5) circle (0.5);

\node at (-1.5, 1.25) {$k, +$};
\node at (1.5, 1.25) {$k, +$};
\end{tikzpicture}, \qquad 
\psi_4^{(0)} =  
    \begin{tikzpicture}[baseline]
\draw[thick] (-2, 1) -- (2, 1);
\draw[scalar, solid, black] (-1.5, 1) -- (0, -1);
\draw[scalarbar, solid, black] (-0.5, 1) -- (0, -1);

\draw[graviton, black] (0.5, 1) -- (0, -1);
\draw[graviton, black] (1.5, 1) -- (0, -1);

\node at (-1.5, 1.25) {$k, +$};
\node at (-0.5, 1.25) {$k, +$};
\node at (0.5, 1.25) {$l, +$};
\node at (1.5, 1.25) {$l, +$};
\end{tikzpicture}
\end{eqn}
 
which correspond to the integral expressions 

\begin{eqn}
    \psi_2^{(1)}&= \int d^3 l \int_{-\infty}^\infty dp \int_0^\infty dz \, \mathcal{V}_{4,ijnm}(z) \frac{(1+k z)^2 e^{-2 k z}}{z}  pJ_{\frac{3}{2}}(p z)^2 \bigg[ \frac{\Pi_{ijnm}}{p^2+l^2} + \tilde T_{ijnm}\bigg] \\ 
    \psi_4^{(0)}&= \int d^3l \; \mathcal{V}_{4,ijnm}(z) \Pi_{ijnm} \frac{(1+k z)^2(1+l
    z)^2}{z^4}e^{-2k z}e^{-2l z}
\end{eqn}
where the $\Pi_{ijnm}$ in $\psi^{(0)}_4$ comes from a polarisation sum. By evaluating the $p$ and $z$ integrals, we find perfect agreement with the result from the shadow action in \eqref{shadow_tadpole}.

\subsection{Dressing rules and double copy} \label{dsdoublecopy}

In section \ref{ssec:tree} we showed that the tree-level 4-point correlator in scalar QED can be obtained from flat space Feynman diagrams by applying certain dressing rules for the transverse and longitudinal parts of the photon propagator. In this section we will derive similar dressing rules for the correlator describing scalars exchanging a graviton. Since the longitudinal part of the exchange is very complicated and does not seem to be as canonical as the transverse part, we will focus our attention on the latter. It is conceivable that the longitudinal part can be reconstructed from the transverse part, as we discuss in the conclusion. Using the transverse dressing rules, we then show that the transverse part of the graviton exchange correlator can be obtained from that of gluon exchange in scalar QCD via a double copy inherited from flat space.

Let us now consider the s-channel contribution to the transverse part of the graviton exchange correlator in \eqref{eq:GravitonUnintegratedTrans} and \eqref{gravityTT}. With a bit of algebra we find that it can be written as
\begin{eqn}
    \mathcal{M}_4^{TT} &= \frac{V_{12}^{ij} V_{34}^{nm}}{4}\int dp \hat\Delta^T_h(k_1,k_2,p) \frac{\Pi_{ijnm}}{p^2+y^2}\hat\Delta^T_h(k_3,k_4,p) \\
    &=\begin{tikzpicture}[baseline]
\begin{feynman}
\draw (-2, 1) -- (-1, 0);
\draw (-2, -1) -- (-1, 0);
\draw (2, 1) -- (1, 0);
\draw (2, -1) -- (1, 0);
\draw[graviton,black] (-1, 0) -- (1, 0);
\node at (0,0.3) {$p$};
\draw[dotted, thick] (-1, 0) -- (0, -1);
\draw[dotted, thick] (1, 0) -- (0, -1);
\draw[->] (-0.5, -0.75) -- (-0.75, -0.5);
\node at (-0.75, -0.75) {$p$};
\draw[<-] (0.5, -0.75) -- (0.75, -0.5);
\node at (0.75, -0.75) {$p$};
\end{feynman}
\end{tikzpicture}
\label{4ptgravitydressing}
\end{eqn}
where the auxiliary propagators are given by
\begin{eqn}\label{diffGR}
   \hd_{h}^T(k_i,k_j,p)&= p k_i^2 k_j^2 
   \hat \p_p \hat \p_{k_i} \hat \p_{k_j} \int_0^\infty ds \, s  \frac{k_{ij}+s}{(s+k_{ij})^2+p^2}\\
   &= p k_i^2 k_j^2 
   \hat \p_p \hat \p_{k_i} \hat \p_{k_j} \int_0^\infty ds \, s \hd_{\gamma}^T(k_{ij}+s,p)
\end{eqn}
where $\hd_{\gamma}^T(k,p)$ is the transverse gluon dressing given in \eqref{QEDconfdress1} and $\hat \p_x f \equiv \p_x \left( \frac{f}{x} \right)$. We have used the notation $k_{ij} = k_i + k_j$. Hence, we can obtain the transverse part of the s-channel contribution to the graviton exchange correlator by simply taking the corresponding Feynman diagram in flat space, applying an auxiliary propagator to each vertex (which carries the non-conserved energy at each vertex in de Sitter space) and integrating over the energy flowing through the propagators. We can then obtain the transverse part of the full correlator by summing over all channels (where the other channels are obtained by permuting the external legs).

Let us now explain how to obtain this from the double copy procedure. First note that the 3-point vertices in \eqref{4ptgravitydressing} can be written as
\begin{eqn}
    V_{12}^{ij}&=k_{1}^{i}k_{2}^{j}+k_{1}^{j}k_{2}^{i}\\
    &= \frac{1}{2}\left(y^{i}y^{j}-\alpha^{i}\alpha^{j}\right)
\end{eqn}
where $\vec{y}=\vec{k}_{1} + \vec k_2$, $\vec{\alpha}=\vec{k}_{1}-\vec{k}_{2}$, and there is a similar expression for the right-hand vertex. Since $y^i \Pi_{ijnm}=0$, we can therefore write \eqref{4ptgravitydressing} as follows:
\begin{eqn}
\mathcal{M}_{4}^{TT}=\frac{\alpha^{i}\alpha^{j}\beta^{n}\beta^{m}\Pi_{ijnm}}{16}\int dp\hat{\Delta}_{h}^{T}\frac{1}{p^{2}+y^{2}}\hat{\Delta}_{h}^{T}
\end{eqn}
where $\vec{\beta} = \vec{k}_{3}-\vec{k}_{4}$. Comparing this to the dressed gluon exchange diagram in \eqref{QEDconfdress1}, we see that the graviton diagram can be obtained by squaring the flat space numerators of the gluonic one and replacing gluonic auxiliary propagators with gravitational ones (recall that Feynman diagrams of scalar QED can be uplifted to those of scalar QCD ). Note that when squaring the numerator, we naively obtain $\alpha^{i}\alpha^{j}\beta^{n}\beta^{m}\left(\pi_{in}\pi_{jm}+\pi_{jn}\pi_{im}\right)$. We then subtract $\pi_{ij} \pi_{mn}$ from the tensor structure in order to reproduce the transverse part of the graviton propagator. In flat space, this additional step does not effect the final result since it cancels out after summing over all three channels (see appendix \ref{app:FlatDCDilaton} for more details). On the other hand, in de Sitter space this subtraction is required in order to obtain the correct result for the transverse part of the graviton exchange correlator. It would be interesting to understand the origin of this additional step more systematically as subtracting the contribution of an unwanted dilaton which arises from squaring a gluon.

\section{Spinning correlators}\label{sec:spinning}

In this section we will look at correlators with spinning external legs. First, we write 3-point gluon and graviton correlators in terms of dressed flat space Feynman diagrams and comment on their double copy structure. Then we recast the 4-point tree-level correlator in Yang-Mills in terms of dressed flat space Feynman diagrams and show that the numerators enjoy a kinematic Jacobi relation analogous to the one found for scalar QCD in \eqref{eq:dSJacobi}.

\subsection{Three points}

Let us begin with the color-ordered 3-point gluon correlator \cite{Maldacena:2011nz,Farrow:2018yni}:
\begin{equation}
\left\langle JJJ\right\rangle =\left(\epsilon_{1}\cdot\epsilon_{2}\epsilon_{3}\cdot p_{1}+{\rm cyclic}\right)k_{123}^{-1}
\end{equation}
where the external polarisations point along the boundary, since we are working in axial gauge, and satisfy $\epsilon_i\cdot \vec{k}_{i}=0$. Note that the prefactor of the energy pole is precisely the 3-point scattering amplitude in flat space. Moreover, the energy pole arises from the dressing rule in \eqref{sqeddressing} specialised to the case where all three legs are external (so $p=0$). We may therefore write the 3-point correlator simply as
\begin{eqn}
\left\langle JJJ\right\rangle =
 \begin{tikzpicture}[baseline]
\draw[gluon, black] (0,0) -- (-1, 1);
\draw[gluon, black] (0,0) -- (-1, -1);
\draw[gluon, black] (0,0) -- (1.5, 0);
\draw[dashed] (0,0) -- (0,-1);
\node at (-1, 1.15) {$1$};
\node at (-1, -1.15) {$2$};
\node at (1.5, 0.25) {$3$};
\end{tikzpicture}
= \Delta_{\gamma}^{T}(k_{123},0)\left(\epsilon_{1}\cdot\epsilon_{2}\epsilon_{3}\cdot p_{1}+{\rm cyclic}\right),
\label{threeptgluon}\end{eqn}
where the diagram is a flat space Feynman diagram dressed with an auxiliary propagator. In the next subsection, we will see that the same dressing rule plays a role at higher points.

Next, let's consider the 3-point graviton correlator \cite{Maldacena:2011nz,Farrow:2018yni}:
\begin{equation}
\left\langle TTT\right\rangle =\left(\epsilon_{1}\cdot\epsilon_{2}\epsilon_{3}\cdot p_{1}+{\rm cyclic}\right)^{2}\left(\frac{k_{1}k_{2}k_{3}}{k_{123}^{2}}+\frac{k_{1}k_{2}+{\rm cyclic}}{k_{123}}-k_{123}\right)
\label{graivty3pt}
\end{equation}
This was obtained in \cite{Maldacena:2011nz} by taking the real part of the 3-point wavefunction. As shown in that reference, the latter exhibits an infrared divergence which cancels out after taking the real part (see section \ref{review} for a review of how to relate wavefunction coefficients to in-in correlators). Moreover, the tensor structure in \eqref{gravity3pt} is the three-point Einstein gravity amplitude in flat space, which is simply the square of the 3-point Yang-Mills amplitude. Hence, the double copy for 3-point correlator tensor structure is directly inherited from flat space, as first pointed out in \cite{Farrow:2018yni}. This property can be made more manifest by writing the 3-point graviton correlator in terms of a dressed flat space Feynman diagram as follows:  
\begin{eqn}\label{gravity3pt}
\left\langle TTT\right\rangle 
=\begin{tikzpicture}[baseline]
\draw[graviton, black] (0,0) -- (-1, 1);
\draw[graviton, black] (0,0) -- (-1, -1);
\draw[graviton, black] (0,0) -- (1.5, 0);
\draw[dashed] (0,0) -- (0,-1);
\node at (-1, 1.15) {$1$};
\node at (-1, -1.15) {$2$};
\node at (1.5, 0.25) {$3$};
\end{tikzpicture}
= \left(\epsilon_{1}\cdot\epsilon_{2}\epsilon_{3}\cdot p_{1}+{\rm cyclic}\right)^{2} \Delta_{hhh}(k_{123}).
\end{eqn}
The second diagram is a flat space Feynman diagram dressed with the auxiliary propagator
\begin{equation}
\Delta_{hhh}(k_{123})=-k_1^2 k_2^2 k_3^2 \hat \p_{k_1} \hat \p_{k_2} \hat \p_{k_3} \Bigg(  \intsinf ds \frac{s }{ s + k_{123}}\Bigg) 
\end{equation}
It is straightforward to show that
\begin{eqn}\label{grav3ptdress}
\Delta_{hhh}(k_{123})= \frac{k_{1}k_{2}k_{3}}{k_{123}^{2}}+\frac{k_{1}k_{2}+{\rm cyclic}}{k_{123}}-k_{123}.
\end{eqn}
Note that the integral is divergent but becomes finite after acting with the derivatives. This dressing rule can also be realised as the leading part of the $p \to 0$ limit of the dressing rule in \eqref{diffGR} with an additional external leg:
\begin{eqn}
\Delta_{hhh}(k_{123})=\lim_{p \to 0} p\bigg(p (k_1 k_2 k_3)^2 \hat\p_p\hat\p_{k_1}\hat\p_{k_2} \hat\p_{k_3} \intsinf ds s\frac{k_{123} + s}{(s + k_{123})^2 + p^2}\bigg).  
\end{eqn}
In summary, we can obtain the the 3-point graviton correlator from the 3-point gluon correlator by squaring the flat space amplitude in the dressed Feynman diagram and then replacing the gluonic dressing rule with a gravitational one. For other perspectives on the double copy of 3-point correlators, see \cite{Jain:2021qcl,Lee:2022fgr,Baumann:2024ttn}. In the next section, we will look at 4-point correlators in Yang-Mills theory from the perspective of dressing. We save an analysis of higher-point graviton correlators from this perspective for future work.

\subsection{Four points}

Now let us consider the color-ordered tree-level 4-point Yang-Mills correlator. Using the shadow formalism, the s-channel contribution is given by a sum of two diagrams:
\begin{eqn}
\braket{JJJJ}^{(s)} 
= \begin{tikzpicture}[baseline]
\draw[gluon, black] (-1, 1) -- (-0.5,0);
\draw[gluon, black] (-1, -1) -- (-0.5,0);
\draw[gluon, black] (1, 1) -- (0.5,0);
\draw[gluon, black] (1, -1) -- (0.5,0);
\draw[gluon, black] (-0.5, 0) -- (0.5, 0);
\draw[thick] (0,0) circle (1.45);
\node at (-1, -1.5) {$1,-$};
\node at (-1, 1.5) {$2,-$};
\node at (1, 1.5) {$3,-$};
\node at (1, -1.5) {$4,-$};
\end{tikzpicture}
+ 
\begin{tikzpicture}[baseline]
\draw[gluon, black] (0,0) -- (120:1.45);
\draw[gluon, black] (0,0) -- (60:1.45);
\draw[gluon, black] (0,0) -- (240:1.45);
\draw[gluon, black] (0,0) -- (-60:1.45);

\draw[thick] (0,0) circle (1.45);
\node at (-1, -1.5) {$1,-$};
\node at (-1, 1.5) {$2,-$};
\node at (1, 1.5) {$3,-$};
\node at (1, -1.5) {$4,-$};
\end{tikzpicture}
\end{eqn}
The analogous diagrams contributing to the color-ordered wavefunction were evaluated in \cite{Armstrong:2023phb} using Feynman rules derived in \cite{Raju:2011mp}. To adapt them to the shadow formalism, we simply split the gluon propagator in the exchange diagram into it transverse and longitudinal components and then flip the sign of the spectral parameter appearing the Bessel functions of the transverse part, which flips the boundary conditions of that component from Dirichlet to Neumann. In the end we obtain
\begin{eqn}
\braket{JJJJ}^{(s)} 
&=  V_{12}^{i}V_{34}^{j}\pi_{ij}\intsinf dz_{1}dz_{2}e^{-k_{12z_{1}}}e^{-k_{34}z_{2}}(z_{1}z_{2})^{1/2}\intinf dpp\frac{J_{-1/2}(pz_{1})J_{-1/2}(pz_{2})}{p^{2}+y^{2}}\\
&+V_{12}^{i}V_{34}^{j}\frac{y_{i} y_{j}}{y^{2}}\intsinf dz_{1}dz_{2}e^{-k_{12z_{1}}}e^{-k_{34}z_{2}}(z_{1}z_{2})^{1/2}\intinf dpp\frac{J_{1/2}(pz_{1})J_{1/2}(pz_{2})}{p^{2}}\\
&+V_c^{(s)}\intsinf dz e^{-k_{1234} z}
\end{eqn}
where
\begin{eqn}
V_{ab}^i &= \e_a \cdot \e_b (\vec k_a - \vec k_b)^i + 2 (\e_a \cdot \vec k_b) \e_b^i - 2 (\e_b \cdot \vec k_a) \e_a^i~.\\
V_c^{(s)} &= (\e_1 \cdot \e_3)(\e_2 \cdot \e_4)-(\e_1 \cdot \e_4)(\e_2 \cdot \e_3).
\end{eqn}
The first line corresponds to the transverse part of the exchange, the second to the longitudinal part, and the third encodes a contact interaction. The $z$ integrals arise from Wick-rotating dS$_4$ to EAdS$_4$.

\begin{figure}[H]
\centering
\begin{tikzpicture}[baseline]
\draw[gluon, black] (-2, 1) -- (-1, 0);
\draw[gluon, black] (-2, -1) -- (-1, 0);
\draw[gluon, black] (2, 1) -- (1, 0);
\draw[gluon, black] (2, -1) -- (1, 0);
\draw[gluon, black] (-1,0) -- (1,0);
\draw[dotted, thick] (-1,0) -- (0,-1);
\draw[dotted, thick] (1,0) -- (0,-1);
\node at (-2, -1.5) {$1$};
\node at (-2, 1.5) {$2$};
\node at (2, 1.5) {$3$};
\node at (2, -1.5) {$4$};
\end{tikzpicture} \hspace{1cm} + \hspace{1cm} \begin{tikzpicture}[baseline]
\draw[gluon, black] (0,0) -- (120:1.45);
\draw[gluon, black] (0,0) -- (60:1.45);
\draw[gluon, black] (0,0) -- (240:1.45);
\draw[gluon, black] (0,0) -- (-60:1.45);
\draw[dotted, thick] (0, 0) -- (0, -1.5);
\node at (-1, -1.5) {$1$};
\node at (-1, 1.5) {$2$};
\node at (1, 1.5) {$3$};
\node at (1, -1.5) {$4$};
\end{tikzpicture} 
\caption{Dressed diagrams contributing to the four-point Yang-Mills correlator. The dressing of the exchange diagram encodes both the transverse and longitudinal dressing rules.}
\label{fig:4ptDressedGluon}
\end{figure}
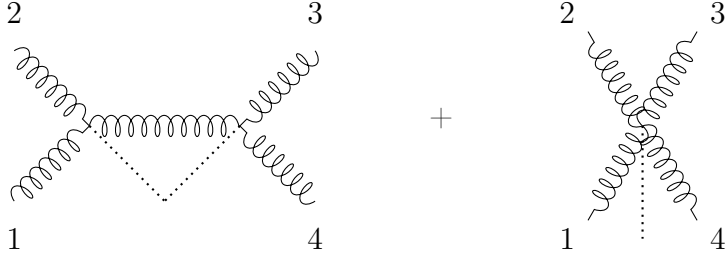

After performing the $z$ integrals, we can write the correlator in terms of dressed flat space diagram, as depicted in Figure \ref{fig:4ptDressedGluon}.
\noindent In more detail, we have
\begin{eqn}\label{gluon4ptdress}
\braket{JJJJ}^{(s)} 
&=  V_{12}^{i}V_{34}^{j}\pi_{ij}\intinf dp\Delta_{\gamma}^{T}(k_{12},p)\Delta_{\gamma}^{T}(k_{34},p)\frac{1}{p^{2}+y^{2}}\\
&+V_{12}^{i}V_{34}^{j}\frac{y_{i} y_{j}}{y^{2}}\intinf dp\Delta_{\gamma}^{L}(k_{12},p)\Delta_{\gamma}^{L}(k_{34},p)\frac{1}{p^{2}}\\
&+V_{c}^{(s)}\Delta_{\gamma}^{T}(k_{1234},0)
\end{eqn}
where $\Delta_\gamma^T$ and $\Delta_\gamma^L$ are the same transverse and longitudinal dressings that appeared in scalar QED. Note that $\Delta_\gamma^T$ also dresses the 4-point contact interaction and the 3-point contact interaction in \eqref{threeptgluon} and  in those cases it reduces to an energy pole. While the transverse dressing rules can be applied to generic diagrams, the longitudinal dressing rule $\Delta_\gamma^L$ is only valid for this diagram and would have to be reworked for higher points using the methods described in appendix E of \cite{Chowdhury:2025ohm}. Finally, note that we can recover the scalar QED result in \eqref{SK4ptQED} via generalised dimensional reduction \cite{Cachazo:2014xea}:
\begin{eqn}
\e_a \cdot \e_b \to 1, \qquad \e_a \cdot k_b \to 0.
\end{eqn}
It would be interesting to extend this to 4-point correlators of fermions \cite{Chowdhury:2024snc}, external gravitons, which we leave for future work.

\subsection*{CK duality}
In section \ref{ssec:dressingrule} we showed that the kinematic numerators of the tree-level 4-point correlator of scalar QCD enjoy a certain kinematic Jacobi relation which reflects an underlying duality between color and kinematics. Let us now show that that this extends to the 4-point Yang-Mills correlator. Starting with \eqref{gluon4ptdress}, we may write the s-channel contribution to the color-ordered correlator as 
\begin{eqn}
\braket{JJJJ}_{(s)} = \intinf  \frac{dp}{ (p^2 + k_{12}^2) (p^2 + k_{34}^2)} \Bigg[ k_{12} k_{34} \frac{V_{12}^i V_{34}^j\pi_{ij}}{p^2 + y^2} + \frac{V_{12}^i V_{34}^j y_i y_j}{y^2} + k_{12} k_{34}  V_{c}^{(s)} \Bigg].
\end{eqn}
As in the scalar QCD case, we pull out the Feynman propagator $(p^2+y^2)^{-1}$ and define the kinematic numerator to be the remaining part of the integrand:
\begin{eqn}
n_{s}(p) = \frac{1}{ (p^2 + k_{12}^2) (p^2 + k_{34}^2)} \Bigg[ k_{12} k_{34} V_{12}^i V_{34}^j\pi_{ij} + (p^2 + y^2) \frac{V_{12}^i V_{34}^jk_i k_j}{y^2} + (p^2 + y^2) k_{12} k_{34}  V_{c}^{(s)} \Bigg]
\end{eqn}
The analogous numerators in the $t$ and $u$ channels are then obtained by permuting the external labels according to \eqref{numchannels}. While the the numerators in all three channels do not directly add to zero as they do in flat space, they do so after integrating over $p$:
\begin{equation}
\int_{-\infty}^{\infty}dp\left(n_{s}(p)+n_{t}(p)+n_{u}(p)\right)=0.
\end{equation}
The interested reader can find the full calculation explicitly verified at \cite{github}. This relation also holds for the kinematic numerators of the wavefunction, similar to the case of scalar QCD described earlier.

We emphasize that for this choice of numerators the Jacobi relation is automatically satisfied without requiring energy conservation and one does not need to introduce the generalized gauge transformations \cite{Armstrong:2020woi}. This was previously seen for calculations of the correlator in Mellin space \cite{Alday:2021odx, Mei:2023jkb} and we now notice it as a novel feature for the correlator in momentum space.

\section{Conclusion} \label{conclusion_sec}
In this paper we revisited the SK formalism for photons, gluons, and gravitons coupled to scalar fields in dS$_4$ and derived local actions in EAdS$_4$ (which we collectively refer to as the shadow formalism) whose Witten diagrams compute in-in correlators of these theories. We described how the residual gauge fixing at the boundary affected the propagators for the longitudinal modes in a non-trivial manner and verified the validity of this approach for a number of examples at tree-level and 1-loop. This also demonstrated its relative simplicity compared to more traditional approaches based on the SK and wavefunction-based formalisms. We then derived a set of dressing rules for uplifting flat space Feynman diagrams to photon and gluon exchange correlators and used them to propose an analogue of colour/kinematics duality for such correlators. Similarly, we found that the transverse traceless part of graviton exchange correlators can also be obtained by dressing flat space Feynman diagrams and derived from gluon exchange diagrams via a simple double copy prescription inherited from flat space.

There are numerous directions for further exploration. Firstly, it would be interesting to extend the dressing rules and double copy to the tree-level 4-point graviton in-in correlator. This is straightforward to do for the transverse traceless part of graviton exchange diagrams but reconstructing the longitudinal part in this way could be challenging. On the other hand, the longitudinal part is identical to that of the 4-graviton wavefunction which has been previously computed in \cite{Bonifacio:2022vwa,Armstrong:2023phb} using bootstrap methods and the double copy. This leads to the broader question of whether the shadow formalism can be combined in a useful way with bootstrap methods such as the cosmological optical theorem \cite{Goodhew:2020hob,Melville:2021lst}, manifestly local test \cite{Jazayeri:2021fvk}, soft limits \cite{Hinterbichler:2013dpa,Chowdhury:2024wwe, Mei:2024sqz}, and conformal Ward identities \cite{Bzowski:2018fql,Bzowski:2019kwd,Armstrong:2020woi,Dymarsky:2013wla} to compute in-in correlators more efficiently. Another important direction would be to develop a more systematic understanding of loop-level in-in correlators in the framework of the shadow formalism. Indeed, in this paper we have only considered simple examples that do not receive contributions from ghosts and we did not attempt to regulate or renormalise them. For recent work on this direction, see for example \cite{Chowdhury:2023arc,Bzowski:2023nef,Jain:2025maa}. Recently, the shadow action of a general higher-derivative scalar effective field theory in dS$_4$ was reconstructed from a boundary perspective by imposing crossing symmetry and certain anomalous dimension constraints on its in-in correlators \cite{Dey:2025kci}. It would be of great interest to use the results of this paper to generalise this analysis to gravitational effective field theories in dS$_4$ and deduce bounds on their coefficients analogous to those in AdS \cite{Caron-Huot:2021enk}. It would also be interesting to extend this formalism to compute correlation functions of gauge invariant operators such as Wilson loops and the field strength tensor \cite{Donnelly:2015hta}.
Finally, it would be be interesting to adapt the formalism developed in this paper to compute other physical observables including out-of-time-order correlators in de Sitter space \cite{Haehl:2017qfl, Haehl:2017eob, Chaudhuri:2018ymp, Loganayagam:2025jmw}. We hope to address these questions in the future. 

\subsubsection*{Acknowledgements}
We would like to thank Debanjan Karan and Babli Khatun for initial collaboration. We also thank Arhum Ansari, Pinaki Banerjee, Sadra Jazayeri, Giorgio Frangi, Ioannis Matthaiakakis, Paul McFadden, Radu Moga, Nilay Kundu, Prashanth Raman, Suvrat Raju, Ivo Sachs, and Omkar Shetye for useful discussions. We especially thank Felix Haehl and Priyadarshi Paul for very useful discussions, and feedback for the material in section \ref{sksection}. AL is supported by the STFC Consolidated Grant ST/X000591/1, CC is supported by the STFC consolidated grant ST/X000583/1, and JM is supported by a Durham Doctoral Teaching Fellowship.

\begin{appendix}

\section{SK propagators \label{app:SKProps}} 
We review the derivation of the SK propagators following the approach of \cite{Weinberg:2005vy}. We describe the consequence of fixing the residual gauge invariance at the turning point and demonstrate how it leads to propagators satisfying Dirichlet boundary conditions. 

\subsection*{Matching conditions}
Consider a single scalar field $\varphi(\eta, \vec x)$ whose boundary value at $\eta = 0$  is denoted by $\phi(\vec x)$. In our conventions $-\infty < \eta < 0$. The correlators are computed at $\eta = 0$ and the boundary  conditions at $\eta \to -\infty(1 - i \e)$ are the standard Hartle-Hawking boundary conditions which implies $\varphi(-\infty(1 - i \e)) = 0$ for $\e \to 0^+$. We evaluate the correlator $\braket{Q(\phi)} = \braket{\Psi|Q(\phi)|\Psi}$ where $Q(\phi)$ is some function of the fields $\phi$, 
\begin{eqn}\label{expecval}
\braket{Q(\phi)} = \int D\phi D\phi' \Psi(\phi) \Psi^*(\phi') \delta(\phi- \phi') Q(\phi) 
\end{eqn}
The wavefunction and its conjugate admit the following path integral representations, 
\begin{eqn}\label{psipsistar}
\Psi(\phi) &=  \int_0^{\varphi^>(0) = \phi} D\varphi^>  e^{i S[\varphi^>]}, \qquad
\Psi^*(\phi') =  \int_0^{\varphi^<(0) = \phi'} D\varphi^<   e^{- i S[\varphi^<]} ~,
\end{eqn}
where the dummy variables $\varphi^>, \varphi^<$  are meant to represent fields on the time-ordered and anti-time-ordered contour, respectively. We also use a shorthand notation $0$ instead of $\varphi^{>/<}_{(-\infty(1 - i \e))} = 0$ for denoting the lower limit of the path integral. Combining \eqref{expecval} and \eqref{psipsistar} we get the following path integral representation for $\braket{Q(\phi)} $: 
\begin{eqn}\label{Qphi2}
\braket{Q(\phi)} = \int D \phi D\phi' \delta(\phi - \phi') \int_{0}^{\varphi^>(0) = \phi} D\varphi^> \int_{0}^{\varphi^<(0) = \phi'} D \varphi^< \ Q(\phi) e^{i S[\varphi^>]} e^{- i S[\varphi^<]} 
\end{eqn}
At this stage it is convenient to spatially discretize the path integral and express the fields $\varphi(\eta, \vec x)$ in terms of discrete mode functions $\varphi(\eta, \vec x) \rightarrow \varphi_n(\eta)$. 
This step is not strictly essential but helps in making some steps more apparent. Hence in these notations the path integral over the field $D\varphi$ is denoted as
\begin{eqn}
D\varphi(\eta, \vec x) \equiv \prod_n d \varphi_n(\eta)
\end{eqn}
In this notation the path integrals over the field at $\eta = 0$, i.e, $ \int D\phi D\phi' \delta(\phi - \phi')$ simply becomes 
\begin{eqn}
\int D\phi D\phi' \delta(\phi - \phi') = \int \prod_n d\varphi^>_n(0)  d\varphi^<_n(0) \prod_m  \delta(\varphi_m^>(0) - \varphi_m^<(0) )
\end{eqn}
Combining these we get the familiar SK representation for \eqref{Qphi2} 
\begin{eqn}
\braket{Q(\phi)} = \int \prod_n d\varphi_n^>(\eta) d\varphi_n^<(\eta) Q(\varphi^>(0)) \prod_m \delta(\varphi^>_m(0) - \varphi^<_m(0)) e^{i S[\varphi^>]} e^{- i S[\varphi^<]}
\end{eqn}
Without the matching condition of the fields at the boundary $\eta = 0$ the two path integrals would be completely decoupled. It is convenient to absorb the matching condition into the exponential via 
\begin{eqn}
\prod_m \delta(\varphi_m^>(0) - \varphi_m^<(0)) &= e^{- \frac{1}{\e}\sum_m (\varphi_m^>(0) - \varphi_m^<(0))^2} \\
&= e^{- \frac{1}{\e} \int d\eta d\eta' \sum_{mm'}\delta_{mm'} \delta(\eta) \delta(\eta') (\varphi_m^>(\eta) - \varphi_m^<(\eta)) (\varphi_{m'}^>(\eta') - \varphi_{m'}^<(\eta'))} \\
&\equiv e^{-\frac1\e \int d\eta d\eta' \sum_{mm'} C_{mm'}(\eta, \eta') (\varphi_m^>(\eta) - \varphi_m^<(\eta)) (\varphi_{m'}^>(\eta') - \varphi_{m'}^<(\eta'))},
\end{eqn}
where in the second step we have introduced some extra integrals to mimic the structure obtained from the standard quadratic action. The normalization of the delta function is unimportant for our discussion. The function $C_{mm'}(\eta, \eta')$ is given as, 
\begin{eqn}\label{Cdef}
C_{mm'}(\eta, \eta') = \delta_{mm'} \delta(\eta) \delta(\eta')
\end{eqn}
The actions $S[\varphi^>]$ and $S[\varphi^<]$ for a scalar field of mass $M$ can also be expressed in the discretized form as 
\begin{eqn}
S[\varphi^>] &=  \int^0_{-\infty} d\eta \frac12 (\nabla^\a \varphi^>)^2 - \frac12 M^2  (\varphi^>)^2 \equiv \sum_{nn'} \intnsinf d\eta d\eta' D^>_{nn'}(\eta, \eta') \varphi^>_n(\eta)  \varphi^>_{n'} (\eta') , \\
S[\varphi^<] &= - \int^0_{-\infty} d\eta \frac12 (\nabla^\a \varphi^<)^2 - \frac12 M^2  (\varphi^<)^2 \equiv \sum_{nn'} \intnsinf d\eta d\eta' D^<_{nn'}(\eta, \eta') \varphi^<_n(\eta)  \varphi^<_{n'} (\eta') , \\
\end{eqn}
where $D^>_{nn'}(\eta, \eta')$ and  $D^<_{nn'}(\eta, \eta')$ are the  discretized version of the standard quadratic operator and are defined with an implicit $\delta(\eta - \eta')$ function in order to resemble \eqref{Cdef}. Hence up to an overall normalization, $\braket{Q(\phi)}$ becomes 
\begin{align}
\braket{Q(\phi)} &= \int \prod_n d \varphi_n^>  d \varphi_n^< Q(\phi) \\
&\times e^{\sum_{nn'} \int d\eta \Big[ i D^>_{nn'}(\eta, \eta') \varphi_n^>(\eta) \varphi_{n'}^>(\eta') - iD^<_{nn'}(\eta, \eta') \varphi_n^<(\eta) \varphi_{n'}^<(\eta') - \frac1\e C_{nn'}(\eta,\eta')(\varphi_n^>(\eta) - \varphi_n^<(\eta)) (\varphi_{n'}^>(\eta') - \varphi_{n'}^<(\eta'))  \Big]}\nonumber
\end{align}
The term inside the square bracket can be written as 
\begin{eqn}
\Big[ \cdots \Big] = \begin{pmatrix}
\varphi^> & \varphi^<
\end{pmatrix} 
\begin{pmatrix}
i D^> - \e^{-1} C & \e^{-1}C \\
\e^{-1} C & - i D^< - \e^{-1}C
\end{pmatrix}
\begin{pmatrix}
\varphi^> \\
\varphi^<
\end{pmatrix} 
\end{eqn}
where we have suppressed the $n, n', \eta, \eta'$ dependences. The propagators for this system are obtained by inverting the matrix above:
\begin{eqn}
\begin{pmatrix}
i D^> - \e^{-1}C & \e^{-1}C \\
\e^{-1} C & - i D^< - \e^{-1} C
\end{pmatrix} \begin{pmatrix}
-i G^{>>} & i G^{><} \\
 i G^{<>} & i G^{<<}
\end{pmatrix} = 
\begin{pmatrix}
1 & 0 \\
0 & 1
\end{pmatrix}.
\end{eqn}
From this equation it is clear that by setting the matching condition $C = 0$, we would not get a mixed propagator. By solving the equation above we obtain the following conditions:
\begin{eqn}
 D^> G^{>>} - \frac{i}{\e} C( G^{>>} + G^{<>}) &= 1, \\
- D^{>} G^{><}+\frac{i}{\e} C ( G^{<<}-G^{><})&=0\\
D^{<} G^{<>}-\frac{i}{\e} C(G^{>>}+ G^{<>})&=0\\ 
 D^{<} G^{<<} + \frac{i}{\e} C(G^{><}-G^{<<})&=1
\end{eqn}
Demanding that the equations are true for arbitrarily small $\e$, we obtain the following constraints: 
\begin{align}
D^> G^{>>} &=  D^< G^{<<} = 1 \label{xy2},\\
C G^{<<} &= C G^{><}, \ C G^{>>} = -C G^{<>}  \label{xy1}, \\
D^> G^{><} &= D^< G^{<>} = 0 \label{xy}
\end{align}
The solutions for the functions, $G^{>>}$ and $G^{<<}$ are given by the inverting \eqref{xy2} and are the standard time-ordered and anti-time-ordered differential propagators 
\begin{eqn}
G_{nn'}^{>>}(\eta, \eta') &= \braket{\mathcal T \phi_n(\eta) \phi_{n'}(\eta')}, \qquad 
G_{nn'}^{<<}(\eta, \eta') = \braket{\bar{\mathcal T} \phi_n(\eta) \phi_{n'}(\eta')},
\end{eqn}
where $\mathcal T$ and $\bar{\mathcal T}$ refer to time-ordering and anti-time-ordering, respectively. The mixed propagators satisfy the homogeneous differential equations \eqref{xy} with the initial condition set by \eqref{xy1} leading to the following solutions (for $\eta, \eta' < 0$, as that is the turning point of the contour), 
\begin{eqn}
G_{nn'}^{><}(\eta, \eta') &= \braket{ \phi_n(\eta) \phi_{n'}(\eta')}, \qquad
G_{nn'}^{<>}(\eta, \eta') = \braket{ \phi_n(\eta') \phi_{n'}(\eta)}.
\end{eqn}
These propagators resemble the usual SK propagators and demonstrates how the matching condition of the fields leads to the mixed propagators. We give explicit forms of the propagators for scalar fields and the transverse part of gauge fields below.

\subsection*{Scalar and transverse propagators}
The position space propagators for the conformally coupled and the massless fields can be represented in this form,
\begin{eqn}\label{Ghankelform1}
G_\nu^{>>}(\eta, \eta') &= (\eta\eta')^{3/2} \intinf \frac{dp}{2\pi} \frac{ p H_{\nu}^{(1)}(p\eta) H_{\nu}^{(2)}(p\eta')}{p^2 - y^2 + i \e},
\end{eqn}
\begin{eqn}\label{Ghankelform2}
\quad G_\nu^{<<}(\eta,\eta') =(\eta\eta')^{3/2} \intinf \frac{dp}{2\pi}  \frac{ p H_{\nu}^{(2)}(p\eta) H_{\nu}^{(1)}(p\eta')}{p^2 - y^2 - i \e}
\end{eqn}
where $\nu = \frac12$ ($\frac32$) for conformally coupled (massless) fields. We perform each of these integrals carefully. First consider $G_\nu^{>>}(\eta)$. For $\nu \in \mathbb R$, and $\eta > \eta'$ the integral is analytic in the upper-half plane whereas for $\eta < \eta'$ it is analytic in the lower-half plane. Hence we can perform the integral by picking up residues. For this we note that the roots of the denominator are given as 
\begin{eqn}
p = y +i \e, \qquad p = - y - i \e 
\end{eqn}
The first case contributes while closing the contour from above and the second case while closing the contour from below. Evaluating the integral via residues then gives
\begin{eqn}
G_\nu^{>>}(\eta, \eta') &= (\eta\eta')^{3/2} \frac{i}{2\pi} \Bigg[ \Theta(\eta - \eta') \intinf \frac{dp}{2\pi}  \frac{p H_{\nu}^{(1)}(p\eta) H_{\nu}^{(2)}(p\eta')}{p^2 - y^2 + i \e} \\
&\qquad \qquad \quad \;+ \Theta(\eta' - \eta)  \intinf \frac{dp}{2\pi} \frac{p H_{\nu}^{(1)}(p\eta) H_{\nu}^{(2)}(p\eta')}{p^2 - y^2 + i \e} \Bigg] \\
&=(\eta\eta')^{3/2}\Big[ \Theta(\eta - \eta') H_\nu^{(2)}(y \eta) H_\nu^{(1)}(y \eta') + \Theta(-\eta) H_\nu^{(1)}(y \eta) H_\nu^{(2)}(y \eta') \Big].
\end{eqn}
Similarly, for $G_{<<}(\eta)$ we get 
\begin{eqn}
G_\nu^{<<}(\eta) = (\eta\eta')^{3/2}\Big[ \Theta(\eta - \eta') H_\nu^{(1)}(y \eta) H_\nu^{(2)}(y \eta') + \Theta(\eta'-\eta) H_\nu^{(2)}(y \eta) H_\nu^{(1)}(y \eta') \Big].
\end{eqn}
For the purpose of finding the propagators for conformally coupled fields, which we couple to gauge fields, we set $\nu= \frac12$ to obtain 
\begin{eqn}
G_\nu^{>>}(\eta, \eta') = \frac{\eta\eta'}{2y} \Big[ \Theta(\eta - \eta') e^{-i y (\eta - \eta')} + \Theta(\eta'-\eta) e^{i y (\eta -\eta')} \Big], 
\end{eqn}
\begin{eqn}
G_\nu^{<<}(\eta, \eta') = \frac{\eta\eta'}{2y} \Big[ \Theta(\eta - \eta') e^{i y (\eta - \eta')} + \Theta(\eta'-\eta) e^{-i y (\eta -\eta')} \Big],
\end{eqn}
which are the standard expressions for the time-ordered and the anti-time-ordered propagators. 

The mixed propagators $G^{><}(\eta,\eta')$ and $G^{<>}(\eta, \eta')$ can be read from the coefficients of the Heaviside Theta functions, 
\begin{eqn}
G^{><}(t,t') &= \frac{\eta\eta'}{2y} e^{i y (\eta - \eta')}, \\
G^{<>}(t,t') &= \frac{\eta\eta'}{2y} e^{i y (\eta' - \eta)}.
\end{eqn}
A similar form of the propagators is obtained for the transverse part of gauge fields with the tensor structure $\pi_{ij}$ (when evaluated in flat space) is shown in \eqref{gaugetransverseprop}.

\subsection*{Decoupling condition}\label{ssec:decoup}
We now discuss the implication of residual gauge fixing at the turning point of the SK path integral. This is applicable for the longitudinal component of the gauge field as discussed in section \ref{sksection}. To avoid a clutter of indices this can be understood by considering two scalar fields $a, \phi$ such that $a$ vanishes at the boundary. The $a$ field mimics the longitudinal component of the gauge field. Thus the only correlators that survive on the boundary are of the field $\phi$:
\begin{eqn}\label{Qphi1}
\braket{Q(\phi)} = \int D \phi Da Q(\phi) \delta(a) |\Psi(\phi, a)|^2 
\end{eqn}
where we have denoted the vanishing of the field $a$ at the boundary with the insertion of the $\delta(a)$ function. This plays the role of the residual gauge fixing condition as described in section \ref{sec:gauge}. The wavefunction $\Psi(\phi, a)$ and its complex conjugate $\Psi^*(\phi, a)$ are now given by
\begin{eqn}
\Psi(\phi, a) &=  \int_0^{\varphi_>(0) = \phi} D\varphi_> \int_0^{ \tilde a_>(0) = a} D\tilde a_> e^{i S[\varphi_>, \tilde a_>]}, \\
\Psi^*(\phi, a) &=  \int_0^{\varphi_<(0) = \phi} D\varphi_<  \int_0^{ \tilde a_<(0) = a} D\tilde a_< e^{- i S[\varphi_<, \tilde a_<]} ~.
\end{eqn}
Plugging these integral representations into \eqref{Qphi1} we get, 
\begin{eqn}
\braket{Q(\phi)} &= \int D \phi Da  \delta(a) \int_0^{\varphi_>(0) = \phi} D\varphi_> \int_0^{\varphi_<(0) = \phi} D\varphi_< \int_0^{\tilde a_<(0) = a} D\tilde a_<  \int_0^{\tilde a_>(0) =a} D\tilde a_>  \\
&\qquad \times e^{i S[\varphi_>, \tilde a_>]}  e^{- i S[\varphi_<, \tilde a_<]} Q(\phi).
\end{eqn}
Performing the integral over $a$ on the support of the $\delta(a)$ we obtain 
\begin{eqn}
\braket{Q(\phi)} &= \int D \phi  \int_0^{\varphi_>(0) = \phi} D\varphi_> \int_0^{\varphi_<(0) = \phi} D\varphi_<  Q(\phi) \\
&\qquad \times   \int_0^{\tilde a_>(0) =0} D\tilde a_>  e^{i S[\varphi_>, \tilde a_>]}  \int_0^{\tilde a_<(0) = 0} D\tilde a_<  e^{- i S[\varphi_<, \tilde a_<]} 
\end{eqn}
Therefore the integrals over $a_>$ and $a_<$ are decoupled and produce no mixed propagator (as the analog of the $C_{mn}$ matrix \eqref{Cdef} is absent here). Since the field $a$ vanishes at the boundary its Green function satisfies the Dirichlet boundary conditions.

\subsection*{Longitudinal propagators}
For a field satisfying Dirichlet boundary conditions, the momentum space propagators are given by replacing the Hankel Functions in \eqref{Ghankelform1} and \eqref{Ghankelform2} with Bessel Functions \cite{Raju:2011mp} which our case are of the form $\sin(p \eta)$,

\begin{eqn}\label{longprop}
 G_D^{>>}(\eta, \eta') = (\eta\eta') i \intinf \frac{dp}{2\pi} \frac{\sin(p \eta) \sin(p \eta')}{p^2 + i \e},
  \qquad 
  G_D^{<<}(\eta, \eta') = - (\eta\eta') i \intinf \frac{dp}{2\pi} \frac{\sin(p \eta) \sin(p \eta')}{p^2 - i \e}
\end{eqn}
Note the similarity between these equations and the Green functions for the longitudinal mode given in \eqref{eq:SKLongProps}. The extra factors of $\frac{y_i y_j}{y^2}$ will not be important for the analysis below and hence we strip them off. 
Explicit computation gives 
\begin{eqn}
G_D^{>>}(\eta, \eta') &= \lim_{\e\to 0}(\eta\eta') \frac{i}{2}\Theta(\eta - \eta') \big[ \eta' - (-1)^{1/4} \eta\eta' \sqrt{\e}   \big] + (\eta\eta')\frac{i}{2}\Theta(\eta' - \eta) \big[ \eta - (-1)^{1/4} \eta\eta' \sqrt{\e}   \big] \\
 &= (\eta\eta') \frac{i}{2} \Big[ \Theta(\eta - \eta') \eta' +  \Theta(\eta' - \eta) \eta\Big] = - G_D^{<<}(\eta, \eta').
\end{eqn}
Since the $a$ fields are decoupled at the boundary as explained in \ref{ssec:decoup} the mixed propagators are zero:
\begin{eqn}
G_D^{><}(\eta, \eta') = G_D^{<>}(\eta, \eta')   = 0~. 
\end{eqn}
From this equation and \eqref{longprop} it is easy to see that these propagators  satisfy the unitarity condition of the Green function \cite{Kamenev:2004jgp}: 
\begin{eqn}
G_D^{>>}(\eta, \eta') + G_D^{<<}(\eta, \eta') + G_D^{><}(\eta, \eta') + G_D^{<>}(\eta, \eta') = 0.
\end{eqn}

\section{Correlators from wavefunctions}\label{app:wavefunc}
In this appendix we review the derivation of the in-in correlators from wavefunction coefficients and derive some of the formulae used in the body of the paper. We remind the reader that we drop overall factors of $k^{-\rho}$ with $\rho = 1$ for conformally coupled scalars/gluons and $\rho = 3$ for massless scalars/gravitons, in the final formulae (see \eqref{4ptwf} for an example). For concreteness, we will mainly focus on scalar QED and briefly comment on the analogous computations in gravity coupled to massless scalars, which are very similar.


Let us consider correlators of the form
\begin{eqn}
&\braket{\Psi| \phi(\vec k_1) \phi^\dagg(\vec k_2) \cdots \phi(\vec k_{2n-1}) \phi^\dagg(\vec k_{2n}) |\Psi}\\
&\qquad\qquad= \int D\phi D\phi^\dagg D\vec A_T |\Psi(\vec A_T, \phi, \phi^\dagg)|^2 \phi(\vec k_1) \phi^\dagg(\vec k_2) \cdots \phi(\vec k_{2n-1}) \phi^\dagg(\vec{k}_{2n}) 
\end{eqn}
where we have set the normalization of the wavefunction to 1. The wavefunction $\Psi(\vec A_T, \phi^\dagg, \phi )$ has the following expansion:
\begin{eqn}\label{WFexpansion}
\Psi(\vec A_T, \phi^\dagg, \phi) 
&=  e^{- \int \psi^{ij}_2(\vec k) A_T^i(\vec k) A_T^j(-\vec k) d^3 k} e^{- \int \psi_2^{\phi^\dagg\phi}(\vec k) \phi^\dagg(-\vec k) \phi(\vec k) d^3 k }\\
&\times \exp \Bigg[ \int \psi_{3i}^{A_T\phi^\dagg\phi}(\vec k_1, \vec k_2, \vec k_3)  A_T^i(\vec k_1)  \phi^\dagg(\vec k_2) \phi(\vec k_3) d^3 k_1 d^3 k_2 d^3 k_3 \delta(\vec k_1 + \vec k_2 + \vec k_3) 
+ \cdots \Bigg],
\end{eqn}
where $\cdots$ denotes the other wavefunction coefficients. 
The quadratic wavefunction coefficients outside the square bracket are 
\begin{eqn}
\psi^{ij}_{2}(\vec k, -\vec k) = \pi^{ij}k, \qquad
\psi_2^{\phi^\dagg \phi}(\vec k, -\vec k) = k. 
\end{eqn}

\subsection*{Tree-level 4-point}
Let us consider the following 4-point function in scalar QED:  
\begin{eqn}\label{4pttreewf}
\braket{ \phi^\dagg_1 \phi_2 \phi^\dagg_3 \phi_4} = \int D\phi^\dagg D\phi D\vec A_T  |\Psi(\vec A_T, \phi^\dagg, \phi)|^2 \phi^\dagg_1 \phi_2 \phi^\dagg_3 \phi_4
\end{eqn}
where we have suppressed the momentum dependence. From the expansion \eqref{WFexpansion} we can deduce the expansion of the square of the wavefunction $|\Psi(\vec A, \phi^\dagg, \phi)|^2$, which can be effectively treated as an action for the path integral whose coefficients are given by the real parts of the wavefunction coefficients. In the present case $\text{Re}\psi_4 = \psi_4$ (see \cite{Thavanesan:2025kyc} for a discussion on more general cases).
To compute the correlator to $O(g^2)$, we need the following terms in the expansion of $|\Psi|^2$: 
\begin{eqn}
|\Psi(\vec A_T, \phi^\dagg, \phi)|^2 &= e^{- \int \psi^{ij}_2(\vec k) A_T^i(\vec k) A_T^j(-\vec k) d^3 k} e^{- \int \psi_2^{\phi^\dagg\phi}(\vec k) \phi^\dagg(-\vec k) \phi(\vec k) d^3 k } \\
&\quad\times \Bigg[
 \int \psi_4^{\phi^\dagg\phi\phi^\dagg\phi}(\vec k_1, \cdots, \vec k_4) \phi^\dagg(\vec k_1) \phi(\vec k_2) \phi^\dagg(\vec k_3)\phi(\vec k_4) d^3 k_1 
\cdots d^3 k_4 \\
&\quad +\frac12 \Big( \int \psi_{3i}^{A_T\phi^\dagg\phi}(\vec k_1, \vec k_2, \vec k_3)  A_T^i(\vec k_1)  \phi^\dagg(\vec k_2) \phi(\vec k_3) d^3 k_1 d^3 k_2 d^3 k_3 \Big)^2 
\Bigg],
\end{eqn}
where we have suppressed the momentum conserving delta function in each term. 
Using Wick contractions, we then find the following two contributions: 
\begin{eqn}\label{4ptwf}
\braket{ \phi^\dagg_1 \phi_2 \phi^\dagg_3 \phi_4} = \frac{1}{\psi_2^{\phi^\dagg\phi}(\vec k_1) \cdots \psi_2^{\phi^\dagg\phi}(\vec k_4) } \Bigg[ \psi_4^{\phi^\dagg \phi \phi^\dagg \phi} + \psi_{3i}^{A_T \phi^\dagg \phi } \psi_{3j}^{A_T \phi^\dagg \phi } \frac{\pi^{ij}}{k}   \Bigg].
\end{eqn}
The expressions for the wavefunctions are given in \eqref{psi4qed} and \eqref{psi3qed}. In practice, we will ignore the prefactor consisting of two-point functions. 

The analogous computation in gravity is very similar. In particular the formula for the 4-point scalar correlator is similar to \eqref{4ptwf} but with more indices:
\begin{eqn}\label{gravtreewf}
\braket{\phi_1 \cdots \phi_4} = \psi_4^{\phi\phi\phi\phi} + \psi_{3ij}^{h_{TT}\phi\phi} \psi_{3mn}^{h_{TT}\phi\phi} \frac{\Pi^{ijmn}}{y^3}
\end{eqn}
where the factor $\frac{\Pi^{ijmn}}{y^3}$ arises from the polarization sums of the transverse-traceless gravitons. 

\subsection*{1-loop tadpole}
Next let's derive the 1-loop tadpole contribution to the scalar 2-point function in \eqref{tadpolewf}. The following terms in \eqref{WFexpansion} will be needed for this purpose: 
\begin{eqn}
|\Psi(\vec A, \phi^\dagg, \phi)|^2 &= e^{- \int \psi^{ij}_2(\vec k) A_T^i(\vec k) A_T^j(-\vec k) d^3 k} e^{- \int \psi_2^{\phi^\dagg\phi}(\vec k) \phi^\dagg(-\vec k) \phi(\vec k) d^3 k } \\
&\quad\times \Bigg[
 \int \psi_{4ij}^{\phi^\dagg\phi A_T A_T}(\vec k_1, \cdots, \vec k_4) \phi^\dagg(\vec k_1) \phi(\vec k_2) A_T^i(\vec k_3) A_T^j(\vec k_4) d^3 k_1 
\cdots d^3 k_4 \\ 
&\qquad + \int \psi_{2}^{(1)\phi^\dagg\phi}(\vec k_1, \vec k_2) \phi^\dagg(\vec k_1) \phi(\vec k_2)  d^3 k_1 
d^3 k_2 \Bigg].
\end{eqn}
The two-point function can then be computed by performing Wick contractions. To $\mathcal{O}(g^2)$ we obtain
\begin{eqn}
\braket{\phi^\dagg_1 \phi_2} = \int d^3 l \Bigg[  \frac{\pi^{ij}}{2l}\psi_{4ij}^{\phi^\dagg \phi A_T A_T}(\vec k, \vec k, \vec l, \vec l)  + \psi_2^{(1) \phi^\dagg\phi}(\vec k, \vec k) \Bigg],
\end{eqn}
where we have ignored the overall factors depending on the 2-point functions of the external momenta. The expression for the wavefunction coefficients are computed in \eqref{tadpolewfcoeff}.

\section{Graviton 2-point functions \label{app:GR}}
In this Appendix, we will  provide more details about the derivation of \eqref{eq:gravLongShadowProp}. Starting with the graviton decomposition $h_{ij}=h^{TT}_{ij} + \chi\delta_{ij} + \partial_{(i}\xi_{j)}$, with $\partial_i h^{TT}_{ij}=0 $ and $h^{TT}_{ii}=0 $,  the action \eqref{eq:axial_graviton} becomes
\begin{eqn}\label{eq:axial_graviton_full_decompose}
S_{\text{free gr}} &= S_{\text{TT}}\Bigl[h^{TT}_{ij} \Bigr] + S_{\chi+\xi}\Bigl[\xi_i,\chi \Bigr],
\end{eqn}
where
\begin{eqn}\label{eq:axial_graviton_tt}
S_{\text{TT}} &= 
\int d^4x \eta^2\Bigl( \frac{1}{4}\partial_\eta h^{TT}_{ij}\partial_\eta h^{TT}_{ij} - \frac{1}{4}\partial_k h^{TT}_{ij}\partial_k h^{TT}_{ij}+\frac{1}{2}\eta^{-2}h^{TT}_{ij}h^{TT}_{ij} \Bigr),
\end{eqn}
and
\begin{eqn}\label{eq:axial_graviton_longitudinal_trace}
S_{\chi+\xi}&=
\int d^4x \eta^2 \bigg[\frac{1}{4}\Bigl(-6\partial_\eta\chi\partial_\eta\chi - 4\partial_\eta\chi\partial_\eta\partial\cdot\xi +\frac{1}{2}\partial_\eta\partial_j\xi_i\partial_\eta\partial_j\xi_i- \frac{1}{2}\partial_\eta\partial\cdot\xi\partial_\eta\partial\cdot\xi  \Bigr)\\&+\frac{1}{2}\eta^{-2}\Bigl(-6 \chi^2 - 4 \chi \partial\cdot\xi +\frac{1}{2} \partial_j\xi_i \partial_j\xi_i- \frac{1}{2} \partial\cdot\xi \partial\cdot\xi  \Bigr) +\frac{1}{2}\partial_i\chi\partial_i\chi\bigg].
\end{eqn}

Next we Wick rotate to EAdS$_4$ following \eqref{eq:WickRotationConvention}. We will only consider the $> $ branch of SK contour for simplicity. The action for this branch becomes
\begin{eqn}\label{eq:axial_graviton_longitudinal_trace_E}
S_{\chi+\xi}^{\text{ E}}&= 
\int d^4x z^2 \bigg[\frac{1}{4}\Bigl(-6\partial_z\chi\partial_z\chi - 4\partial_z\chi \partial_z\partial\cdot\xi +\frac{1}{2} \partial_z\partial_j\xi_i \partial_z\partial_j\xi_i- \frac{1}{2} \partial_z\partial\cdot\xi \partial_z\partial\cdot\xi  \Bigr)\\&+\frac{1}{2}z^{-2}\Bigl(-6 \chi^2 - 4 \chi \partial\cdot\xi +\frac{1}{2} \partial_j\xi_i \partial_j\xi_i- \frac{1}{2} \partial\cdot\xi \partial\cdot\xi  \Bigr) -\frac{1}{2}\partial_i\chi\partial_i\chi\bigg].
\end{eqn}
The action can be further simplified if we decompose $\xi_i=\hat{\xi}_i + \partial_i\sigma $:
\begin{eqn}
S_{\chi+\xi}^{\text{ E}}&= 
\int d^4x z^2 \bigg[\frac{1}{4}\Bigl(-6\partial_z\chi\partial_z\chi - 4\partial_z\chi \partial_z\partial^2\sigma +\frac{1}{2} \partial_z\partial_j\hat{\xi}_i \partial_z\partial_j\hat{\xi}_i \Bigr)\\&+\frac{1}{2}z^{-2}\Bigl(-6 \chi^2 - 4 \chi \partial^2\sigma +\frac{1}{2} \partial_j\hat{\xi}_i \partial_j\hat{\xi}_i\Bigr) -\frac{1}{2}\partial_i\chi\partial_i\chi\biggr].
\end{eqn}
Recalling that the longitudinal modes obey Dirichlet boundary conditions, we find that
\begin{eqn}
\langle \chi^>(\vec{x},z) \chi^>(\vec{x}^{\prime},z^{\prime}) \rangle_{\text{E}} &=0\\
\langle \chi^>(\vec{x},z) \sigma^>(\vec{x}^{\prime},z^{\prime}) \rangle_{\text{E}} &=\int \frac{pdpd^3y}{(2\pi)^3}e^{\text{i}\vec{y}\cdot(\vec{x}-\vec{x}^{\prime}) }\frac{J_{3/2}(pz)J_{3/2}(pz^{\prime})}{y^2p^2}\\
\langle \sigma^>(\vec{x},z) \sigma^>(\vec{x}^{\prime},z^{\prime}) \rangle_{\text{E}} &=\int \frac{pdpd^3y}{(2\pi)^3}e^{\text{i}\vec{y}\cdot(\vec{x}-\vec{x}^{\prime}) }\frac{J_{3/2}(pz)J_{3/2}(pz^{\prime}) (y^2+3p^2)}{y^4p^4}\\
\langle \hat{\xi}^>_i(\vec{x},z) \hat{\xi}^>_j(\vec{x}^{\prime},z^{\prime}) \rangle_{\text{E}} &=\int \frac{pdpd^3y}{(2\pi)^3}e^{\text{i}\vec{y}\cdot(\vec{x}-\vec{x}^{\prime}) }\frac{4J_{3/2}(pz)J_{3/2}(pz^{\prime})  \pi_{ij}}{y^2p^2}. 
\end{eqn}
Recall that $h^L_{ij}=\chi\delta_{ij} + \partial_{(i}\xi_{j)}=\chi\delta_{ij} + \partial_{(i}\hat{\xi}_{j)} + \partial_i\partial_j\sigma$. From this we deduce the following 2-point functions of longitudinal modes:
\begin{eqn}\label{eq:axial_graviton_long_prop}
&\langle h_{ij}^{L>}(\vec{x},z)h_{kl}^{L>}(\vec{x}^{\prime},z^{\prime})  \rangle_E\\ 
&=\int  \frac{pdpd^3y}{(2\pi)^3}e^{\text{i}\vec{y}\cdot(\vec{x}-\vec{x}^{\prime}) }J_{3/2}(pz)J_{3/2}(pz^{\prime})\bigg[\frac{1}{k^2p^2}\Bigl(\pi_{ik}y_j y_l+\pi_{il}y_j y_k + \pi_{jk}y_iy_l+\pi_{jl}y_i y_k \\& -y_iy_j\pi_{kl}-y_k y_l\pi_{ij} \Bigr)  +y_i y_j y_k y_l \frac{(y^2+p^2)}{y^4p^4}     \bigg].
\end{eqn}
The 2-point functions of $h_{ij}^{L<}$ take a similar form and $\braket{h^{L>}h^{L<}}_E =\braket{h^{L<}h^{L>}}_E =0$. Moreover, the 2-point functions of the transverse-traceless modes of the graviton are straightforward to derive and are given by 
\begin{eqn}
\braket{h_{ij}^{TT>}(\vec x, z) h_{mn}^{TT>}(\vec x', z')}_E &= \Pi_{ijmn} \frac{ (z z')^{1/2}}{2} \int \frac{p dp d^3 y}{(2\pi)^3} e^{i \vec y \cdot (\vec x - \vec x')}  \frac{H_{3/2}^{(2)}(p z) H_{3/2}^{(1)}(p z')}{p^2 + y^2}, \\
\braket{h_{ij}^{TT<} (\vec x, z)h_{mn}^{TT<}(\vec x', z')}_E &= \Pi_{ijmn} \frac{ (z z')^{1/2}}{2} \int \frac{p dp d^3 y}{(2\pi)^3} e^{i \vec y \cdot (\vec x - \vec x')}  \frac{H_{3/2}^{(1)}(p z) H_{3/2}^{(2)}(p z')}{p^2 + y^2}, \\
\braket{h_{ij}^{TT>}(\vec x, z) h_{mn}^{TT<}(\vec x', z')}_E &= 0,\\ 
\braket{h_{ij}^{TT<}(\vec x, z) h_{mn}^{TT>}(\vec x', z')}_E &= 0. 
\end{eqn}
To obtain the 2-point functions of the shadow fields, we perform the following field redefinitions which unmix the 2-point functions of the transverse-traceless modes:
\begin{eqn}
h^-_{ij}&=-\frac{h^{>}_{ij}+h^{<}_{ij}}{2},\qquad h^+_{ij}=-i\frac{h^{>}_{ij}-h^{<}_{ij}}{2}.
\end{eqn}
After implementing this change of basis, we obtain the two point functions given in \eqref{eq:gravLongShadowProp}.

\section{Flat space double copy \label{app:FlatDCDilaton}}
In this Appendix, we will review the double copy relating tree-level 4-point scattering amplitudes of massless scalars exchanging gluons and gravitons in flat space. Sine the amplitudes are gauge-invariant we will work in de Donder gauge for simplicity. Recall that in de Donder gauge ($g^{\mu\nu} \Gamma^\rho_{\mu\nu} = 0$ where $\Gamma^\mu_{\nu\rho}$ is the Christoffel symbol) the graviton propagator is
\begin{eqn}
    G_{\mu \nu \sigma \rho} = \frac{\eta_{\mu \rho} \eta_{\nu \sigma}+\eta_{\mu \sigma} \eta_{\nu \rho} - \eta_{\mu \nu} \eta_{\sigma \rho}}{p^2} \equiv \frac{T_{\mu \nu \sigma \rho}}{p^2}
\end{eqn}
and the scalar-graviton three-point vertex is 
\begin{eqn}
    \begin{tikzpicture}[baseline=(v.base)]
    \begin{feynman}
    \vertex (v) ; 
    \vertex [above left = 20pt and 20pt of v](a) {\(p\)};
    \vertex [below left = 20pt and 20pt of v](b){\(q\)};
    \vertex [right = 28pt of v](c);

    \diagram{
    (a) --  (v) -- (b),
    (v) --[graviton] (c),
    };
    \end{feynman}
\end{tikzpicture}=\frac{\kappa}{2}\bigg(p_\mu q_\nu + p_\nu q_\mu - \eta_{\mu \nu} p^\lambda q_\lambda\bigg) \equiv V^{qp}_{\mu \nu}.
\end{eqn}
For the four-point exchange, we wish to write this in terms of the colour-ordered scalar QCD vertex factors and the exchanged momentum: 
\begin{eqn}
    \alpha^\mu &= (p_1 - p_2)^\mu, \\
    \beta^\mu &= (p_3-p_4)^\mu, \\
    y^\mu &= (p_1+p_2)^\mu=-(p_3+p_4)^\mu.
\end{eqn}
To do this, note that for four-momenta in flat space we have
\begin{eqn}
    \eta^{\mu \nu} k_1^\mu k_2^\nu &= \frac{1}{2} \eta^{\mu \nu} y_\mu y_\nu = -\frac{1}{2}\eta^{\mu \nu} \alpha_\mu \alpha_\nu, \\
    \eta^{\mu \nu} k_3^\mu k_4^\nu &= \frac{1}{2} \eta^{\mu \nu} y_\mu y_\nu = -\frac{1}{2}\eta^{\mu \nu} \beta_\mu \beta_\nu, \\
    k_{1\mu} k_{2\nu} + k_{1 \nu} k_{2\mu}&= \frac{1}{2}\big(y_\mu y_\nu - \alpha_\mu \alpha_\nu \big).
\end{eqn}
Hence,
\begin{eqn}
    V^{12}_{\mu \nu} &= \frac{\kappa}{4} \bigg( y_\mu y_\nu - \alpha_\mu \alpha_\nu + \eta_{\mu \nu}\alpha^2\bigg),\\
    V^{34}_{\mu \nu} &= \frac{\kappa}{4} \bigg( y_\mu y_\nu - \beta_\mu \beta_\nu + \eta_{\mu \nu}\beta^2\bigg),
\end{eqn}
and the s-channel contribution to the amplitude is 
\begin{eqn}
    \mathcal{A}_{4,\rm{GR}}^{(s)}=\begin{tikzpicture}[baseline=(v1.base)]
    \begin{feynman}
    \vertex (v1) ; 
    \vertex [above left = 20pt and 20pt of v1](a) {\(p_1\)};
    \vertex [below left = 20pt and 20pt of v1](b){\(p_2\)};
    \vertex [right = 28pt of v1](v2);
    \vertex [above right = 20pt and 20pt of v2](c) {\(p_3\)};
    \vertex [below right = 20pt and 20pt of v2](d){\(p_4\)}; 

    \diagram{
    (a) -- [scalar] (v1) -- [scalar](b),
    (v) --[graviton] (v2),
    (c) -- [scalar] (v2) -- [scalar](d),
    };
    \end{feynman}
\end{tikzpicture}= V^{12}_{\mu \nu} \frac{T^{\mu \nu \sigma \rho}}{s}V^{34}_{\sigma \rho}.
\end{eqn}
After some algebra making use of the massless external kinematics, we find that 
\begin{eqn}
    V_{\mu \nu}^{12}(y,\alpha) T^{\mu \nu \rho \sigma}V_{\rho \sigma}^{34}(y,\beta)&= \frac{\kappa^2}{16}\Big(\alpha^\mu \alpha^\nu \mathcal{T}_{\mu \nu \sigma \rho}\beta^\sigma \beta^\rho - \alpha^2 \beta^2 \Big)\\
    &=\frac{\kappa^2}{16 }\alpha^\mu \alpha^\nu \big(\eta_{\mu \rho }\eta_{\nu \sigma }+\eta_{\mu \sigma}\eta_{\nu \rho}-2\eta_{\mu \nu}\eta_{\rho \sigma} \big) \beta^\rho \beta^\sigma.
    \label{eq:FlatLorentzSimpGraviton}
\end{eqn}
Let's now compare \eqref{eq:FlatLorentzSimpGraviton} with the square of the s-channel scalar QCD numerator
\begin{eqn}
    n_s^2 = \alpha^\mu \alpha^\nu \big(\eta_{\mu \rho }\eta_{\nu \sigma }+\eta_{\mu \sigma}\eta_{\nu \rho}\big) \beta^\rho \beta^\sigma.
\end{eqn}
Clearly, the third term in  \eqref{eq:FlatLorentzSimpGraviton} is not generated by this procedure. The reason is that when we square a gluon, this gives a 2-form and dilaton in addition to a graviton and the third term in \eqref{eq:FlatLorentzSimpGraviton} is cancelled by the dilaton exchange arising from the double copy. Hence, we must subtract the dilaton contribution in order to obtain a graviton propagator in each channel. On the other hand, this contribution vanishes after summing over all three channels since $\alpha_\mu \alpha^\mu = \beta_{\mu} \beta^{\mu}= -s$, so after dividing the third term in \eqref{eq:FlatLorentzSimpGraviton} by $s$ and summing over all three channels we get $s+t+u=0$. Hence, we still get a graviton exchange amplitude even if we do not explicitly remove the dilaton contribution after taking the double copy of a gluon amplitude. In contrast, this contribution does not cancel out after summing over channels in dS$_4$ so we must explicitly subtract it when implementing the double copy of scalar QCD correlators, as described in section \ref{dsdoublecopy}.    

\section{Relation to non-local formulations} \label{app:nonLocal}
In this Appendix we relate our local shadow formulations of photons and gravitons coupled to scalars to non-local formulations which are more commonly used in the literature \cite{Maldacena:2002vr, Maldacena:2011nz, Seery:2006vu, Seery:2008ax, Bonifacio:2022vwa, McFadden:2009fg, Bzowski:2011ab, McFadden:2010vh}. 
We note that local formulations are in-principle more convenient for computing higher-point  correlation functions (see \cite{Kabat:2012av} for a discussion of computing higher-point vertices using a non-local action).


\subsection*{QED}
First we derive a non-local shadow action for scalar QED by using the Gauss law to solve for the longitudinal component of the gauge field. This is equivalent to integrating out the longitudinal component of the field from the path integral. By restricting to the linear sector in the action \eqref{eq:SQEDShadowFull} we get \cite{Liu:1998ty}
\begin{eqn}
S =- \frac12 \int \frac{d\eta d^3 x}{\eta^4} A_{Ti} (\partial^2) A_{T}^i - \frac12 \int \frac{d\eta d^3 x}{\eta^4} J^0(z, \vec x) \frac{1}{\p^2} J^0(\eta, \vec x)
\end{eqn}
The first term denotes the transverse part of the gauge field and the second term arises by integrating out the longitudinal mode and is a contact term with $J^{(0)}(\eta, \vec x)$ given as
\begin{eqn}
J^{0}(\eta, \vec x) &=  \phi^\dagg \p_\eta \phi - \phi \p_\eta \phi^\dagg.
\end{eqn}
For scalar QED the current is quadratic in the field and hence this a 4-point interaction term. This term contains an inverse Laplacian and hence is non-local.

From this we can easily obtain a shadow action by performing the procedure described in section \ref{ssec:GaugeShadow} resulting in the following non-local action:
\begin{eqn}
S_{\rm non-local}^{\rm shadow} &= -\frac12  \int \frac{dz d^3 x}{z^4} \Big[ A_{Ti}^{-} (\partial^2) A_i^{-T} -  A_{Ti}^{+} (\partial^2) A_i^{+ T} \Big]\\
&\qquad + ig \eta^{\mu\nu} \bigg[A_\mu^+\big({\phi^-}^\dagger\p_\nu \phi^+-\phi^+ \p_\nu{\phi^-}^\dagger   + {\phi^+}^\dagger \p_\nu \phi^- - \phi^- \p_\nu {\phi^+}^\dagger\big)\\
     &+A_\mu^-\big({\phi^+}^\dagger\p_\nu \phi^+-\phi^+ \p_\nu{\phi^+}^\dagger   -(  {\phi^-}^\dagger \p_\nu \phi^- - \phi^- \p_\nu{\phi^-}^\dagger)\big) \bigg]\\
& \qquad - \frac12 \int \frac{dz d^3 x}{z^4} \Big[ \big( J_{-+}^0(z, \vec x) + J_{+-}^0(z, \vec x) \big) \frac{1}{\p^2} \big( J_{-+}^0(z, \vec x) + J_{+-}^0(z, \vec x) \big) \\
&\qquad \qquad - \big( J_{--}^0(z, \vec x) + J_{++}^0(z, \vec x) \big) \frac{1}{\p^2} \big( J_{--}^0(z, \vec x) + J_{++}^0(z, \vec x) \big)  \Big],\label{nonlocalshadow}
\end{eqn}
where we use the notation
\begin{eqn}\label{currentnonlocal}
J^0_{ab} = \phi^{a\dagg} \p_z \phi^b - \phi^{b} \p_z \phi^{a\dagg}
\end{eqn}
with $a, b = \pm$. The same action is obtained if one integrates out the longitudinal degrees of freedom from the full shadow action \eqref{eq:SQEDShadowFull}.

Next let us compare the 4-point photon exchange correlator with external scalars obtained using the non-local action in \eqref{nonlocalshadow} to what we obtain using the local action in section \ref{sec:gauge}. Note that we only consider $-$ fields on the external legs, as before. 
Using the action \eqref{nonlocalshadow}, the s-channel contribution to the correlator is given by a sum of two terms: 
\begin{eqn}
\braket{\phi_1 \phi^\dagg_2 \phi_3 \phi^\dagg_4}_{(s)}  = C_T + C_L
\end{eqn}
where $C_T$ and $C_L$ denote the transverse and longitudinal parts of the photon exchange, respectively. The transverse part is given by
\begin{eqn}
C_T &= \a^i \b^j \pi_{ij} \intsinf dz_1 dz_2 e^{- k_{12} z_1} e^{- k_{34} z_2} \intinf dp \frac{\cos(p z_1) \cos(p z_2)}{p^2 + y^2} \\
&= \a^i \b^j \pi_{ij} \Big( \frac{1}{(k_{12} + k_{34} )(k_{12} + y)(k_{34} + y) } + \frac{1}{(k)(k_{12} + y)(k_{34} + y) } \Big),
\end{eqn}
which is exactly the transverse part of the correlator found in section \ref{ssec:GaugeShadow}. The longitudinal part is given by a single diagram with four external legs:
\begin{eqn}\label{4ptnonlocalshadow}
C_4^L = \intsinf d z j_{--}^z(\vec k_1, \vec k_2, z)  \frac{1}{y^2}  j_{--}^z(\vec k_3, \vec k_4, , z),
\end{eqn}
where the function $j_{--}^z(z)$ is obtained from \eqref{currentnonlocal} by plugging in the form of the bulk-boundary propagators and is given as
\begin{eqn}
j^z_{--}(\vec k_a, \vec k_b, z) = (k_a - k_b)e^{- k_{ab}z}.
\end{eqn}
By performing the integral in $z$ in \eqref{4ptnonlocalshadow}, 
\begin{eqn}
C_4^L = \frac{(k_1 - k_2)(k_3 - k_4)}{k_{12} + k_{34}},
\end{eqn}
we find an explicit match with the longitudinal part of the answer computed using the local form of the action given in \eqref{QED4ptresult1}.

However it is also possible to show more formally how the form of the answer found using the local form matches with the answer from the non-local form using integration by parts (IBP).  To do this we start with the expression for the longitudinal part obtained using the local form of the action as given in section \ref{ssec:GaugeShadow}: 
\begin{eqn}
  C_{4,local}^L &= \int_{zx} J^i(\vec x, z) G_{ij}^L(\vec x, z; \vec x', z') J^i(\vec x', z') 
\end{eqn}
where 
\begin{eqn}
G^L_{ij}(\vec x, z; \vec x', z') &= \int d^3 k e^{i \vec k \cdot (\vec x - \vec x')} \intinf \frac{dp \sin(p z) \sin(p z')}{p^2}  \frac{y_i y_j}{y^2}. 
\end{eqn}
The exact form of $J^i$ will not be relevant here, but we will only be using the conservation equation 
\begin{eqn}\label{conservationeq1}
\p_i J^i = - \p_z J^z,
\end{eqn}
where $\p_i \equiv \p_{x_i}$. We now use the following identity: 
\begin{eqn}
 &\int d^3 y e^{i \vec y \cdot (\vec x - \vec x')} \intinf \frac{dp \sin(p z) \sin(p z')}{p^2 }  \frac{y_i y_j}{y^2}  
 =- \p_{x_i} \p_{x'_j} \int d^3 y e^{i \vec y \cdot (\vec x - \vec x')} \intinf \frac{dp \sin(p z) \sin(p z')}{p^2 y^2},
\end{eqn}
and also use the shorthand notation $\int_{xz}$ to denote 
\begin{eqn*}
\int_{xz} = \int d^3 x d^3 x' \intsinf dz dz'
\end{eqn*}
Thus $C_{4, local}^L$ becomes 
\begin{eqn}
C_{4, local}^L &= - \int_{zx} J^i(\vec x, z)  J^i(\vec x', z')   \p_{x_i} \p_{x'_j} \int d^3 y e^{i \vec y \cdot (\vec x - \vec x')} \intinf \frac{dp \sin(p z) \sin(p z')}{p^2 y^2}.
\end{eqn}
After performing IBP and using  \eqref{conservationeq1} we get
\begin{eqn}
C_{4, local}^L =  - \int_{zx} \p_{z} J^z(\vec x, z)   \p_{z'}  J^z(\vec x', z')  \int d^3 y e^{i \vec y \cdot (\vec x - \vec x')} \intinf \frac{dp \sin(p z) \sin(p z')}{p^2 y^2}.
\end{eqn}
Moreover repeated use of IBP gives
\begin{eqn}
C_{4, local}^L  &=  \int_{zx} J^z(\vec x, z)  \frac{1}{\p'^2}  J^z(\vec x', z')  \delta(\vec x- \vec x')\delta(z - z') \\
 &=  \int d^3 x d z J^z(\vec x, z)  \frac{1}{\p^2}  J^z(\vec x, z),
\end{eqn}
recovering the non-local form of the answer \eqref{4ptnonlocalshadow}.

\subsection*{Gravity}
Let us now sketch how to match graviton exchange correlators with external scalars obtained using local and non-local formulations. A key step is to use stress-tensor conservation equation. The stress tensor of the massless scalar is given as 
\begin{eqn}
T_{\mu\nu} = \frac12 \p_\mu\phi \p_\nu\phi - \frac12 \eta_{\mu\nu} (\p\phi)^2
\end{eqn}
and this satisfies the conservation equations
\begin{eqn}
\p_j (\sqrt{g} T^{j i}) &= - z^2 \p_z \Big( \frac{\sqrt{g} T^{zi}}{z^2} \Big)\\
\sqrt{g}\delta_{ij} T^{ij} &= z^2 \p_z\Big( \frac{\sqrt{g}}{z} T^{zz}\Big) -  z \p_i \big( \sqrt{g} T^{iz} \big)~.
\label{eq:StressConservation}
\end{eqn}
Moreover, let us recall that the longitudinal part of the wavefunction coefficient $\psi_4$ is the only contribution to the longitudinal part of the correlator, since the polarisation sum in the $\frac{\psi_3 \psi_3}{\psi_2}$ term makes it transverse and traceless \eqref{gravtreewf}. We therefore split $\psi_4$ as 
\begin{eqn}
    \psi_4 = \psi_4^{TT}+\psi_4^{TL}+\psi_4^{LL}
\end{eqn}
where
\begin{eqn}
 \psi_4^{TL} &=  \int_{zx} T^{ij}(\vec x, z) G_{ijab}^{TL}(\vec x, z; \vec x', z') T^{ab}(\vec x', z')  ,  \\
 \psi_4^{LL} &=  \int_{zx} T^{ij}(\vec x, z) G_{ijab}^{LL}(\vec x, z; \vec x', z') T^{ab}(\vec x', z') ,
\end{eqn}
and
\begin{eqn}
G_{ijab}^{TL}(\vec x, z; \vec x', z') &=  \int d^3 y e^{i \vec y \cdot (\vec x - \vec x')} (z z')^{-3/2}\intinf dp J_{3/2}(p z) J_{3/2}(p z')  \frac{y_i y_j \pi_{ab} + \text{perms.}}{y^2} \frac{1}{p}\\
G_{ijab}^{LL}(\vec x, z; \vec x', z') &=  \int d^3 y e^{i \vec y \cdot (\vec x - \vec x')} (zz')^{-1/2}\intinf dp J_{3/2}(p z) J_{3/2}(p z')  \frac{y_i y_j y_a y_b}{y^4} \frac{p^2 + y^2}{p^3}.
\end{eqn}
We note that $\psi_4^{TT}$ can be mapped to the transverse-traceless part of the in-in correlator by flipping the sign on the spectral parameter in the Bessel functions of the bulk-to-bulk propagators, so we focus our attention on the remaining terms. Using IBP and \eqref{eq:StressConservation}, after a tedious calculation we find that
\begin{eqn}
    \psi_4^{TL}+\psi_4^{LL}= R_1+R_2+R_3,
\end{eqn}
where 
\begin{eqn}
R_1 &= \intsinf \frac{dz}{z^2}   T_{zb}(z)  \frac{1}{\p^2} T_{zb}(z) , \\
R_2 &= \intsinf \frac{dz}{z} \p_a T_{za}(z) \frac{1}{\p^2}  T_{zz}(z), \\
R_3 &= \intsinf dz \frac{1}{z^2} \p_a T_{za}(z) \frac{1}{\p^4} \p_b T_{zb}(z).
\end{eqn}
From the local form of the action, we found that the s-channel contribution to the graviton exchange correlator is
\begin{eqn}\label{GRcorr}
\braket{\phi_1\phi_2 \phi_3 \phi_4}_{(s)} = M_4^{TT} + \tilde M_4
\end{eqn}
where $M_4^{TT}$ is given by \eqref{gravityTT} and $\tilde M_4$ is given by \eqref{eq:grav_s_channel_long_part}. The first  term in \eqref{GRcorr}, $M_4^{TT}$, encodes the transverse-traceless part of the graviton exchange and matches $\widehat W + \widehat G$ in equations 4.25 and 5.5-5.8 of \cite{Ghosh:2014kba}. This shows how the shadow formalism naturally encodes the transverse-traceless piece into a single term. The second term in  \eqref{GRcorr}, $\tilde M_4$, matches with the $R$ terms discussed above, which are given in equations 4.27 - 4.30 of \cite{Ghosh:2014kba}. See the relevant Mathematica file in \cite{github} for more details.

\section{Relation to in-out formalism}
In \cite{Chowdhury:2025ohm}, certain dressing rules were derived for uplifting scattering amplitudes in flat space to in-in correlators  for a variety of scalar theories in dS$_4$ and in this paper we extend this story to photons, gluons, and gravitons coupled to scalar fields. It has also been pointed out that  certain in-in correlators be computed using an in-out formalism \cite{Donath:2024utn}. In this Appendix we point out how one can derive the dressing rules from the in-out formalism in certain cases. 

\subsection*{Review the In-Out formalism}
We provide a lightning review of the propagators in the in-out formalism in flat space and we refer the reader to \cite{Kamenev:2004jgp, Donath:2024utn} for details. The in-in correlator is defined as,
\begin{eqn}\label{inindefn1}
\braket{\phi(\vec k_1) \cdots \phi(\vec k_n)} =  \braket{\Psi|\phi(\vec k_1) \cdots \phi(\vec k_n)|\Psi} 
\end{eqn}
where $\vec{k}$ are spatial momenta, the state $\ket\Psi$ is the vacuum state and is obtained via $U(0, -\infty) \ket{0_F}$ where $\ket{0_F}$ is the Fock vacuum state and $U = \mathcal T e^{- i \int_{-\infty}^t H_{int}(t') dt'}$. The fields $\phi(\vec k)$ are inserted at $t = 0$. The basic idea behind the in-out formalism is that there exists a Fock vacuum at $ t \to \infty$ through which one can back-evolve in time to obtain the ket $\ket{\Psi}$. This  allows one to rewrite \eqref{inindefn1} as a single time path integral, 
\begin{eqn}
\braket{\Psi|\phi(\vec k_1) \cdots \phi(\vec k_n)|\Psi} &= \braket{0_F| \mathcal T \Bigl\{ U(-\infty, \infty) \phi(\vec k_1) \cdots \phi(\vec k_n)  \Bigr\} |0_F} 
\end{eqn}
in the absence of subtleties at the turning point in the integral at $t =0$. 
Hence the propagators appearing in the computation are same as the ones for the un-amputated scattering amplitude. The bulk-bulk propagator is given as the (time-ordered) Feynman propagator,
\begin{eqn}\label{bulk-bulk-scalars}
G(\eta_1, \eta_2, k) =\intinf \frac{d\omega}{\omega^2 - k^2+ i \e} e^{-i \omega (\eta_1 - \eta_2)}
\end{eqn}
where $k=\vec{k}$. The bulk-boundary propagators can be obtained as a limit of the bulk-bulk propagator 
\begin{eqn}\label{bulk-boundary-scalars}
B(\eta, k) = \lim_{\eta' \to 0} G(\eta, \eta', k) = \intinf \frac{d\omega}{\omega^2 - k^2 + i\e} e^{-i \omega \eta}  = e^{-i k \eta}.
\end{eqn}
While the bulk-boundary propagator can be written via plane waves (as shown in the final equality of \eqref{bulk-boundary-scalars}), we shall find it convenient to use the expression in the intermediate step. 

\subsection*{Relation to dressing rules}
We begin by considering a 4-point function in massless $\phi^3$ theory in flat space for illustration. From the rules of the propagator given above, the $s-$channel contribution to the four point function $\braket{\psi|\phi(\vec k_1) \phi(\vec k_2) \phi(\vec k_3) \phi(\vec k_4)|\psi}$ is given as 
\begin{eqn}
\braket{\phi_1  \phi_2  \phi_3 \phi_4}_{(s)} = \prod_{i = 1}^4 \intinf \frac{d\omega_i }{\omega_i^2 - k_i^2} \intinf \frac{d\omega}{\omega^2 - y^2} \intinf d\eta_1 d\eta_2 e^{-i (\omega_1 + \omega_2) \eta_1} e^{-i (\omega_3 + \omega_4) \eta_2} e^{-i\omega(\eta_1 - \eta_2)}
\end{eqn}
where we have suppressed $i \epsilon$ for simplicity.  Performing the time-integrals gives the momentum conserving delta function,
\begin{eqn}
\braket{\phi_1  \phi_2  \phi_3 \phi_4}_{(s)} &= \prod_{i = 1}^4 \intinf \frac{d\omega_i }{\omega_i^2 - k_i^2} \intinf \frac{d\omega}{\omega^2 - y^2} \delta(\omega_1 + \omega_2 + \omega)  \delta(\omega_3 + \omega_4 - \omega)\\
&= \prod_{i = 1}^4 \intinf \frac{d\omega_i }{\omega_i^2 - k_i^2} \frac{1}{(\omega_1 + \omega_2)^2 - y^2}  \delta(\omega_3 + \omega_4 - \omega_1 - \omega_2)\\
&= \prod_{i = 1}^3 \intinf \frac{d\omega_i }{\omega_i^2 - k_i^2} \frac{1}{(\omega_1 + \omega_2 - \omega_3)^2 - k_4^2}  \frac{1}{(\omega_1 + \omega_2)^2 - y^2}
\end{eqn}
We have suppressed the $i\e$ for brevity. By performing the integrals we obtain
\begin{eqn}
\braket{\phi_1  \phi_2  \phi_3 \phi_4}_{(s)} &= \frac{1}{(k_{12} + k_{34})(k_{12} + k) (k_{34} + k)} + \frac{1}{k (k_{12} + k) (k_{34} + k)}
\end{eqn}
While in this simple example the integrals could be performed by brute force, it will be instructive to perform a variable transform in the intermediate step to obtain an alternate formula for general diagrams. 

From the example above it is clear that for any off-shell Feynman diagram at 4-points the time integrals will result in energy conservation:
\begin{eqn}\label{Cinout1}
\braket{\phi_1  \phi_2  \phi_3 \phi_4}_{(s)} &= \prod_{n = 1}^4 \int \frac{d\omega_n}{\omega_n^2 + k_n^2 } f(\omega_L) \delta(\omega_L + \omega_R), 
\end{eqn}
where $\omega_L = \omega_1 + \omega_2$, $\omega_R = \omega_3 + \omega_4 $, and $f(\omega_L) $ is a rational function. We have Wick rotated the energies in this step to make contact with the formulas in previous section. Such off-shell diagrams have been studied in the literature, eg, \cite{Cespedes:2025dnq, Salcedo:2022aal}.
A few comments are in order: 
\begin{enumerate}
\item The function $f(\omega)$ depends on the energies only via $\omega_L$. This is because in general $f(x)$ arises from a product of bulk-bulk propagators (of the form \eqref{bulk-bulk-scalars}) and they are always going to be a function of the four momentum $(\omega, \vec k)$. For an off-shell Feynman diagram there is no relation between $\omega$ and $\vec k$. However, since the integrals over time (which are exponential functions) are running from $(-\infty, +\infty)$ there is still going to be four-momentum $(\omega, \vec k)$ conservation. Therefore any general bulk-bulk propagator can always be written as a function of $(\omega_{vertex}, \vec k_{vertex})$  where this denotes the sum of off-shell four momenta entering a vertex. For the case of the single exchange, at the left vertex we have $(\omega_{vertex}, \vec k_{vertex}) = (\omega_1 + \omega_2, \vec k_1 + \vec k_2)$~.

\item The 4-point calculation above straightforwardly generalizes to any number of points.

\item This expression also holds for loop integrands as it is valid for any rational $f(\omega_L)$. 

\item  This is true for all interactions including derivative interactions, gauge fields and gravity in axial gauges, but the cases with a single derivative (w.r.t $\partial_z$) in the interaction term and half-integer spins will have to be dealt with separately. 

\end{enumerate}

We now use the formula \eqref{Cinout1} to derive the dressing rules of $\phi^3$ theory in flat space (which can be mapped  by a Weyl transformation to dS$_4$) first obtained in \cite{Chowdhury:2025ohm}. Note that this theory can be mapped to massless $\phi^4$ theory in half of flat space via a Weyl transformation. Consider the explicit form of the correlator given in \eqref{Cinout1}: 
\begin{eqn}\label{Cinout}
\braket{\phi_1  \phi_2  \phi_3 \phi_4}_{(s)} &= \prod_{n = 1}^4 \int \frac{d\omega_n}{\omega_n^2 + k_n^2 } f(\omega_L) \delta(\omega_L + \omega_R) \\
&=  \int \frac{d\omega_1}{\omega_1^2 + k_1^2 } \cdots  \frac{d\omega_4}{\omega_4^2 + k_4^2 } f(\omega_L) \delta(\omega_L + \omega_R) 
\end{eqn}
where $\omega_L = \omega_1 + \omega_2$ and $\omega_R = \omega_3 + \omega_4 $~. We first analyze the terms dependent on $\omega_1$ and $\omega_2$:
\begin{eqn}
\intinf \frac{d\omega_1 d\omega_2}{(\omega_1^2 + k_1^2)(\omega_2^2 + k_2^2)} f(\omega_L).
\end{eqn}
Since the function $f(\omega_L) = f(\omega_1 +\omega_2)$ is independent of $\omega_1 - \omega_2$ we perform the following change of variables:
\begin{eqn}
\omega_L &= \omega_1 + \omega_2, \qquad 
\bar \omega_L = \omega_1 - \omega_2 , 
\end{eqn}
Performing a similar manipulation for $\omega_3, \omega_4$, results in the following expression for \eqref{Cinout}:
\begin{eqn}
&\braket{\phi_1  \phi_2  \phi_3 \phi_4}_{(s)} \\
&= 2^2\intinf \frac{d\omega_L d\omega_R d\bar\omega_L d\bar\omega_R \ \delta(\omega_L + \omega_R)}{\big( (\omega_L + \bar \omega_L)^2 + 4 k_1^2 \big)\big( (\omega_L - \bar \omega_L)^2 + 4 k_2^2 \big) \big( (\omega_R + \bar \omega_R)^2 + 4 k_3^2 \big)\big( (\omega_R - \bar \omega_R)^2 + 4 k_4^2 \big)} f(\omega_L).
\end{eqn}
Performing the integrals over $\bar\omega_L, \bar\omega_R$ then gives
\begin{eqn}\label{dress-scalar1}
\braket{\phi_1  \phi_2  \phi_3 \phi_4}_{(s)} = \frac{\pi^2}{16} \frac{1}{k_1 k_2 k_3 k_4} \intinf d\omega_L d\omega_R \frac{k_{12} k_{34}\delta(\omega_L + \omega_R)}{(\omega_L^2 + k_{12}^2) (\omega_R^2 + k_{34}^2)} f(\omega_L).
\end{eqn}
For the tree level single exchange this can be be diagrammatically expressed as 
\begin{eqn}
\braket{\phi_1  \phi_2  \phi_3 \phi_4}_{(s)} = \begin{tikzpicture}[baseline]
\draw (0, 0) -- (2, 0);
\draw (-1, 1) -- (0, 0);
\draw (-1, -1) -- (0, 0);

\draw (3, 1) -- (2, 0);
\draw (3, -1) -- (2, 0);

\node at (-1, 1.25) {$k_2$};
\node at (-1, -1.25) {$k_1$};

\node at (3, 1.25) {$k_3$};
\node at (3, -1.25) {$k_4$};

\draw[dashed] (0, 0) -- (1, -2);
\draw[dashed] (2, 0) -- (1, -2);

\node at (0, -1) {$\omega_L$};
\node at (2, -1) {$\omega_R$};

\end{tikzpicture}
\end{eqn}
where the dashed propagators denote the kernels in \eqref{dress-scalar1} including the energy-conserving delta function $\delta(\omega_L + \omega_R)$, thus recovering the dressing rules for conformally coupled $\phi^4$ theory in \cite{Chowdhury:2025ohm}. A similar argument allows us to derive the dressing rules for photons and gluons in dS$_4$, which can be trivially mapped to half of Minkowski space via a Weyl transformation due to classical conformal invariance. When doing so, one has to take care of the residual gauge invariance at the boundary which can be used to fix Dirichlet boundary conditions for the longitudinal components as discussed in section \ref{sksection}.

\end{appendix}

\newpage
\bibliographystyle{JHEP}
\bibliography{references}

@article{Jain:2025maa,
    author = "Jain, Diksha and Pajer, Enrico and Tong, Xi",
    title = "{Unitary and Analytic Renormalisation of Cosmological Correlators}",
    eprint = "2509.02696",
    archivePrefix = "arXiv",
    primaryClass = "hep-th",
    month = "9",
    year = "2025"
}

@inproceedings{Dixon:1996wi,
    author = "Dixon, Lance J.",
    title = "{Calculating scattering amplitudes efficiently}",
    booktitle = "{Theoretical Advanced Study Institute in Elementary Particle Physics (TASI 95): QCD and Beyond}",
    eprint = "hep-ph/9601359",
    archivePrefix = "arXiv",
    reportNumber = "SLAC-PUB-7106",
    pages = "539--584",
    month = "1",
    year = "1996"
}

@article{Alday:2021odx,
    author = "Alday, Luis F. and Behan, Connor and Ferrero, Pietro and Zhou, Xinan",
    title = "{Gluon Scattering in AdS from CFT}",
    eprint = "2103.15830",
    archivePrefix = "arXiv",
    primaryClass = "hep-th",
    doi = "10.1007/JHEP06(2021)020",
    journal = "JHEP",
    volume = "06",
    pages = "020",
    year = "2021"
}

@article{Armstrong:2020woi,
    author = "Armstrong, Connor and Lipstein, Arthur E. and Mei, Jiajie",
    title = "{Color/kinematics duality in AdS$_{4}$}",
    eprint = "2012.02059",
    archivePrefix = "arXiv",
    primaryClass = "hep-th",
    doi = "10.1007/JHEP02(2021)194",
    journal = "JHEP",
    volume = "02",
    pages = "194",
    year = "2021"
}

@article{Albayrak:2020fyp,
    author = "Albayrak, Soner and Kharel, Savan and Meltzer, David",
    title = "{On duality of color and kinematics in (A)dS momentum space}",
    eprint = "2012.10460",
    archivePrefix = "arXiv",
    primaryClass = "hep-th",
    doi = "10.1007/JHEP03(2021)249",
    journal = "JHEP",
    volume = "03",
    pages = "249",
    year = "2021"
}

@article{Chowdhury:2023arc,
    author = "Chowdhury, Chandramouli and Lipstein, Arthur and Mei, Jiajie and Sachs, Ivo and Vanhove, Pierre",
    title = "{The subtle simplicity of cosmological correlators}",
    eprint = "2312.13803",
    archivePrefix = "arXiv",
    primaryClass = "hep-th",
    reportNumber = "LMU-ASC 37/23, IPhT-t23/119",
    doi = "10.1007/JHEP03(2025)007",
    journal = "JHEP",
    volume = "03",
    pages = "007",
    year = "2025"
}

@article{Chowdhury:2025ohm,
    author = "Chowdhury, Chandramouli and Lipstein, Arthur and Marshall, Joe and Mei, Jiajie and Sachs, Ivo",
    title = "{Cosmological Dressing Rules}",
    eprint = "2503.10598",
    archivePrefix = "arXiv",
    primaryClass = "hep-th",
    reportNumber = "LMU-ASC 03/25",
    month = "3",
    year = "2025"
}

@article{Sleight:2020obc,
    author = "Sleight, Charlotte and Taronna, Massimo",
    title = "{From AdS to dS exchanges: Spectral representation, Mellin amplitudes, and crossing}",
    eprint = "2007.09993",
    archivePrefix = "arXiv",
    primaryClass = "hep-th",
    doi = "10.1103/PhysRevD.104.L081902",
    journal = "Phys. Rev. D",
    volume = "104",
    number = "8",
    pages = "L081902",
    year = "2021"
}

@article{Schaub:2023scu,
    author = "Schaub, Vladimir",
    title = "{Spinors in (Anti-)de Sitter Space}",
    eprint = "2302.08535",
    archivePrefix = "arXiv",
    primaryClass = "hep-th",
    doi = "10.1007/JHEP09(2023)142",
    journal = "JHEP",
    volume = "09",
    pages = "142",
    year = "2023"
}

@article{Lee:2022fgr,
    author = "Lee, Hayden and Wang, Xinkang",
    title = "{Cosmological double-copy relations}",
    eprint = "2212.11282",
    archivePrefix = "arXiv",
    primaryClass = "hep-th",
    doi = "10.1103/PhysRevD.108.L061702",
    journal = "Phys. Rev. D",
    volume = "108",
    number = "6",
    pages = "L061702",
    year = "2023"
}

@article{Jain:2021qcl,
    author = "Jain, Sachin and John, Renjan Rajan and Mehta, Abhishek and Nizami, Amin A. and Suresh, Adithya",
    title = "{Double copy structure of parity-violating CFT correlators}",
    eprint = "2104.12803",
    archivePrefix = "arXiv",
    primaryClass = "hep-th",
    doi = "10.1007/JHEP07(2021)033",
    journal = "JHEP",
    volume = "07",
    pages = "033",
    year = "2021"
}

@article{Sleight:2025dmt,
    author = "Sleight, Charlotte and Taronna, Massimo",
    title = "{(Non-)Conserved Currents and Cosmological Correlators}",
    eprint = "2509.18888",
    archivePrefix = "arXiv",
    primaryClass = "hep-th",
    month = "9",
    year = "2025"
}

@article{Bzowski:2020kfw,
    author = "Bzowski, Adam and McFadden, Paul and Skenderis, Kostas",
    title = "{Conformal correlators as simplex integrals in momentum space}",
    eprint = "2008.07543",
    archivePrefix = "arXiv",
    primaryClass = "hep-th",
    doi = "10.1007/JHEP01(2021)192",
    journal = "JHEP",
    volume = "01",
    pages = "192",
    year = "2021"
}

@article{Arkani-Hamed:2023kig,
    author = "Arkani-Hamed, Nima and Baumann, Daniel and Hillman, Aaron and Joyce, Austin and Lee, Hayden and Pimentel, Guilherme L.",
    title = "{Differential equations for cosmological correlators}",
    eprint = "2312.05303",
    archivePrefix = "arXiv",
    primaryClass = "hep-th",
    doi = "10.1007/JHEP09(2025)009",
    journal = "JHEP",
    volume = "09",
    pages = "009",
    year = "2025"
}

@article{Arkani-Hamed:2023bsv,
    author = "Arkani-Hamed, Nima and Baumann, Daniel and Hillman, Aaron and Joyce, Austin and Lee, Hayden and Pimentel, Guilherme L.",
    title = "{Kinematic Flow and the Emergence of Time}",
    eprint = "2312.05300",
    archivePrefix = "arXiv",
    primaryClass = "hep-th",
    doi = "10.1103/dsjm-tckw",
    journal = "Phys. Rev. Lett.",
    volume = "135",
    number = "3",
    pages = "031602",
    year = "2025"
}

@article{Melville:2021lst,
    author = "Melville, Scott and Pajer, Enrico",
    title = "{Cosmological Cutting Rules}",
    eprint = "2103.09832",
    archivePrefix = "arXiv",
    primaryClass = "hep-th",
    doi = "10.1007/JHEP05(2021)249",
    journal = "JHEP",
    volume = "05",
    pages = "249",
    year = "2021"
}

@article{Arkani-Hamed:2024jbp,
    author = "Arkani-Hamed, Nima and Figueiredo, Carolina and Vaz{\~a}o, Francisco",
    title = "{Cosmohedra}",
    eprint = "2412.19881",
    archivePrefix = "arXiv",
    primaryClass = "hep-th",
    doi = "10.1007/JHEP11(2025)029",
    journal = "JHEP",
    volume = "11",
    pages = "029",
    year = "2025"
}

@article{Roehrig:2020kck,
    author = "Roehrig, Kai and Skinner, David",
    title = "{Ambitwistor strings and the scattering equations on AdS$_{3}${\texttimes}S$^{3}$}",
    eprint = "2007.07234",
    archivePrefix = "arXiv",
    primaryClass = "hep-th",
    doi = "10.1007/JHEP02(2022)073",
    journal = "JHEP",
    volume = "02",
    pages = "073",
    year = "2022"
}

@article{Eberhardt:2020ewh,
    author = "Eberhardt, Lorenz and Komatsu, Shota and Mizera, Sebastian",
    title = "{Scattering equations in AdS: scalar correlators in arbitrary dimensions}",
    eprint = "2007.06574",
    archivePrefix = "arXiv",
    primaryClass = "hep-th",
    doi = "10.1007/JHEP11(2020)158",
    journal = "JHEP",
    volume = "11",
    pages = "158",
    year = "2020"
}

@article{Gomez:2021qfd,
    author = "Gomez, Humberto and Jusinskas, Renann Lipinski and Lipstein, Arthur",
    title = "{Cosmological Scattering Equations}",
    eprint = "2106.11903",
    archivePrefix = "arXiv",
    primaryClass = "hep-th",
    doi = "10.1103/PhysRevLett.127.251604",
    journal = "Phys. Rev. Lett.",
    volume = "127",
    number = "25",
    pages = "251604",
    year = "2021"
}

@article{Herderschee:2022ntr,
    author = "Herderschee, Aidan and Roiban, Radu and Teng, Fei",
    title = "{On the differential representation and color-kinematics duality of AdS boundary correlators}",
    eprint = "2201.05067",
    archivePrefix = "arXiv",
    primaryClass = "hep-th",
    reportNumber = "LCTP-22-01",
    doi = "10.1007/JHEP05(2022)026",
    journal = "JHEP",
    volume = "05",
    pages = "026",
    year = "2022"
}

@article{Sleight:2019mgd,
    author = "Sleight, Charlotte",
    title = "{A Mellin Space Approach to Cosmological Correlators}",
    eprint = "1906.12302",
    archivePrefix = "arXiv",
    primaryClass = "hep-th",
    doi = "10.1007/JHEP01(2020)090",
    journal = "JHEP",
    volume = "01",
    pages = "090",
    year = "2020"
}

@article{Bern:2019prr,
    author = "Bern, Zvi and Carrasco, John Joseph and Chiodaroli, Marco and Johansson, Henrik and Roiban, Radu",
    title = "{The duality between color and kinematics and its applications}",
    eprint = "1909.01358",
    archivePrefix = "arXiv",
    primaryClass = "hep-th",
    reportNumber = "CERN-TH-2019-135, UCLA/TEP/2019/104, NUHEP-TH/19-11, UUITP-35/19,
  NORDITA 2019-079",
    doi = "10.1088/1751-8121/ad5fd0",
    journal = "J. Phys. A",
    volume = "57",
    number = "33",
    pages = "333002",
    year = "2024"
}

@article{Jazayeri:2021fvk,
    author = "Jazayeri, Sadra and Pajer, Enrico and Stefanyszyn, David",
    title = "{From locality and unitarity to cosmological correlators}",
    eprint = "2103.08649",
    archivePrefix = "arXiv",
    primaryClass = "hep-th",
    doi = "10.1007/JHEP10(2021)065",
    journal = "JHEP",
    volume = "10",
    pages = "065",
    year = "2021"
}

@article{Hinterbichler:2013dpa,
    author = "Hinterbichler, Kurt and Hui, Lam and Khoury, Justin",
    title = "{An Infinite Set of Ward Identities for Adiabatic Modes in Cosmology}",
    eprint = "1304.5527",
    archivePrefix = "arXiv",
    primaryClass = "hep-th",
    doi = "10.1088/1475-7516/2014/01/039",
    journal = "JCAP",
    volume = "01",
    pages = "039",
    year = "2014"
}

@article{Mei:2024sqz,
    author = "Mei, Jiajie and Mo, Yuyu",
    title = "{From on-shell amplitude in AdS to cosmological correlators: gluons and gravitons}",
    eprint = "2410.04875",
    archivePrefix = "arXiv",
    primaryClass = "hep-th",
    month = "10",
    year = "2024"
}

@article{Caloro:2023cep,
    author = "Caloro, Francesca and McFadden, Paul",
    title = "{$\mathcal{A}$-hypergeometric functions and creation operators for Feynman and Witten diagrams}",
    eprint = "2309.15895",
    archivePrefix = "arXiv",
    primaryClass = "hep-th",
    month = "9",
    year = "2023"
}

@article{Chowdhury:2024wwe,
    author = "Chowdhury, Chandramouli and Lipstein, Arthur and Mei, Jiajie and Mo, Yuyu",
    title = "{Soft limits of gluon and graviton correlators in Anti-de Sitter space}",
    eprint = "2407.16052",
    archivePrefix = "arXiv",
    primaryClass = "hep-th",
    doi = "10.1007/JHEP10(2024)070",
    journal = "JHEP",
    volume = "10",
    pages = "070",
    year = "2024"
}

@article{Dymarsky:2013wla,
    author = "Dymarsky, Anatoly",
    title = "{On the four-point function of the stress-energy tensors in a CFT}",
    eprint = "1311.4546",
    archivePrefix = "arXiv",
    primaryClass = "hep-th",
    doi = "10.1007/JHEP10(2015)075",
    journal = "JHEP",
    volume = "10",
    pages = "075",
    year = "2015"
}

@article{Bzowski:2018fql,
    author = "Bzowski, Adam and McFadden, Paul and Skenderis, Kostas",
    title = "{Renormalised CFT 3-point functions of scalars, currents and stress tensors}",
    eprint = "1805.12100",
    archivePrefix = "arXiv",
    primaryClass = "hep-th",
    doi = "10.1007/JHEP11(2018)159",
    journal = "JHEP",
    volume = "11",
    pages = "159",
    year = "2018"
}

@article{Bzowski:2019kwd,
    author = "Bzowski, Adam and McFadden, Paul and Skenderis, Kostas",
    title = "{Conformal $n$-point functions in momentum space}",
    eprint = "1910.10162",
    archivePrefix = "arXiv",
    primaryClass = "hep-th",
    doi = "10.1103/PhysRevLett.124.131602",
    journal = "Phys. Rev. Lett.",
    volume = "124",
    number = "13",
    pages = "131602",
    year = "2020"
}

@article{Dey:2025kci,
    author = "Dey, Parijat and Huang, Zhongjie and Lipstein, Arthur",
    title = "{de Sitter locality from conformal field theory}",
    eprint = "2508.15627",
    archivePrefix = "arXiv",
    primaryClass = "hep-th",
    month = "8",
    year = "2025"
}

@article{Zhou:2021gnu,
    author = "Zhou, Xinan",
    title = "{Double Copy Relation in AdS Space}",
    eprint = "2106.07651",
    archivePrefix = "arXiv",
    primaryClass = "hep-th",
    doi = "10.1103/PhysRevLett.127.141601",
    journal = "Phys. Rev. Lett.",
    volume = "127",
    number = "14",
    pages = "141601",
    year = "2021"
}

@article{Caron-Huot:2021enk,
    author = "Caron-Huot, Simon and Mazac, Dalimil and Rastelli, Leonardo and Simmons-Duffin, David",
    title = "{AdS bulk locality from sharp CFT bounds}",
    eprint = "2106.10274",
    archivePrefix = "arXiv",
    primaryClass = "hep-th",
    doi = "10.1007/JHEP11(2021)164",
    journal = "JHEP",
    volume = "11",
    pages = "164",
    year = "2021"
}

@article{Baumann:2024ttn,
    author = "Baumann, Daniel and Mathys, Gr{\'e}goire and Pimentel, Guilherme L. and Rost, Facundo",
    title = "{A new twist on spinning (A)dS correlators}",
    eprint = "2408.02727",
    archivePrefix = "arXiv",
    primaryClass = "hep-th",
    doi = "10.1007/JHEP01(2025)202",
    journal = "JHEP",
    volume = "01",
    pages = "202",
    year = "2025"
}

@article{Chowdhury:2021nxw,
	archiveprefix = {arXiv},
	author = {Chowdhury, Chandramouli and Godet, Victor and Papadoulaki, Olga and Raju, Suvrat},
	eprint = {2107.14802},
	month = {7},
	primaryclass = {hep-th},
	title = {{Holography from the Wheeler-DeWitt equation}},
	year = {2021}}

@article{DeWitt:1967yk,
	author = {DeWitt, Bryce S.},
	doi = {10.1103/PhysRev.160.1113},
	editor = {Fang, Li-Zhi and Ruffini, R.},
	journal = {Phys. Rev.},
	pages = {1113--1148},
	title = {{Quantum Theory of Gravity. 1. The Canonical Theory}},
	volume = {160},
	year = {1967}}

@article{Raju:2011mp,
	archiveprefix = {arXiv},
	author = {Raju, Suvrat},
	doi = {10.1103/PhysRevD.83.126002},
	eprint = {1102.4724},
	journal = {Phys. Rev. D},
	pages = {126002},
	primaryclass = {hep-th},
	reportnumber = {HRI-ST-1103},
	title = {{Recursion Relations for AdS/CFT Correlators}},
	volume = {83},
	year = {2011}}

@article{Bzowski:2013sza,
	archiveprefix = {arXiv},
	author = {Bzowski, Adam and McFadden, Paul and Skenderis, Kostas},
	doi = {10.1007/JHEP03(2014)111},
	eprint = {1304.7760},
	journal = {JHEP},
	pages = {111},
	primaryclass = {hep-th},
	title = {{Implications of conformal invariance in momentum space}},
	volume = {03},
	year = {2014}}

@article{Arkani-Hamed:2017fdk,
	archiveprefix = {arXiv},
	author = {Arkani-Hamed, Nima and Benincasa, Paolo and Postnikov, Alexander},
	eprint = {1709.02813},
	month = {9},
	primaryclass = {hep-th},
	title = {{Cosmological Polytopes and the Wavefunction of the Universe}},
	year = {2017}}

@article{Weinberg:2005vy,
	archiveprefix = {arXiv},
	author = {Weinberg, Steven},
	doi = {10.1103/PhysRevD.72.043514},
	eprint = {hep-th/0506236},
	journal = {Phys. Rev. D},
	pages = {043514},
	reportnumber = {UTTG-01-05},
	title = {{Quantum contributions to cosmological correlations}},
	volume = {72},
	year = {2005}}

@article{Maldacena:2002vr,
	archiveprefix = {arXiv},
	author = {Maldacena, Juan Martin},
	doi = {10.1088/1126-6708/2003/05/013},
	eprint = {astro-ph/0210603},
	journal = {JHEP},
	pages = {013},
	title = {{Non-Gaussian features of primordial fluctuations in single field inflationary models}},
	volume = {05},
	year = {2003}}

@article{Benincasa:2022gtd,
	archiveprefix = {arXiv},
	author = {Benincasa, Paolo},
	doi = {10.1142/S0217751X22300101},
	eprint = {2203.15330},
	month = {3},
	primaryclass = {hep-th},
	title = {{Amplitudes meet Cosmology: A (Scalar) Primer}},
	year = {2022}}

@article{Albayrak:2019asr,
	archiveprefix = {arXiv},
	author = {Albayrak, Soner and Chowdhury, Chandramouli and Kharel, Savan},
	doi = {10.1007/JHEP10(2019)274},
	eprint = {1904.10043},
	journal = {JHEP},
	pages = {274},
	primaryclass = {hep-th},
	title = {{New relation for Witten diagrams}},
	volume = {10},
	year = {2019}}

@article{Raju:2012zs,
	archiveprefix = {arXiv},
	author = {Raju, Suvrat},
	doi = {10.1103/PhysRevD.85.126008},
	eprint = {1201.6452},
	journal = {Phys. Rev. D},
	pages = {126008},
	primaryclass = {hep-th},
	reportnumber = {HRI-ST-1202},
	title = {{Four Point Functions of the Stress Tensor and Conserved Currents in AdS$_4$/CFT$_3$}},
	volume = {85},
	year = {2012}}

@article{Arkani-Hamed:2015bza,
	archiveprefix = {arXiv},
	author = {Arkani-Hamed, Nima and Maldacena, Juan},
	eprint = {1503.08043},
	month = {3},
	primaryclass = {hep-th},
	title = {{Cosmological Collider Physics}},
	year = {2015}}

@article{Arkani-Hamed:2018kmz,
	archiveprefix = {arXiv},
	author = {Arkani-Hamed, Nima and Baumann, Daniel and Lee, Hayden and Pimentel, Guilherme L.},
	doi = {10.1007/JHEP04(2020)105},
	eprint = {1811.00024},
	journal = {JHEP},
	pages = {105},
	primaryclass = {hep-th},
	title = {{The Cosmological Bootstrap: Inflationary Correlators from Symmetries and Singularities}},
	volume = {04},
	year = {2020}}

@article{Raju:2012zr,
	archiveprefix = {arXiv},
	author = {Raju, Suvrat},
	doi = {10.1103/PhysRevD.85.126009},
	eprint = {1201.6449},
	journal = {Phys. Rev. D},
	pages = {126009},
	primaryclass = {hep-th},
	reportnumber = {HRI-ST-1201},
	title = {{New Recursion Relations and a Flat Space Limit for AdS/CFT Correlators}},
	volume = {85},
	year = {2012}}

@article{Baumann:2020dch,
    author = "Baumann, Daniel and Duaso Pueyo, Carlos and Joyce, Austin and Lee, Hayden and Pimentel, Guilherme L.",
    title = "{The Cosmological Bootstrap: Spinning Correlators from Symmetries and Factorization}",
    eprint = "2005.04234",
    archivePrefix = "arXiv",
    primaryClass = "hep-th",
    doi = "10.21468/SciPostPhys.11.3.071",
    journal = "SciPost Phys.",
    volume = "11",
    pages = "071",
    year = "2021"
}

@article{Liu:1998ty,
    author = "Liu, Hong and Tseytlin, Arkady A.",
    title = "{On four point functions in the CFT / AdS correspondence}",
    eprint = "hep-th/9807097",
    archivePrefix = "arXiv",
    reportNumber = "IMPERIAL-TP-97-98-060",
    doi = "10.1103/PhysRevD.59.086002",
    journal = "Phys. Rev. D",
    volume = "59",
    pages = "086002",
    year = "1999"
}

@article{DiPietro:2021sjt,
    author = "Di Pietro, Lorenzo and Gorbenko, Victor and Komatsu, Shota",
    title = "{Analyticity and unitarity for cosmological correlators}",
    eprint = "2108.01695",
    archivePrefix = "arXiv",
    primaryClass = "hep-th",
    reportNumber = "CERN-TH-2021-118",
    doi = "10.1007/JHEP03(2022)023",
    journal = "JHEP",
    volume = "03",
    pages = "023",
    year = "2022"
}

@article{Bzowski:2023nef,
    author = "Bzowski, Adam and McFadden, Paul and Skenderis, Kostas",
    title = "{Renormalisation of IR divergences and holography in de Sitter}",
    eprint = "2312.17316",
    archivePrefix = "arXiv",
    primaryClass = "hep-th",
    month = "12",
    year = "2023"
}

@article{Goodhew:2020hob,
    author = "Goodhew, Harry and Jazayeri, Sadra and Pajer, Enrico",
    title = "{The Cosmological Optical Theorem}",
    eprint = "2009.02898",
    archivePrefix = "arXiv",
    primaryClass = "hep-th",
    doi = "10.1088/1475-7516/2021/04/021",
    journal = "JCAP",
    volume = "04",
    pages = "021",
    year = "2021"
}

@article{Maldacena:2011nz,
    author = "Maldacena, Juan M. and Pimentel, Guilherme L.",
    title = "{On graviton non-Gaussianities during inflation}",
    eprint = "1104.2846",
    archivePrefix = "arXiv",
    primaryClass = "hep-th",
    reportNumber = "PUPT-2371",
    doi = "10.1007/JHEP09(2011)045",
    journal = "JHEP",
    volume = "09",
    pages = "045",
    year = "2011"
}

@article{Donath:2024utn,
    author = "Donath, Yaniv and Pajer, Enrico",
    title = "{The in-out formalism for in-in correlators}",
    eprint = "2402.05999",
    archivePrefix = "arXiv",
    primaryClass = "hep-th",
    doi = "10.1007/JHEP07(2024)064",
    journal = "JHEP",
    volume = "07",
    pages = "064",
    year = "2024"
}

@article{Chowdhury:2024snc,
    author = "Chowdhury, Chandramouli and Chowdhury, Pratyusha and Moga, Radu N. and Singh, Kajal",
    title = "{Loops, recursions, and soft limits for fermionic correlators in (A)dS}",
    eprint = "2408.00074",
    archivePrefix = "arXiv",
    primaryClass = "hep-th",
    doi = "10.1007/JHEP10(2024)202",
    journal = "JHEP",
    volume = "10",
    pages = "202",
    year = "2024"
}

@article{Seery:2006vu,
    author = "Seery, David and Lidsey, James E. and Sloth, Martin S.",
    title = "{The inflationary trispectrum}",
    eprint = "astro-ph/0610210",
    archivePrefix = "arXiv",
    doi = "10.1088/1475-7516/2007/01/027",
    journal = "JCAP",
    volume = "01",
    pages = "027",
    year = "2007"
}

@article{Bonifacio:2022vwa,
    author = "Bonifacio, James and Goodhew, Harry and Joyce, Austin and Pajer, Enrico and Stefanyszyn, David",
    title = "{The graviton four-point function in de Sitter space}",
    eprint = "2212.07370",
    archivePrefix = "arXiv",
    primaryClass = "hep-th",
    doi = "10.1007/JHEP06(2023)212",
    journal = "JHEP",
    volume = "06",
    pages = "212",
    year = "2023"
}

@article{Armstrong:2023phb,
    author = "Armstrong, C. and Goodhew, H. and Lipstein, A. and Mei, J.",
    title = "{Graviton trispectrum from gluons}",
    eprint = "2304.07206",
    archivePrefix = "arXiv",
    primaryClass = "hep-th",
    doi = "10.1007/JHEP08(2023)206",
    journal = "JHEP",
    volume = "08",
    pages = "206",
    year = "2023"
}

@article{Ghosh:2014kba,
    author = "Ghosh, Archisman and Kundu, Nilay and Raju, Suvrat and Trivedi, Sandip P.",
    title = "{Conformal Invariance and the Four Point Scalar Correlator in Slow-Roll Inflation}",
    eprint = "1401.1426",
    archivePrefix = "arXiv",
    primaryClass = "hep-th",
    reportNumber = "ICTS-2013-23, TIFR-TH-13-31",
    doi = "10.1007/JHEP07(2014)011",
    journal = "JHEP",
    volume = "07",
    pages = "011",
    year = "2014"
}

@book{Hatfield:1992rz,
    author = "Hatfield, B.",
    title = "{Quantum field theory of point particles and strings}",
    year = "1992"
}

@article{Chakraborty:2023los,
    author = "Chakraborty, Tuneer and Chakravarty, Joydeep and Godet, Victor and Paul, Priyadarshi and Raju, Suvrat",
    title = "{Holography of information in de Sitter space}",
    eprint = "2303.16316",
    archivePrefix = "arXiv",
    primaryClass = "hep-th",
    doi = "10.1007/JHEP12(2023)120",
    journal = "JHEP",
    volume = "12",
    pages = "120",
    year = "2023"
}

@article{Kamenev:2004jgp,
    author = "Kamenev, Alex",
    title = "{Many-body theory of non-equilibrium systems}",
    eprint = "cond-mat/0412296",
    archivePrefix = "arXiv",
    month = "12",
    year = "2004"
}

@article{Kabat:2012av,
    author = "Kabat, Daniel and Lifschytz, Gilad",
    title = "{CFT representation of interacting bulk gauge fields in AdS}",
    eprint = "1212.3788",
    archivePrefix = "arXiv",
    primaryClass = "hep-th",
    doi = "10.1103/PhysRevD.87.086004",
    journal = "Phys. Rev. D",
    volume = "87",
    number = "8",
    pages = "086004",
    year = "2013"
}

@article{Seery:2008ax,
    author = "Seery, David and Sloth, Martin S. and Vernizzi, Filippo",
    title = "{Inflationary trispectrum from graviton exchange}",
    eprint = "0811.3934",
    archivePrefix = "arXiv",
    primaryClass = "astro-ph",
    doi = "10.1088/1475-7516/2009/03/018",
    journal = "JCAP",
    volume = "03",
    pages = "018",
    year = "2009"
}

@article{McFadden:2009fg,
    author = "McFadden, Paul and Skenderis, Kostas",
    title = "{Holography for Cosmology}",
    eprint = "0907.5542",
    archivePrefix = "arXiv",
    primaryClass = "hep-th",
    reportNumber = "ITF-22",
    doi = "10.1103/PhysRevD.81.021301",
    journal = "Phys. Rev. D",
    volume = "81",
    pages = "021301",
    year = "2010"
}

@article{Bzowski:2011ab,
    author = "Bzowski, Adam and McFadden, Paul and Skenderis, Kostas",
    title = "{Holographic predictions for cosmological 3-point functions}",
    eprint = "1112.1967",
    archivePrefix = "arXiv",
    primaryClass = "hep-th",
    doi = "10.1007/JHEP03(2012)091",
    journal = "JHEP",
    volume = "03",
    pages = "091",
    year = "2012"
}

@article{McFadden:2010vh,
    author = "McFadden, Paul and Skenderis, Kostas",
    title = "{Holographic Non-Gaussianity}",
    eprint = "1011.0452",
    archivePrefix = "arXiv",
    primaryClass = "hep-th",
    reportNumber = "ITFA-10-23",
    doi = "10.1088/1475-7516/2011/05/013",
    journal = "JCAP",
    volume = "05",
    pages = "013",
    year = "2011"
}

@article{Cespedes:2025dnq,
    author = "Cespedes, Sebastian and Jazayeri, Sadra",
    title = "{The massive flat space limit of cosmological correlators}",
    eprint = "2501.02119",
    archivePrefix = "arXiv",
    primaryClass = "hep-th",
    doi = "10.1007/JHEP07(2025)032",
    journal = "JHEP",
    volume = "07",
    pages = "032",
    year = "2025"
}

@article{Salcedo:2022aal,
    author = "Salcedo, Santiago Agui and Lee, Mang Hei Gordon and Melville, Scott and Pajer, Enrico",
    title = "{The Analytic Wavefunction}",
    eprint = "2212.08009",
    archivePrefix = "arXiv",
    primaryClass = "hep-th",
    doi = "10.1007/JHEP06(2023)020",
    journal = "JHEP",
    volume = "06",
    pages = "020",
    year = "2023"
}

@article{Senatore:2009cf,
    author = "Senatore, Leonardo and Zaldarriaga, Matias",
    title = "{On Loops in Inflation}",
    eprint = "0912.2734",
    archivePrefix = "arXiv",
    primaryClass = "hep-th",
    doi = "10.1007/JHEP12(2010)008",
    journal = "JHEP",
    volume = "12",
    pages = "008",
    year = "2010"
}

@article{McFadden:2011CC3,
   title={Cosmological 3-point correlators from holography},
   volume={2011},
   ISSN={1475-7516},
   url={http://dx.doi.org/10.1088/1475-7516/2011/06/030},
   DOI={10.1088/1475-7516/2011/06/030},
   number={06},
   journal={Journal of Cosmology and Astroparticle Physics},
   publisher={IOP Publishing},
   author={McFadden, Paul and Skenderis, Kostas},
   year={2011},
   month=jun, pages={030–030} }

@article{Maldacena:2011ngi,
   title={On graviton non-gaussianities during inflation},
   volume={2011},
   ISSN={1029-8479},
   url={http://dx.doi.org/10.1007/JHEP09(2011)045},
   DOI={10.1007/jhep09(2011)045},
   number={9},
   journal={Journal of High Energy Physics},
   publisher={Springer Science and Business Media LLC},
   author={Maldacena, Juan M. and Pimentel, Guilherme L.},
   year={2011},
   month=sep }

@article{Cachazo:2014xea,
    author = "Cachazo, Freddy and He, Song and Yuan, Ellis Ye",
    title = "{Scattering Equations and Matrices: From Einstein To Yang-Mills, DBI and NLSM}",
    eprint = "1412.3479",
    archivePrefix = "arXiv",
    primaryClass = "hep-th",
    doi = "10.1007/JHEP07(2015)149",
    journal = "JHEP",
    volume = "07",
    pages = "149",
    year = "2015"
}

@article{Farrow:2018yni,
    author = "Farrow, Joseph A. and Lipstein, Arthur E. and McFadden, Paul",
    title = "{Double copy structure of CFT correlators}",
    eprint = "1812.11129",
    archivePrefix = "arXiv",
    primaryClass = "hep-th",
    doi = "10.1007/JHEP02(2019)130",
    journal = "JHEP",
    volume = "02",
    pages = "130",
    year = "2019"
}

@article{Mei:2023jkb,
    author = "Mei, Jiajie",
    title = "{Amplitude Bootstrap in (Anti) de Sitter Space And The Four-Point Graviton from Double Copy}",
    eprint = "2305.13894",
    archivePrefix = "arXiv",
    primaryClass = "hep-th",
    month = "5",
    year = "2023"
}

@article{Calzetta:1986ey,
    author = "Calzetta, E. and Hu, B. L.",
    title = "{Closed Time Path Functional Formalism in Curved Space-Time: Application to Cosmological Back Reaction Problems}",
    reportNumber = "MdDP-PP-86-189",
    doi = "10.1103/PhysRevD.35.495",
    journal = "Phys. Rev. D",
    volume = "35",
    pages = "495",
    year = "1987"
}

@article{Ansari:2024pgq,
    author = "Ansari, Arhum and Banerjee, Pinak and Dhivakar, Prateksh and Jain, Sachin and Kundu, Nilay",
    title = "{Inflationary non-Gaussianities in alpha vacua and consistency with conformal symmetries}",
    eprint = "2403.10513",
    archivePrefix = "arXiv",
    primaryClass = "hep-th",
    doi = "10.1007/JHEP10(2024)147",
    journal = "JHEP",
    volume = "10",
    pages = "147",
    year = "2024"
}

@article{Witten:2022xxp,
    author = "Witten, Edward",
    title = "{A note on the canonical formalism for gravity}",
    eprint = "2212.08270",
    archivePrefix = "arXiv",
    primaryClass = "hep-th",
    doi = "10.4310/ATMP.2023.v27.n1.a6",
    journal = "Adv. Theor. Math. Phys.",
    volume = "27",
    number = "1",
    pages = "311--380",
    year = "2023"
}

@article{Chakraborty:2025izq,
    author = "Chakraborty, Tuneer and H, Ashik and Raju, Suvrat",
    title = "{Cosmological correlators in gravitationally-constrained de Sitter states}",
    eprint = "2507.15926",
    archivePrefix = "arXiv",
    primaryClass = "hep-th",
    month = "7",
    year = "2025"
}

@article{Kuchar:1970mu,
    author = "Kuchar, K.",
    title = "{Ground state functional of the linearized gravitational field}",
    doi = "10.1063/1.1665133",
    journal = "J. Math. Phys.",
    volume = "11",
    pages = "3322--3334",
    year = "1970"
}

@article{Chakraborty:2023yed,
    author = "Chakraborty, Tuneer and Chakravarty, Joydeep and Godet, Victor and Paul, Priyadarshi and Raju, Suvrat",
    title = "{The Hilbert space of de Sitter quantum gravity}",
    eprint = "2303.16315",
    archivePrefix = "arXiv",
    primaryClass = "hep-th",
    doi = "10.1007/JHEP01(2024)132",
    journal = "JHEP",
    volume = "01",
    pages = "132",
    year = "2024"
}

@article{Schwinger:1960qe,
    author = "Schwinger, Julian S.",
    title = "{Brownian motion of a quantum oscillator}",
    doi = "10.1063/1.1703727",
    journal = "J. Math. Phys.",
    volume = "2",
    pages = "407--432",
    year = "1961"
}

@article{Keldysh:1964ud,
    author = "Keldysh, L. V.",
    title = "{Diagram Technique for Nonequilibrium Processes}",
    doi = "10.1142/9789811279461_0007",
    journal = "Sov. Phys. JETP",
    volume = "20",
    pages = "1018--1026",
    year = "1965"
}

@article{Haehl:2025zfn,
    author = "Haehl, Felix and Rangamani, Mukund",
    title = "{Schwinger{\textendash}Keldysh Formalism}",
    doi = "10.1007/978-3-031-90352-6_3",
    journal = "Lect. Notes Phys.",
    volume = "1041",
    pages = "89--129",
    year = "2025"
}

@article{Haehl:2016pec,
    author = "Haehl, Felix M. and Loganayagam, R. and Rangamani, Mukund",
    title = "{Schwinger-Keldysh formalism. Part I: BRST symmetries and superspace}",
    eprint = "1610.01940",
    archivePrefix = "arXiv",
    primaryClass = "hep-th",
    doi = "10.1007/JHEP06(2017)069",
    journal = "JHEP",
    volume = "06",
    pages = "069",
    year = "2017"
}

@article{Sleight:2021plv,
    author = "Sleight, Charlotte and Taronna, Massimo",
    title = "{From dS to AdS and back}",
    eprint = "2109.02725",
    archivePrefix = "arXiv",
    primaryClass = "hep-th",
    doi = "10.1007/JHEP12(2021)074",
    journal = "JHEP",
    volume = "12",
    pages = "074",
    year = "2021"
}

@article{MdAbhishek:2025dhx,
    author = "Abhishek, Md. and Sleight, Charlotte and Taronna, Massimo",
    title = "{Cosmological Correlators in Gauge Theory and Gravity from EAdS}",
    eprint = "2509.09536",
    archivePrefix = "arXiv",
    primaryClass = "hep-th",
    month = "9",
    year = "2025"
}

@article{Heckelbacher:2022hbq,
    author = "Heckelbacher, Till and Sachs, Ivo and Skvortsov, Evgeny and Vanhove, Pierre",
    title = "{Analytical evaluation of cosmological correlation functions}",
    eprint = "2204.07217",
    archivePrefix = "arXiv",
    primaryClass = "hep-th",
    reportNumber = "IPhT-t22/02, LMU-ASC 13/22",
    doi = "10.1007/JHEP08(2022)139",
    journal = "JHEP",
    volume = "08",
    pages = "139",
    year = "2022"
}

@article{Thavanesan:2025kyc,
    author = "Thavanesan, Ayngaran",
    title = "{No-go Theorem for Cosmological Parity Violation}",
    eprint = "2501.06383",
    archivePrefix = "arXiv",
    primaryClass = "hep-th",
    month = "1",
    year = "2025"
}

@article{Haehl:2017qfl,
    author = "Haehl, Felix M. and Loganayagam, R. and Narayan, Prithvi and Rangamani, Mukund",
    title = "{Classification of out-of-time-order correlators}",
    eprint = "1701.02820",
    archivePrefix = "arXiv",
    primaryClass = "hep-th",
    doi = "10.21468/SciPostPhys.6.1.001",
    journal = "SciPost Phys.",
    volume = "6",
    number = "1",
    pages = "001",
    year = "2019"
}

@article{Haehl:2017eob,
    author = "Haehl, Felix M. and Loganayagam, R. and Narayan, Prithvi and Nizami, Amin A. and Rangamani, Mukund",
    title = "{Thermal out-of-time-order correlators, KMS relations, and spectral functions}",
    eprint = "1706.08956",
    archivePrefix = "arXiv",
    primaryClass = "hep-th",
    doi = "10.1007/JHEP12(2017)154",
    journal = "JHEP",
    volume = "12",
    pages = "154",
    year = "2017"
}

@article{Chaudhuri:2018ymp,
    author = "Chaudhuri, Soumyadeep and Chowdhury, Chandramouli and Loganayagam, R.",
    title = "{Spectral Representation of Thermal OTO Correlators}",
    eprint = "1810.03118",
    archivePrefix = "arXiv",
    primaryClass = "hep-th",
    doi = "10.1007/JHEP02(2019)018",
    journal = "JHEP",
    volume = "02",
    pages = "018",
    year = "2019"
}

@misc{github,
  howpublished = {\url{https://github.com/JMarshall-Dur/12-25-Cosmo-Gauge-Fields}},
}

@article{Donnelly:2015hta,
    author = "Donnelly, William and Giddings, Steven B.",
    title = "{Diffeomorphism-invariant observables and their nonlocal algebra}",
    eprint = "1507.07921",
    archivePrefix = "arXiv",
    primaryClass = "hep-th",
    reportNumber = "NSF-KITP-15-133",
    doi = "10.1103/PhysRevD.93.024030",
    journal = "Phys. Rev. D",
    volume = "93",
    number = "2",
    pages = "024030",
    year = "2016",
    note = "[Erratum: Phys.Rev.D 94, 029903 (2016)]"
}

@article{Loganayagam:2025jmw,
    author = "Loganayagam, R. and Shetye, Omkar",
    title = "{Influence phase of a dS observer. Part II. Electromagnetism}",
    eprint = "2503.00135",
    archivePrefix = "arXiv",
    primaryClass = "hep-th",
    doi = "10.1007/JHEP08(2025)027",
    journal = "JHEP",
    volume = "08",
    pages = "027",
    year = "2025"
}
\end{document}